\documentclass{WileyMSP-template}

\usepackage{hyperref}
\usepackage{braket}
\usepackage{amsmath}
\usepackage[english]{babel}
\usepackage{color,soul}
\usepackage{commath}
\usepackage{fixltx2e}

\begin{document}

\rhead{\includegraphics[width=2.5cm]{vch-logo.png}}

\title{Harnessing the Quantum Behavior of Spins on Surfaces}

\maketitle


\author{Yi Chen$^\dagger$}
\author{Yujeong Bae$^{\dagger,*}$}
\author{Andreas J. Heinrich}



\begin{affiliations}
Center for Quantum Nanoscience, Institute for Basic Science (IBS), Seoul 03760, Korea\\
Department of Physics, Ewha Womans University, Seoul 03760, Korea\\
$^\dagger$ These authors contributed equally.\\
$^*$ Email Address: bae.yujeong@qns.science

\end{affiliations}



\keywords{Quantum sensing, spins on surfaces, quantum simulation, quantum manipulation, scanning tunneling microscopy, quantum nanoscience}

\begin{abstract}



The desire to control and measure individual quantum systems such as atoms and ions in a vacuum has led to significant scientific and engineering developments in the past decades that form the basis of today's quantum information science. Single atoms and molecules on surfaces, on the other hand, are heavily investigated by physicists, chemists, and material scientists in search of novel electronic and magnetic functionalities. These two paths crossed in 2015 when it was first clearly demonstrated that individual spins on a surface can be coherently controlled and read out in an all-electrical fashion. The enabling technique is a combination of scanning tunneling microscopy (STM) and electron spin resonance (ESR), which offers unprecedented coherent controllability at the Angstrom length scale. This review aims to illustrate the essential ingredients that allow the quantum operations of single spins on surfaces. Three domains of applications of surface spins, namely quantum sensing, quantum control, and quantum simulation, are discussed with physical principles explained and examples presented. Enabled by the atomically-precise fabrication capability of STM, single spins on surfaces might one day lead to the realization of quantum nanodevices and artificial quantum materials at the atomic scale.

\end{abstract}


\section{Introduction}

The last half-century has witnessed tremendous advances in the control and detection of individual quantum systems. Ions electromagnetically trapped in vacuum, for example, provide effective two-level systems that enable sophisticated quantum protocols via high-fidelity optical initialization, manipulation, and readout \cite{doi:10.1063/1.5088164}. Single photons with quantum information encoded in their internal degrees of freedom such as polarization and positions can be individually generated, controlled, and detected \cite{Flamini_2018}. In the solid state, Josephson-junction-based superconducting circuits provide a highly competitive platform \cite{Arute:2019ts} where two singled-out energy levels can be controlled and read out via microwave photons \cite{doi:10.1146/annurev-conmatphys-031119-050605}. Spins in solid-state materials constitute another family of individual quantum systems where spin states with naturally quantized energy levels can be manipulated via magnetic fields. Physical realizations of solid-state spin systems include gate-defined quantum dots, single dopants in semiconductors such as phosphorus donors in silicon, and single defects in insulators such as nitrogen-vacancy centers in diamond \cite{RevModPhys.85.961, Chatterjee:2021tn,Wolfowicz:2021uc}. Harnessing quantum resources at the nanoscale gives birth to the discipline of quantum-coherent nanoscience that may bring forth useful quantum nanodevices ``at the bottom'' \cite{Heinrich2021,PlentyofRoom}. \hfill\break

In this review, we focus on a new class of quantum spin systems, individual atomic and molecular spins on surfaces, which has the potential to produce quantum functionalities at the atomic scale. An STM is used to access this tiny length scale, where a sharp metallic tip is positioned in nanometer proximity to spin carriers on a surface to probe their properties through tunneling electrons \cite{Chen:2008vn}. Experiments with single atomic spins on material surfaces date back to the early days of low-temperature STM, when Yu-Shiba-Rusinov states and a Kondo resonance were measured in individual magnetic atoms on bulk superconductors \cite{Yazdani:1997wc} and noble metals \cite{Madhavan567,PhysRevB.61.9990}, respectively. STM has also been extensively used on single molecular spins to characterize molecular structures \cite{GIMZEWSKI1987267, Hipps:1996tz, Lu:1996td}, probe and modify their electronic and magnetic properties \cite{Aidi:2005wv, Coronado:2020vm, Moreno-Pineda:2021wt}, and investigate their classical and quantum applications \cite{Moreno-Pineda:2021wt, sessoli2000, Gaita-Arino:2019ti}. 
Following early STM experiments on single spins, the desire to further control and magnetically image individual spins has prompted the developments of new local probe techniques such as spin-polarized STM \cite{Wiesendanger2009}, inelastic-tunneling-based spin-flip spectroscopy \cite{Heinrich466}, electrical pump-probe measurements \cite{Loth1628}, and various innovative forms of force and magnetic microscopy \cite{Rugar2004, Kaiser:2007wq,Vasyukov:2013ve,Grinolds:2013wq}. 
\hfill\break

An exciting development in recent years is the incorporation of coherent spin control in atomic-scale microscopy. The need to coherently manipulate and measure individual spins at this length scale necessitates all-electrical protocols with Angstrom precision. These stringent requirements are met by drawing on powerful methodologies from material and quantum sciences, i.e., STM's abilities to build nanostructures atom-by-atom and selectively sense individual spin-carrying atoms, as well as ESR's ability to coherently control electron spin states via electromagnetic waves. This unique combination has so far enabled the quantum control of single electron spins of atoms \cite{Baumann417,Yang509, PhysRevB.103.155405, Seiferteabc5511, Veldman964} and molecules \cite{Zhang:2021te}, as well as the manipulation of single nuclear spins through hyperfine interactions \cite{Yang:2018aa}. Quantum coherence can be increased using singlet-triplet states of a coupled spin system \cite{Baeeaau4159}, allowing the observation of a free coherent evolution in the singlet-triplet basis \cite{Veldman964}. The simultaneous control of two electron spins in an engineered atomic structure has recently been realized, shedding light on the debated ESR-STM driving mechanism and showing the potential for multi-spin quantum protocols on a surface \cite{2021arXiv210809880P}. \hfill\break

Quantum sensing with ESR-STM harnesses its unprecedented combination of Angstrom-scale spatial resolution and tens of nanoelectronvolt (neV) energy resolution. Unlike traditional STM-based spectroscopy where the energy resolution is limited to about 0.5 meV at 1 K due to the thermal Fermi-Dirac broadening of tunneling electrons \cite{PhysRev.165.821, Song2010, Ast2016, PhysRevLett.107.076804}, the energy resolution of ESR-STM is not limited by electronic thermal broadening because the energy of the tunneling electrons is not the measured quantity. Instead, the energy in ESR-STM is benchmarked against the frequency of a supplied radio-frequency (RF) electromagnetic wave, which drives the spin resonance of a surface spin and regulates the tunnel current flow through tunneling mangetoresistance. The energy resolution of ESR-STM is therefore given by the performance of this ``spin regulator'', limited only by the coherence properties of the surface spin. Using this sensitive ESR-STM measurement scheme, quantum sensors based on single atomic spins on surfaces have been used to reveal the dipolar magnetic fields from single-atom magnets at atomic proximity \cite{Choi:2017aa,Natterer:2017aa,Singha:2021tj} as well as binding-site-dependent hyperfine interactions \cite{Willke336}. Using single spin sensors on a surface or a tip, we expect that valuable insight can be provided into the microscopic magnetic interactions in low-dimensional materials, strongly correlated systems, and spintronic and magnetic devices. \hfill\break

Quantum simulation with artificial nanomaterials built atom-by-atom serves as another key application. Various atomic and molecular spins can act as building blocks for bottom-up spin nanostructures \cite{Khajetoorians:2019wq,RevModPhys.91.041001}. Local interactions between spins can be controlled by the atomically precise positioning with STM and quantified by the neV-resolved ESR-STM spectroscopy. Global interactions can be provided by substrates, including substrate-mediated exchange interactions and substrate-induced spin-orbit couplings or superconducting pairing interactions. The magnetic field of the STM tip provides a convenient tuning knob that only affects the local spin under the apex of the tip and has an adjustable field strength that may even exceed the external magnetic field.
\hfill\break

This review is organized as follows. In section 2, we introduce the essential ingredients that enable the coherent control and readout of single spins on surfaces. Sections 3 to 5 focus on different domains of applications using spins on surfaces. Section 3 discusses quantum sensing using single atomic spins on surfaces. Section 4 describes the efforts to improve the quantum coherence and extend quantum control from single to multiple spins on surfaces. Section 5 discusses quantum simulation using artificially constructed 1D and 2D spin arrays on surfaces. The final section 6 provides concluding remarks.

\section{Key Ingredients for Coherent Single Spin Experiments on Surfaces}

\label{section: 2}

In this section, we review the key ingredients that enable the successful performance of ESR-STM at the single spin level (for a summary, see Table \ref{table:ingredients}). After a brief introduction to spins and spin resonance in section \ref{section: 2.1}, section \ref{section: 2.2} provides examples of coherently controlled surface spins and their quantum functionalities, and section \ref{section: 2.3} outlines the experimental setup required for performing these experiments. Sections \ref{section: 2.4}, \ref{section:driving}, and \ref{section: detection} focus on three main developments that enabled the harnessing of quantum behavior of single surface spins, namely spin initialization (section \ref{section: 2.4}), spin control (section \ref{section:driving}), and spin detection (section \ref{section: detection}). Extending from a single spin to multi-spin structures, section \ref{section:spin-spin interaction} discusses a key advantage of the STM-based approach, i.e., the atomically precise control of spin-spin interactions via atom manipulation. Finally, section \ref{section: tipfield} highlights a unique tunability of STM-based setups, that is, a sizeable, highly local magnetic field from the STM tip. 

\subsection{Spins as Quantum Information Carriers}

\label{section: 2.1}

Spins, as natural physical observables with discrete energy levels, have been regarded as a promising platform for realizing quantum functionalities since the dawn of quantum information science \cite{PhysRevA.57.120, Kane:1998wh}. This section explains the fundamental concept underlying the use of spins as quantum information carriers. Consider the simplest case, an idealized electron spin-1/2 placed in a static magnetic field $\boldsymbol{B}_{\mathrm{ext}}= (0, 0, -B_{\mathrm{ext}})$ along the $-z$-direction. The spin Hamiltonian reads
\begin{equation}
    H = -\boldsymbol{m} \cdot \boldsymbol{B}_{\mathrm{ext}} = -g \mu_{\mathrm{B}} B_{\mathrm{ext}} S_z.
    \label{eq: 1steq}
\end{equation}
Here $\boldsymbol{m} = -g \mu_{\mathrm{B}} \boldsymbol{S}$ is the spin's magnetic moment, $g$ is the magnitude of the electron spin's $g$ factor, $\mu_{\mathrm{B}}$ is the Bohr magneton, and the spin operators are taken as unitless (without $\hbar$) throughout this review. The static magnetic field splits the spin $\ket{\uparrow}$ and $\ket{\downarrow}$ states, or equivalently, the $\ket{0}$ and $\ket{1}$ states, by the Zeeman energy $g \mu_\mathrm{B} B_{\mathrm{ext}} = \hbar \omega_0$, where $\omega_0$ is called the Larmor frequency. A general spin-1/2 state can be visualized as a vector in a Bloch sphere (Figure \ref{fig:atoms}a, left), parametrized as 
\begin{equation}
    \ket{\psi} = \cos \frac{\theta}{2} \ket{0} + e^{i\phi} \sin \frac{\theta}{2} \ket{1},
    \label{eq: blochsphere}
\end{equation}
where $\theta$ and $\phi$ are the spherical coordinates that indicate the spin direction. The use of spin-1/2 as qubits then relies on the creation of a well-defined initial state such as $\ket{0}$, the control of the spin to reach an arbitrary state $ \ket{\psi}$, and the detection of the spin state, for example, by projection onto a certain axis \cite{DiVincenzo:2000vw}. \hfill\break

The control of spin states is typically achieved by an oscillating magnetic field
\begin{equation}
    \boldsymbol{B}_1 = 2 B_1 \cos (\omega t + \alpha) \boldsymbol{e}_X,
    \label{eq: B1field}
\end{equation}
where the magnetic field has an angular frequency $\omega$ near the Larmor frequency of the spin ($\omega \approx \omega_0$) in the RF range, $\alpha$ is the phase of the magnetic field, and $t$ is the time. Throughout this review, we use $\boldsymbol{e}_{X, Y, Z}$ and $\boldsymbol{e}_{x, y, z}$ to indicate the unit vectors in the lab frame and in the rotating frame, respectively (see below). Inserting both $\boldsymbol{B}_{\mathrm{ext}}$ and $\boldsymbol{B}_1$ fields into the Hamiltonian in Equation \ref{eq: 1steq} yields the Schr$\ddot{\mathrm{o}}$dinger equation
\begin{equation}
  i\hbar \frac{d \ket{\psi(t)}}{dt} = \frac{\hbar}{2} \begin{bmatrix}
    -\omega_0 & 2 \Omega \cos (\omega t + \alpha) \\
    2 \Omega \cos (\omega t + \alpha) & \omega_0
  \end{bmatrix}
  \ket{\psi(t)},
  \label{eq: ESR-Lab-SE-exact}
\end{equation}
where we have converted the $B_1$ field strength into an angular frequency $\Omega = g \mu_\mathrm{B} B_1/\hbar$. The oscillating field can be decomposed into a rotating wave and a counter-rotating wave as
\begin{equation}
    \boldsymbol{B}_1 = 2 B_1 \cos (\omega t + \alpha) \boldsymbol{e}_X = B_1[\cos (-\omega t - \alpha) \boldsymbol{e}_X + \sin (-\omega t - \alpha) \boldsymbol{e}_Y] + B_1[\cos (\omega t + \alpha) \boldsymbol{e}_X + \sin (\omega t + \alpha) \boldsymbol{e}_Y].
    \label{eq: RWA}
\end{equation}
For a nearly resonant field ($\omega \approx \omega_0$), the first, rotating component of the $B_1$ field mostly follows the spin's Larmor precession and acts to induce an additional spin rotation about the rotating field axis. This causes a periodic oscillation of the spin state between $\ket{0}$ and $\ket{1}$, which is the origin of the spin resonance. The second, counter-rotating field component in Equation \ref{eq: RWA} rotates at a very fast rate of $\omega + \omega_0$ relative to the spin's Larmor precession, and its effect can often be ignored (known as the rotating wave approximation) \cite{Levitt:2015wh}. The Schr$\ddot{\mathrm{o}}$dinger equation under the rotating wave approximation reads
\begin{equation}
  i\hbar \frac{d \ket{\psi(t)}}{dt} = \frac{\hbar}{2} \begin{bmatrix}
    -\omega_0 & \Omega e^{i (\omega t + \alpha)} \\
    \Omega e^{-i (\omega t + \alpha)} &  \omega_0
  \end{bmatrix}
  \ket{\psi(t)}.
  \label{ESR-Lab-SE}
\end{equation}
The physics of spin resonance becomes clearest in a reference frame that itself rotates at the angular frequency $\omega$ of the oscillating $B_1$ field. The spin state in this rotating frame is 
\begin{equation}
    \ket{\tilde{\psi}(t)} = \exp (-i \omega t S_z)\ket{ \psi(t)},
    \label{eq: RotatingFrame}
\end{equation}
whose time evolution is governed by a \textit{time-independent} Hamiltonian
\begin{equation}
i\hbar \frac{d \ket{\tilde{\psi}(t)}}{dt} = 
    \frac{\hbar}{2} \begin{bmatrix}
    -\omega_0 + \omega & \Omega e^{i \alpha} \\
    \Omega e^{-i \alpha} & \omega_0 - \omega
  \end{bmatrix}
  \ket{\tilde{\psi}(t)}.
\end{equation}
At resonance ($\omega = \omega_0$), the diagonal terms of the Hamiltonian vanish, and the off-diagonal terms $\Omega e^{\pm i \alpha}$ dominate the time evolution by periodically flipping the spin between states $\ket{0}$ and $\ket{1}$. Mathematically, consider an initial state $\ket{\tilde{\psi}(0)} = \ket{0}$ under a resonant RF wave at $\omega = \omega_0$. The spin state after time $\tau$ becomes
\begin{equation}
  \ket{\tilde{\psi}(\tau)} = \cos (\frac{1}{2} \Omega \tau) \ket{0} + e^{i(3\pi/2 - \alpha)} \sin (\frac{1}{2} \Omega \tau) \ket{1},
\end{equation}
which corresponds to a spin state at an angle $\theta = \Omega \tau$ and $\phi = 3\pi/2 - \alpha$ in the Bloch sphere in the rotating frame. A $B_1$ pulse at the resonance frequency thus controls the spin state by rotating it around an axis in the $xy$ plane (Figure \ref{fig:atoms}a), where the field strength and the pulse duration jointly determine the rotation angle $\Omega \tau$, and the field phase $\alpha$ determines the rotation axis. If the spin population in state $\ket{0}$ is measured at the end of the pulse, a periodic oscillation will appear
\begin{equation}
  P_{\ket{0}}(\tau) = \cos^2 (\frac{1}{2} \Omega \tau) = \frac{1}{2} + \frac{1}{2} \cos (\Omega \tau).
\end{equation}
The oscillation of populations under (nearly) resonant driving is known as the Rabi oscillation. At $\Omega \tau = \pi/2$ and $\alpha = 0$, for example, the pulse acts to rotate the spin state by 90$^\circ$ around the $x$ axis and turn both populations of states $\ket{0}$ and $\ket{1}$ to 50\%. At off resonance ($\omega \ne \omega_0$), the rotation axis is tilted away from the $xy$ plane, and the spin rotation rate, generally known as the Rabi rate, becomes  
\begin{equation}
  \Omega_{\mathrm{R}} = \sqrt{\Omega^2 + (\omega-\omega_0)^2}.
\end{equation}
Finally, if we choose to view the spin evolution in the original lab frame (which does not rotate), the spin motion needs to be combined with a precession of an angular velocity $-\omega$ around the $z$ axis (see Equation \ref{eq: RotatingFrame}). \hfill\break

After spin manipulation, readout can be performed by projecting the spin into $z$ direction (as in quantum dot experiments \cite{Elzerman:2004wt,Morello:2010vm}) or into the $xy$ plane (as in ensemble ESR experiments) via spin-to-charge or spin-to-photon conversions. In an ESR-STM setup, as we shall see in section \ref{section: detection},  the measured signals contain both spin-$z$ and $xy$ contributions, which can be distinguished by their different lineshapes. \hfill\break 

In reality, an atomic spin, even when isolated, differs from the above idealistic case in that (1) both spin and orbital angular momenta are present and coupled through spin-orbit coupling, (2) multiple electrons in different orbitals can contribute to the atomic spin, and (3) many nuclei carry a nuclear spin that couples to the electron spin via the hyperfine interaction. In addition, if the spin is placed in a molecular or solid-state environment, its orbital properties can be significantly modified by the surrounding ligands \cite{abragam2012electron}. Despite these complications, an effective spin Hamiltonian that approximates orbital effects as spin operators is often sufficient for describing the spin properties \cite{abragam2012electron}. The effective Hamiltonian for a single spin can be generally written as
\begin{equation}
    H = \mu_\mathrm{B} \boldsymbol{B}_{\mathrm{ext}} \cdot \mathbf{g} \cdot \boldsymbol{S} + \boldsymbol{S} \cdot \mathbf{D} \cdot  \boldsymbol{S} + \mu_{\mathrm{B}} \boldsymbol{B}_{\mathrm{ext}} \cdot \mathbf{g}_{n} \cdot \boldsymbol{I} + \boldsymbol{S} \cdot \mathbf{A} \cdot \boldsymbol{I} + \boldsymbol{I} \cdot \mathbf{P} \cdot \boldsymbol{I},
    \label{eq: spin-Hamiltonian}
\end{equation}
where $\boldsymbol{S}$ is an effective spin operator, $\boldsymbol{B}_{\mathrm{ext}}$ is the external magnetic field, $\mathbf{g}$ is the $g$-factor tensor, and $\mathbf{D}$ symbolically represents the magnetic anisotropy that lifts the spin degeneracy due to orbital contributions. Tensors are denoted in a bold, non-italic font throughout this review, while vectors are denoted in a bold, italic font. The latter three nucleus-related terms represent the nuclear Zeeman energy, the hyperfine interaction, and the electric quadrupole interaction, respectively, and they will be discussed in more detail in section \ref{section: nuclear-sensing}. Spin resonance of a realistic spin can be understood as the resonant transition between two spin eigenstates where the transition matrix element is non-zero (i.e., containing spin components allowed by selection rules). \hfill\break

The detailed form of the magnetocrystalline anisotropy term $\boldsymbol{S} \cdot \mathbf{D} \cdot \boldsymbol{S}$ can be obtained through group theory analyses, for example, by using the Stevens operators $\sum_{k,q} B_k^q O_k^q (S)$ \cite{abragam2012electron, Stevens_1952, Rudowicz_2004}. The commonly used axial ($D$) and rhombic ($E$) anisotropy terms are related to the second-order Stevens coefficients by $D= 3 B_2^0$ and $E= B_2^2$. The corresponding Hamiltonian terms are defined as
\begin{equation}
    B_2^0 O_2^0 = D [S_z^2 - \frac{1}{3}S(S+1)], \ \ B_2^2 O_2^2 = E(S_x^2 - S_y^2),
    \label{eq: stevens}
\end{equation}
where $x$, $y$, and $z$ indicate the principal axes of the anisotropy tensor, $\mathbf{D}$. A list of $D$ (and $E$, when applicable) values for spins mentioned in this review can be found in Table \ref{table:spins}. The role of the axial anisotropy term ($D$) is to split the spin levels based on $|S_z|$. For example, spins with $D < 0$ have an easy axis along $z$, i.e., the lowest energy states have the maximal $|S_z|$ ($S_z = \pm S$) while higher energy states are spin states with smaller $|S_z|$. This state configuration creates an effective energy barrier (the ``anisotropy barrier'') for spin-reversal between the two lowest-energy states, hence increasing the spin lifetime \cite{Gatteschi:2011uy} (an anisotropy barrier is depicted above the Fe atom in Figure \ref{fig:atoms}a). On top of the effect of axial anisotropy, a non-zero rhombic term ($E$) can provide further splitting of the spin states. In addition, it mixes the spin composition of different states, allowing for additional spin-reversal mechanisms such as under-barrier transitions and quantum tunneling of magnetization \cite{Gatteschi:2011uy}. Due to the lack of mirror symmetry at a surface, spins on surfaces experience at least the axial anisotropy. The magnetocrystalline anisotropy, however, cannot split conjugated doublets in half-integer spin atoms due to the time-reversal symmetry \cite{abragam2012electron}. As a result, states in spin-1/2 systems are not split by anisotropy terms (another way to see this is that for spin-1/2, $S_i^2$ is just a constant (with $i= x, y, z$) and may be ignored).

\subsection{Examples of Coherently Controlled Spins on Surfaces}

\label{section: 2.2}

Single electron spins on surfaces, unlike many other quantum systems, can exist in a wide range of hosts, including transition metal and rare earth adatoms, molecules with magnetic atomic centers or radicals, and spin defects. Commonly used atomic species frequently have isotopes with nonzero nuclear spins, adding to the diversity of surface spins. Most experiments conducted in the ESR-STM setting to date, however, have focused on adatoms, partially owing to surface scientists' familiarity with them. The preparation of spin-carrying adatoms on surfaces follows standard surface-science procedures in which single magnetic atoms and molecules are vacuum-deposited onto a cold substrate in the STM or a cooling stage. In practice, a substrate temperature of less than 100 K is usually sufficient to prevent atoms from clustering. \hfill\break

A specific substrate, two-monolayer (2ML) MgO on Ag(100), has been the surface of choice for ESR-STM for good reason. The thin MgO insulator acts as a decoupling layer between the spins and the metallic substrate, reducing interactions and decoherence, while MgO of 2ML to 4ML thickness is still conductive enough for stable STM operations \cite{Paul:2017aa}. MgO/Ag(100) can also be relatively easily grown into thin crystalline films that allow for atom manipulation \cite{PhysRevLett.87.276801}. Although an early report of ESR-STM \cite{Baumann417} attributed the ESR driving to MgO's crystal fields, subsequent theoretical and experimental investigations show that the role of crystal fields can be replaced by the tip's magnetic field gradient, and the latter likely dominates \cite{PhysRevB.96.205420} (a more complete discussion on the proposed ESR-STM driving mechanisms can be found in section \ref{section:driving}). ESR-STM driving should then impose no special requirements on the substrates.\hfill\break

Other substrates are under active investigation for use in ESR-STM. A suitable substrate should maintain spin integrity, provide sufficient isolation from the decoherence sources (such as substrate conduction electrons), allow sufficient ESR driving, and permit atom manipulation for multi-spin studies. Among passivated metal surfaces \cite{Chen_2008}, 1ML Cu$_2$N on Cu(100) is a popular substrate for spin-related studies (see, for example, section \ref{section: 5.1}), yet ESR-STM for adatoms on Cu$_2$N has not been realized. The reason could be due to Cu$_2$N's lower decoupling effect compared to MgO, decoherence caused by substrate nuclear spins (nearly all Cu and N atoms have nonzero nuclear spins), or simply insufficient experimental trials. The decoupling effects of Cu$_2$N and MgO can be estimated by considering the scattering of a surface spin by conduction electrons starting and ending in the metal substrates, which induces an effective conductance of $\sim 3 \, \mu \mathrm{S}$ for Cu-binding-site Mn on Cu$_2$N \cite{Loth:2010aa}, two orders of magnitude higher than that of Fe on 2ML MgO ($\sim 40 \, \mathrm{nS}$) \cite{Paul:2017aa} (however, we note that the adatoms on N binding sites of Cu$_2$N can have much longer lifetime than those on Cu binding sites \cite{Max:vs}). NaCl is another frequently used decoupling substrate for molecules. Although we found that atoms can often be embedded into the NaCl lattice due to its relatively large lattice constant, we expect NaCl to be a promising substrate for spin-carrying molecules. New types of substrates that better isolate the surface spins from low-energy excitations of substrates are worth future investigation. These substrates may include bulk semiconductors such as silicon where excellent coherence has been demonstrated \cite{RevModPhys.85.961}, superconductors or correlated insulators where no low-energy excitations exist below a threshold energy, and novel platforms such as van der Waals heterostructures where electrical contacts and isolation layers can be individually engineered. Furthermore, surface spins other than adatoms, such as spin defects \cite{Wolfowicz:2021uc} or skyrmions \cite{PhysRevLett.127.067201,Romming636}, may introduce new spin control and detection strategies on these new substrates (Table \ref{table:ingredients}).

\hfill\break

\begin{figure}[pthb]
  \centering
  \includegraphics[width=1 \linewidth]{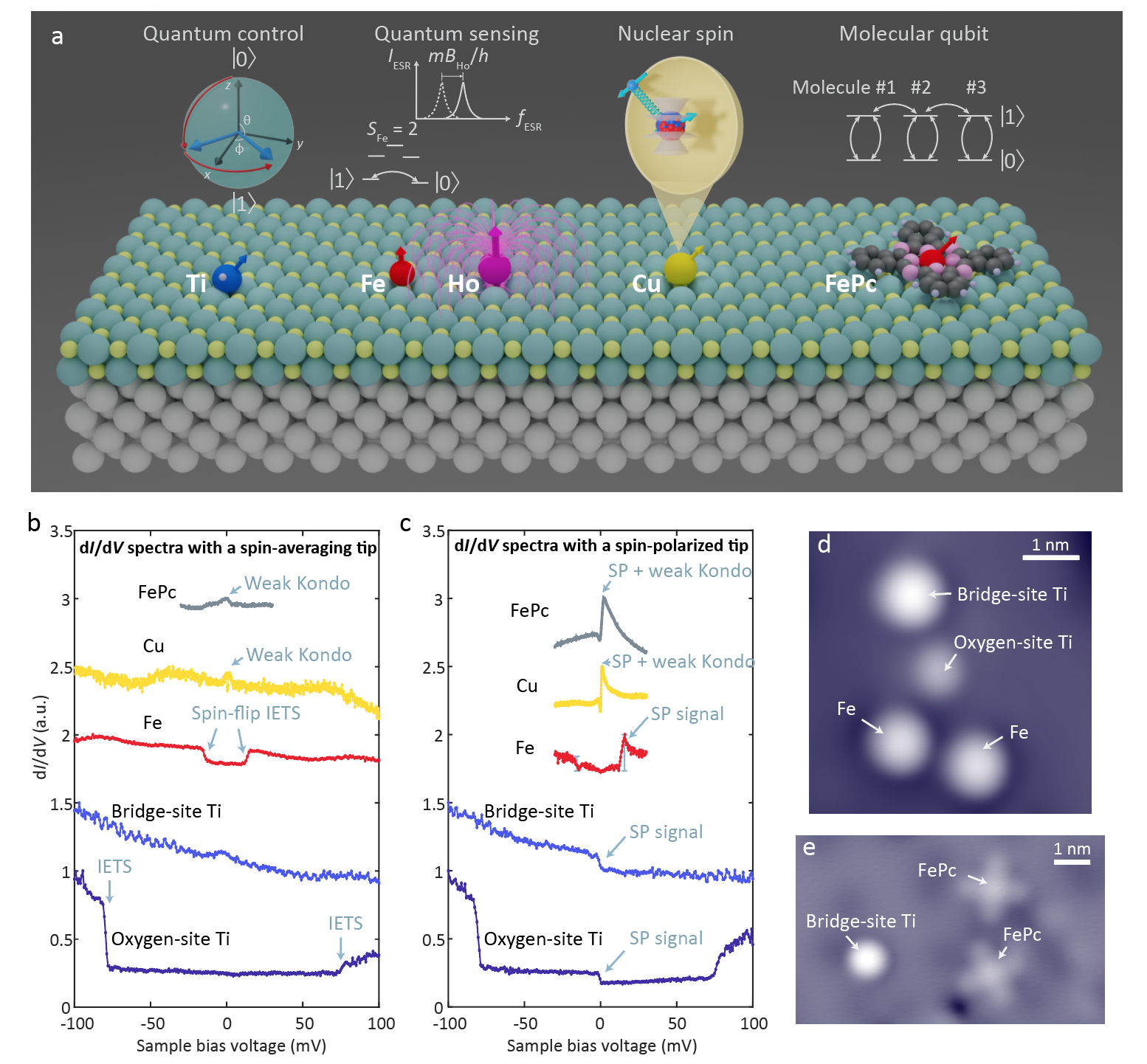}%
  \caption{Coherently controlled spins on 2ML MgO/Ag(100). a) Quantum functionalities of surface spins. Left: Quantum control of a single atomic spin-1/2 localized in a hydrogenated Ti atom. Middle left: Quantum sensing of magnetic fields from a single-atom magnet (Ho) using a spin-2 Fe atom. Middle right: Control of coupled electron and nuclear spins in a Cu atom. Right: Spin-1/2 FePc molecule and its potential applications in a molecular quantum network. O (Mg) atoms of the substrate are depicted in yellow (green). b) Identification of typical surface spins on MgO using STM d$I$/d$V$ spectra with spin-averaging tips (i.e., non-spin-polarized tips) at a typical magnetic field ($\le$ 1 T) for ESR-STM measurements. Successive d$I$/d$V$ spectra are shifted by +0.5 a.u. for clarity. c) Same as (b) but with spin-polarized (SP) tips. The tip's spin polarization (a critical factor in its performance as an ESR tip) can be characterized by the heights of the IETS steps, which lie at around the zero bias for spin-1/2 systems (all except Fe) and at the spin-flip energies for higher spins (Fe, see text). Whether the IETS conductance is higher at positive bias or negative bias indicates the relative orientation between the tip spin and the surface spin, see section \ref{section: QM detection} and Ref. \cite{Yang:2018aa}. Successive d$I$/d$V$ spectra are shifted by +0.5 a.u. for clarity. d,e) Examples of STM topograph of isolated Ti atoms, Fe atoms, and FePc molecules on MgO.}
  \label{fig:atoms}
\end{figure}

Figure \ref{fig:atoms}a depicts a gallery of single spins on 2ML MgO/Ag(100) (with the spin properties summarized in Table \ref{table:spins}). Single hydrogenated Ti atoms, hereafter referred to as Ti, host spin-1/2 and provide a two-level quantum system where continuous-wave (CW) \cite{PhysRevLett.119.227206} and pulsed spin manipulation \cite{Yang509} were demonstrated (as sketched on the left side of Figure \ref{fig:atoms}a). In contrast, single Fe atoms on MgO host spin-2 with a large magnetic moment ($\sim 5.44 \mu_\mathrm{B}$ \cite{Choi:2017aa}) that is fixed perpendicular to the sample plane by a large easy-axis anisotropy $D = -4.7$ meV (as reflected in its d$I$/d$V$ spectra in Figure \ref{fig:atoms}b,c, see below) \cite{PhysRevLett.115.237202}. Despite the anisotropy barrier, an ESR transition of Fe can be driven between the two lowest energy levels due to intermixing of the spin eigenstates \cite{Baumann417}. Because of these properties, Fe is a sensitive quantum sensor of local out-of-plane magnetic fields. Fe sensors have been used to investigate the magnetic fields of nearby single-atom magnets including Ho \cite{Natterer:2017aa}, Dy \cite{Singha:2021tj}, or a second Fe atom \cite{Choi:2017aa}. Single Cu atoms, on the other hand, provide a model system with strongly coupled electronic and nuclear spins where the nuclear spin can be initialized by pumping the electron spin \cite{Yang:2018aa}. Finally, as shown on the right side of Figure \ref{fig:atoms}a, single-molecular spin-1/2 localized in FePc has the potential to be used as a molecular qubit \cite{Zhang:2021te}. The aforementioned building blocks have been used to construct a variety of artificial spin structures, each of which makes use of different quantum functionalities of the surface spins. These structures are summarized in Table \ref{table:experiment} and will be the subjects of sections 3--5. \hfill\break

Different atomic species on 2ML MgO/Ag can be readily identified using STM topographic imaging and d$I$/d$V$ spectroscopy as shown in Figure \ref{fig:atoms}b--e. Ti atoms deposited on 2ML MgO/Ag substrate have two preferential binding sites: the oxygen site (i.e., on top of an oxygen atom of MgO) and the bridge site (i.e., between two oxygen atoms). Under typical scanning conditions of $V_\mathrm{DC} = 0.1$ V and $I_\mathrm{t} = $ 20 pA, a bridge-site Ti atom has a large apparent height (around 1.9 \AA), whereas an oxygen-site Ti atom has a lower local density of states and thus a lower apparent height (around 1.0 \AA) (Figure \ref{fig:atoms}d) (throughout this review, a positive bias voltage means a positive voltage applied to the sample side). Ti atoms can be laterally manipulated along the MgO lattice directions by parking the tip 1.5 to 2 lattice constants ahead of Ti at a setpoint of $V_\mathrm{DC} = 0.1$ V and $I_\mathrm{t} = $ 20 pA and then approaching the tip towards the MgO surface by 0.3 nm at around $V_\mathrm{DC} = 0.3$ V to pull Ti towards the tip. In STM d$I$/d$V$ spectroscopy, a bridge-site Ti atom shows a rather featureless d$I$/d$V$ spectrum with a gradual decrease of d$I$/d$V$ at increasing bias voltage, while an oxygen-site Ti atom exhibits strong inelastic tunneling spectroscopic (IETS) steps at around $\pm 80$ mV that may be related to the excitation of a vibrational mode or into a higher-lying orbital state (Figure \ref{fig:atoms}b). IETS steps generally occur when the STM bias exceeds a threshold voltage corresponding to a bosonic excitation such as phonons and spin flips, while an additional inelastic tunneling channel opens up and contributes to the tunnel current \cite{REED200846, You:2017ua}. Under a spin-polarized tip (Figure \ref{fig:atoms}c), Ti atoms at both binding sites exhibit d$I$/d$V$ IETS steps at around zero bias voltage. These IETS steps arise from spin flips between the $\ket{\pm 1/2}$ spin states of Ti, which only require their Zeeman energy difference of $\sim$100 $\mu$eV at a typical magnetic field for ESR-STM measurements and should thus appear at $V_\mathrm{DC} = \pm 100 \  \mu$eV, essentially the zero bias voltage (100 $\mu$eV corresponds to about 24 GHz, or an out-of-plane field of about 0.86 T on bridge-site Ti). At a higher magnetic field, the two IETS steps of Ti can be separately resolved, but at lower fields typical for ESR-STM, they merge into one feature that we loosely refer to as a zero-bias step \cite{PhysRevLett.122.227203}. It turns out that the difference of the IETS step heights at positive and negative bias voltages (or equivalently, the height of the zero-bias step) is a good indicator of the tip's spin polarization (see section \ref{section: QM detection} and Ref. \cite{Loth_2010}). A good ESR-STM tip for Ti requires a strong spin polarization of the tip, which needs a zero-bias step height of at least 20\% of the total d$I$/d$V$ signal strength as a rule of thumb (as marked in Figure \ref{fig:atoms}c). ESR-STM can drive the spin resonance of both oxygen- and bridge-site Ti on 2ML MgO. Because oxygen-site Ti has a lower local density of states than bridge-site Ti, the tip is typically closer to the former, resulting in a higher Rabi rate but also stronger tip-induced magnetic fields and decoherence. \hfill\break
    
Fe atoms are another commonly used atomic species, particularly for quantum sensing and the preparation of spin-polarized tips. Fe on 2ML MgO/Ag typically sits on the oxygen sites. In STM topography, Fe's apparent height (around 1.5 \AA) is between the bridge- and oxygen-site Ti, and an Fe atom exhibits a distinctive ``dark halo'' around it under a sharp STM tip (Figure \ref{fig:atoms}d). In a d$I$/d$V$ spectrum, Fe on 2ML MgO has clear IETS steps at $\pm 14$ mV due to transitions to anisotropy-induced higher-energy spin levels (Figure \ref{fig:atoms}a,b) \cite{Paul:2017aa}. Under a spin-polarized STM tip (Figure \ref{fig:atoms}c), the Fe's IETS step heights at 14 mV and $- 14$ mV become unequal, and the step height difference reflects the tip's spin polarization along Fe's spin direction (out of the sample plane), see section \ref{section: QM detection} and Ref. \cite{Loth_2010}. Fe can be picked up onto the tip by approaching the tip towards the surface by 0.2--0.7 nm at around 0.6 V. Fe can often be dropped off from the tip by approaching it towards the surface by 0.7 nm or deeper at around $-0.6$ V. This technique is known as vertical manipulation \cite{Spinelli_2015,Khajetoorians:2019wq}. Three to ten Fe atoms on the tip are often enough to create strong spin polarization of the tip and allow ESR-STM measurements. \hfill\break

Cu atoms and FePc molecules are two more examples of spin-1/2 systems. Cu on 2ML MgO/Ag has an apparent height of about 3.2 \AA \ under typical scanning conditions, whereas FePc molecules are distinguishable by their four-lobe shape (Figure \ref{fig:atoms}d,e). Despite having a weak zero-bias Kondo-like resonance in their d$I$/d$V$ spectra (Figure \ref{fig:atoms}b), both systems host spin-1/2 and allow ESR transitions \cite{Yang:2018aa, Zhang:2021te}. Under a spin-polarized tip, Cu and FePc spectra show IETS steps near zero bias owing to the same mechanism as Ti, which, combined with the weak Kondo feature, yields interesting spectral shapes (Figure \ref{fig:atoms}c). Furthermore, $^{63}$Cu and $^{65}$Cu isotopes contain nuclear spins of 3/2 which are strongly coupled to Cu's electron spins (see section \ref{section: nuclear-sensing} and Ref. \cite{Yang:2018aa}). \hfill\break


\begin{table}
 \caption{Key ingredients for coherent single spin control on a surface}
  \begin{tabular}[thbp]{@{}ccc@{}}
    \hline
       & \textbf{Current status} & \textbf{Future directions}  \\
    \hline
    & Atoms and molecules evaporated  &  Various substrates: bulk semiconductors, \\
    \textbf{Spin preparation} &  onto a thin insulating film  &  2D materials, superconductors; \\
    & $\to$ \textit{Requires preparation chamber}  & More forms of spins: spin defects, skyrmions \\
    \hline
    & Thermal;  &  Spin pumping schemes \\
    \textbf{Spin initialization} & Spin-transfer torque & using other energy levels  \\
    & $\to$ \textit{Requires low temperature, magnetic field} & \\
    \hline
    & RF $E$-field applied to a tip or an antenna,  & Using $B$-field gradient from single-atom magnets;  \\
    \textbf{Spin control} & converted to RF $B$-field in the tip's $B$-field gradient  & Better engineered antenna   \\
    & $\to$ \textit{Requires high-frequency cabling} & \\
    \hline
    & Tunnel magnetoresistive readout &  Force detection;\\
    \textbf{Spin detection} & using spin-polarized STM  & Optical detection;  \\
    & $\to$ \textit{Requires spin-polarized STM tip} & Single-shot readout \\
    \hline
  \end{tabular}
  \label{table:ingredients}
\end{table}

\subsection{Experimental Apparatus}
\label{section: 2.3}
\begin{figure}[pthb]
  \centering
  \includegraphics[width=1 \linewidth]{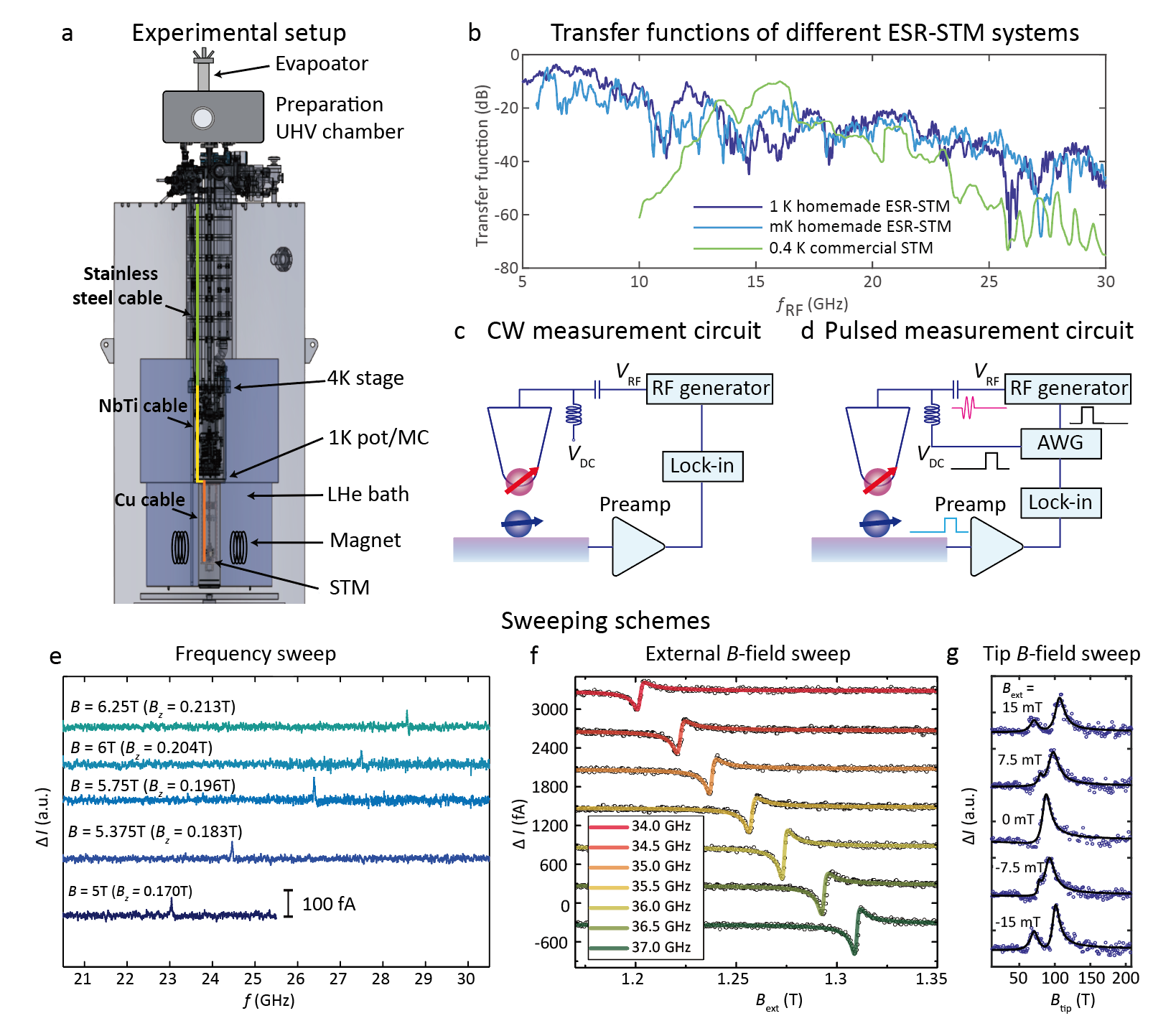}%
  \caption{Experimental apparatus for STM-based coherent spin experiments. a) Design of an ESR-STM system that includes a ultrahigh vacuum (UHV) chamber for sample preparation, a liquid helium (LHe) cryostat for thermalization, and a superconducting magnet. High-frequency coaxial cables composed of different materials are used between different temperature stages. The STM head is attached to the coldest stage, which is typically a 1K pot or a mixing chamber (MC). b) RF transfer functions measured in several ESR-STM systems (1 K homemade ESR-STM: homemade Joule-Thomson refrigerator, American Magnetics Inc. cryostat, 7 T single-axis magnet mostly in the sample plane; mK homemade ESR-STM: Janis JDR-250 dilution refrigerator, Cryomagnetics 9 T + 6 T vector magnet; 0.4 K commercial STM: Unisoku USM1300, 6 T + 5 T vector magnet). In the transfer functions, 0 dB corresponds to zero attenuation from the RF cabling. c,d) Schematics of electronics for continuous-wave (CW) (c) and pulsed (d) ESR-STM experiments. A bias tee is used to combine DC and RF bias voltages before applying them to the tunnel junction. The RF voltage is pulse modulated using a lock-in amplifier in the CW scheme or an arbitrary waveform generator (AWG) in the pulsed scheme. e--g) Three sweeping schemes of ESR-STM measurements: RF frequency sweeps (e), external magnetic field sweeps (f), and tip's magnetic field sweeps (g) (during the sweep of one parameter, the other two are fixed). (e) Reproduced with permission.\textsuperscript{\cite{Baumann417}} 2015, AAAS. (f) Reproduced with permission.\textsuperscript{\cite{Seifert2020}} 2020, APS. (g) Reproduced with permission.\textsuperscript{\cite{Singha:2021tj}} 2021, Springer Nature.
  }
  \label{fig:setup}
\end{figure}

The requirement for preparing, initializing, controlling, and detecting individual spins on a surface places demands on the experimental apparatus. The instrument is typically a low-temperature (typically 1 Kelvin) STM equipped with a moderate magnet (typically 1 Tesla), GHz-frequency coaxial cables, and 

low-temperature evaporation capabilities, as summarized in Table \ref{table:ingredients}. Some examples of ESR-STM instruments can be found in Refs. \cite{doi:10.1063/1.5065384,Seifert2020,doi:10.1063/5.0040011, drost2021combining}. As shown in Figure \ref{fig:setup}a, at the heart of the apparatus is an STM that allows access to individual spins on a surface. A spin-polarized STM tip is used for individual spin resonance driving (section \ref{section:driving}), individual spin readout (section \ref{section: detection}), and optionally spin-torque initialization of the spin state (section \ref{section: 2.4}). When studying spins on passivated substrates, a spin-polarized STM tip can be conveniently prepared by picking up typically three to ten magnetic atoms (e.g., Fe or Mn) onto the tip as discussed in the previous section. Alternative forms of spin-polarized STM tips include a tip made of a ferromagnetic or an antiferromagnetic wire or a metal tip coated with a thin magnetic layer \cite{Wiesendanger2009}. \hfill\break 

A static magnet, built around the STM, establishes the Zeeman energy splitting of the spin levels. A magnetic field of around 1.5 T is usually sufficient because most ESR-STM experiments are conducted at resonance frequencies below 40 GHz limited by the cables' RF transmission and the output range of RF signal generators (Figure \ref{fig:setup}b). A single-axis magnetic field, either in or out of the sample plane, should suffice for many studies, especially on organic radicals where the angle-dependent $g$-factor anisotropy is small. A multi-axis vector magnetic field, on the other hand, is convenient for studying systems such as transition metal atoms on a surface, where the $g$-factor anisotropy and the magnetocrystalline anisotropy can be significant \cite{abragam2012electron}. In coupled-spin structures with small magnetocrystalline anisotropy, a vector magnetic field can be used to modify the dipolar spin-spin interaction by changing the spin directions (see section \ref{section:spin-spin interaction}). Another practical benefit of an adjustable field angle is that it can be used to maximize the ESR signal of a given spin-polarized tip. This is because the ESR signal intensity can vary strongly with the field angle due to changes in the driving strength and the detection sensitivity \cite{PhysRevB.104.174408}.
 \hfill\break

A cryostat that houses the STM and the magnet is critical for the stable operation of the system. Equally importantly, a cryostat provides a low spin temperature for thermal initialization of the spin states (see section \ref{section: 2.4}). ESR-STM has so far been performed at temperatures ranging from 0.05 K \cite{PhysRevB.103.155405} to 5 K \cite{Seiferteabc5511} and is possible up to at least 10 K (unpublished), a temperature achievable in most low-temperature STMs. \hfill\break

To enable the coherent control of the spin states, coaxial cables with high transmission in the GHz range are used to connect room-temperature feedthroughs down to the tip, the sample, or an antenna in the STM head. Here, we use a homemade ESR-STM (equipped with a dilution refrigerator, Janis JDR-250) as an example to illustrate the high-frequency cabling used in the ESR-STM setup:
\begin{itemize}
    \item From the room-temperature flange to the `4K' stage, heat transfer must be minimized. We thus use stainless steel semi-rigid cables (Coax Japan, SC-119/50-SSS-SS) terminated by SMA connectors (Pasternack, PE4116).
    \item From the `4K' stage to the mixing chamber, we use superconducting NbTi cables (Coax Japan, SC-160/50-NbTi-NbTi, with a critical temperature of around 10 K) which provide high RF transmission while maintain a weak heat transfer.
    \item From the mixing chamber to the STM head, we use semi-rigid copper cables (Coax Japan, SC-219/50-SC) for high thermal conduction.
    \item Finally, for the wire going to the tip, we have an additional section of copper flexible wire (Cooner wire, CW2040-3650P) to allow the free motion of the STM coarse approach walker. To connect the flexible and semi-rigid copper cables, a special connector (Rosenberger 19S105-500L5) is used.
\end{itemize}
The connectors between different sections of the cables are usually rigidly placed on the corresponding cold stages. It is advantageous to add RF attenuators at the connections to reduce RF noise coming from the higher-temperature sides \cite{Krinner:2019ws}. \hfill\break

The transmission of high-frequency cables can be characterized by transfer function measurements. The details of transfer function measurements using an STM can be found in Ref. \cite{doi:10.1063/1.4955446}. In essence, one can use the RF broadening of a sharp d$I$/d$V$ feature to evaluate the RF voltage supplied to the tunnel junction. By comparing the RF output power of a signal generator and the RF voltage reaching the tunnel junction at different frequencies, one can obtain the frequency-dependent loss (i.e., the transfer function) of the RF cabling system. Using a well-calibrated transfer function, the RF signal generator's power output can be adjusted to supply a constant RF voltage to the tunnel junction over a wide frequency range, which is required for accurate frequency-sweep ESR-STM measurements (see below).  \hfill\break

In Figure \ref{fig:setup}b, we compare the RF transfer functions of several ESR-STM setups. The relatively poor transmission in one system (`Commercial') is due to an imperfect design of the connectors and the use of long, lossy flexible cables. Other homemade cabling systems provide satisfactory transmission over a wide frequency range. All of the systems in Figure \ref{fig:setup}b use STM tip cables for RF transimission, which necessitates a section of flexible coaxial cable for STM walker motion or for the operation of internal damping system such as springs. These flexible RF cables cause significant RF loss (limiting the available RF frequency up to about 30 GHz) and, consequently, heat generation near the STM junction. An antenna-based design in which the RF power is capacitively coupled to the tunnel junction partially overcomes these shortcomings \cite{Seifert2020}. Higher RF voltages in a wider frequency window (up to 100 GHz) can be applied to the STM junction using an antenna (not shown) \cite{Seifert2020, drost2021combining}. \hfill\break

RF components outside of the cryostat are experiment-specific. Figure \ref{fig:setup}c depicts the setup for continuous-wave ESR-STM experiments (which also works for transfer function measurements). Here a bias tee (e.g., SigaTek SB15D2 or SHF BT45R) is used to combine the DC bias voltage with the RF output of a signal generator (e.g., Keysight E8267d) before sending the signal to the STM. A lock-in amplifier is often employed to improve the signal-to-noise ratio by chopping the RF signal on and off (at a typical frequency of 100 Hz) \cite{Baumann417}. The STM feedback loop can be engaged in these experiments as long as its response time is set to be longer than the lock-in period. As another example, Figure \ref{fig:setup}d depicts the setup for pulsed ESR-STM experiments. Here an arbitrary waveform generator (e.g., Tektronix AWG70000B) is used to trigger the RF signal generator to produce nanosecond-long RF pulses. The arbitrary waveform generator can also generate DC bias pulses for the STM readout. 
\hfill\break

There are three ways to effectively scan across the magnetic resonance frequency in an ESR-STM measurement, namely the frequency sweeps, the external magnetic field sweeps, and the tip's magnetic field sweeps (Figure \ref{fig:setup}e--g), each having its own advantages and disadvantages:
\begin{itemize}
    \item The frequency-sweep method (Figure \ref{fig:setup}e) takes advantage of the broadband RF transmission across the tunnel junction. No mechanical motion is involved in a frequency sweep, resulting in no additional vibrational noise in the tunnel junction. The frequency-sweep ESR spectra are simple to interpret. The frequency-sweep method, however, necessitates a calibrated, high-quality transfer function over a wide frequency range, which may be difficult to realize for systems with poor transmission properties or employing resonator designs. 
    \item In an external magnetic field sweep (Figure \ref{fig:setup}f), the RF wave is fixed at a specific frequency while the external magnetic field is swept across the magnetic resonance. This is the method of choice when the RF wave is sufficiently strong only within a narrow frequency range, such as in resonator-based setups. In STM, sweeping the external magnetic field requires extra caution because it may cause mechanical instabilities from the superconducting magnet or piezoelectric elements. The tip's magnetization might also be altered during an external field sweep, which can affect both the ESR driving and the detection. Despite these challenges, the authors of Refs. \cite{PhysRevB.103.155405, Seiferteabc5511, Seifert2020} were able to successfully demonstrate the use of external magnetic field sweeps in ESR-STM.
    \item A tip's magnetic field sweep (Figure \ref{fig:setup}g) is a unique tool in ESR-STM due to the presence of a spatially varying tip's magnetic field (see Section \ref{section: tipfield} and Ref. \cite{PhysRevLett.122.227203}). The concept is similar to that of an external field sweep, except that instead of sweeping the external magnet, the spatial separation between the magnetic tip and the surface spin is swept, which can significantly change the magnetic field experienced by the spin \cite{PhysRevLett.122.227203, Yan:2015tp}. This method is typically used as a quick preparatory step to characterize the tip properties and the spin resonance frequencies, but it can also be used to collect high-quality data \cite{Singha:2021tj, Willke:2019ab}. A cross-reference to the two preceding methods is required to convert the tip-sample separation into the magnetic field of the tip. If a magnetic tip has bistable magnetization, the tip's magnetic field sweep can reveal two (rather than one) resonance peaks, slightly complicating the interpretation (Figure \ref{fig:setup}g) \cite{Singha:2021tj}. 
\end{itemize}

\begin{table}[tbhp]
 \caption{Examples of well-characterized spins on surfaces. A negative $D$ value indicates easy-axis anisotropy. $\parallel$ and $\perp$ denote that the corresponding quantity's direction is parallel or perpendicular to the sample surface, respectively. $\parallel ^O$ denotes MgO's in-plane O-impurity-O direction (see Figure \ref{fig:atoms}a), $\parallel ^N$ denotes Cu$_2$N's in-plane N-impurity-N direction, and $\parallel ^H$ denotes Cu$_2$N's other in-plane, hollow-impurity-hollow direction (see Figure \ref{fig:Cu2N}).
 Note$^{*}$: it remains unclear at the moment whether the 80-meV IETS step of oxygen-site Ti is related to a higher-lying orbital state or a vibrational mode. Note$^{**}$: the magnetic moment of bridge-site Ti is $0.96 \ \mu_\mathrm{B}$ along the other in-plane direction \cite{PhysRevB.104.174408}. Note$^{***}$: Higher-order anistropy terms mix the Fe spin states, allowing the ESR transition of Fe on MgO to occur \cite{Baumann417}. Other anisotropy terms for Ho \cite{PhysRevLett.121.027201} and Dy \cite{Singha:2021tj} on MgO are also known. On Cu$_2$N, the copper binding site lacks four-fold symmetry and induces transverse anisotropy $E$ of 0.007 meV for Mn and 0.31 meV for Fe \cite{Hirjibehedin1199}.
 } 
  \begin{tabular}[bthp]{@{}llllllllll@{}}
    \hline
    \textbf{Atom}  & \textbf{Site}  & \textbf{Spin} & $\abs{\boldsymbol{m}}  (\mu_\mathrm{B})$ & $D$ (meV) & \textbf{ESR} & \textbf{Manipulation} & \textbf{IETS} (meV) & \textbf{Kondo} & \textbf{Ref.}  \\
    \hline
    \multicolumn{5}{l}{\textbf{2ML MgO on Ag(100) substrate}} \\
    Ti  & Oxygen & 1/2  & 0.835$^\parallel$, 0.305$^\perp$ & 0 & $\ket{\pm 1/2}$ & Lateral & 80$^*$, 0 & No & \cite{PhysRevB.103.155405, PhysRevLett.119.227206} \\ 
    Ti & Bridge & 1/2 & 0.83$^{\parallel O}$, 0.99$^{\perp \ **}$  & 0 & $\ket{\pm 1/2}$ & Lateral & 0 & No &  \cite{Baeeaau4159, PhysRevB.104.174408} \\ 
    Fe & Oxygen & 2 & 5.44$^\perp$  & $-4.7^\perp$ & $\ket{\pm 2}^{***}$ & Vertical & 14 & No & 
    \cite{Choi:2017aa, PhysRevLett.115.237202}\\ 
    Cu & Oxygen & 1/2 & 0.99 & 0  & $\ket{\pm 1/2}$ & Difficult & 0 & Weak & \cite{Yang:2018aa} \\
    FePc & Oxygen & 1/2 & 1.058 & 0 & $\ket{\pm 1/2}$ & Difficult & 0 & Weak & \cite{Zhang:2021te} \\
    Ho & Oxygen & 8 & 10.1$^\perp$ & $-2.5^\perp$  & No & Vertical & No & No &  \cite{Natterer:2017aa, PhysRevLett.121.027201}  \\
    Dy & Oxygen & 15/2 & 9.9$^\perp$ & $-4.5^\perp$ & No & Vertical & No & No &  \cite{Singha:2021tj}  \\ 
    \multicolumn{5}{l}{\textbf{1ML Cu$_2$N on Cu(100) substrate}} \\
    Mn  & Copper & 5/2 & 4.75 & $-0.039^\perp$  & N/A & Vertical & 0.2 & No & \cite{Hirjibehedin1199, Hirjibehedin1021}\\
    Fe  & Copper & 2 & 4.22$^{\parallel N}$ & $-1.55^{\parallel N}$  & N/A & Vertical & 0.2, 3.8, 5.7  & No & \cite{Hirjibehedin1199}\\
    Co  & Copper & 3/2 & 3.3 & 2.75$^{\parallel H}$ & N/A & Vertical & 5.5 & $\ket{\pm 1/2}$ & \cite{Otte:2008aa} \\
    Ti  & Copper & 1/2 & N/A & 0 & N/A & Vertical & N/A & $\ket{\pm 1/2}$ & \cite{Otte:2008aa} \\
    \multicolumn{5}{l}{\textbf{Pt(111) substrate}} \\
    Fe  & hcp & 5/2 & 5 & 0.08$^\perp$ & N/A & Lateral & 0.19 & No & \cite{PhysRevLett.111.157204} \\
    Fe  & fcc & 5/2 & 6 & $-0.19^\perp$ & N/A & Lateral & 0.75 & No & \cite{PhysRevLett.111.157204}\\
    \hline
  \end{tabular}
  \label{table:spins}
\end{table}

\begin{figure}[thb]
  \centering
  \includegraphics[width=1 \linewidth]{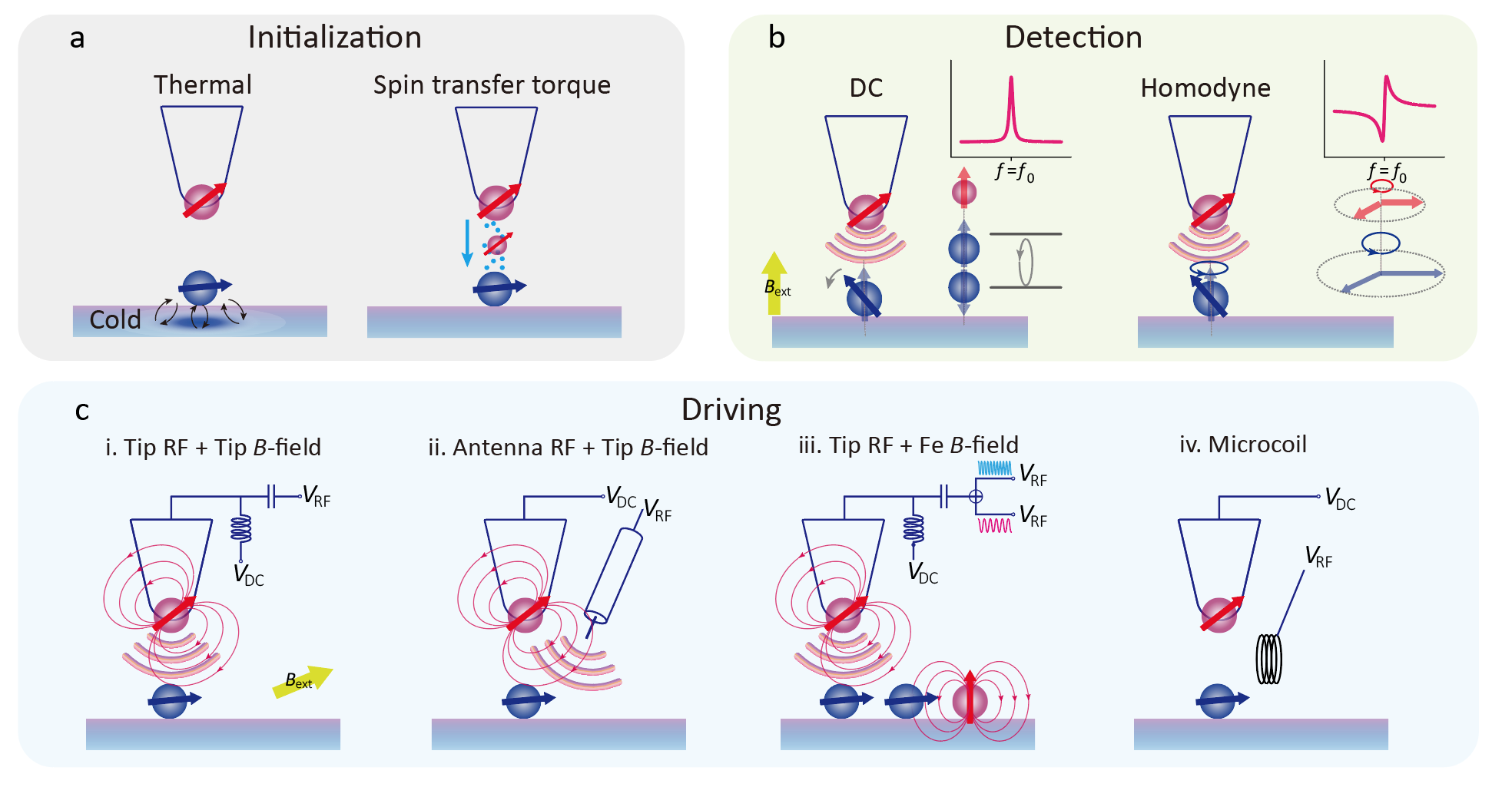}%
  \caption{Initialization, detection, and driving schemes of individual spins using ESR-STM. a) Surface spin initialization schemes using either a low temperature bath (left) or spin-polarized tunnel current (right). b) Single spin detection schemes in ESR-STM. Left: STM DC-bias readout of the surface spin polarization along its quantization axis that produces a symmetric, Lorentzian lineshape. Right: STM RF-bias homodyne readout of the surface spin's precession in a plane perpendicular to the quantization axis that produces an antisymmetric lineshape. c) Single spin driving schemes in ESR-STM. A strong, local RF magnetic field (i--iii) can be generated by combining an RF electrical field across the tunnel junction (applied to the tip or an antenna) with a spatially-varying static magnetic field (from a magnetic tip or a nearby single-atom magnet). Alternatively, a global RF magnetic field can be generated through a microcoil (iv).
  }
  \label{fig:ESRSTMessentials}
\end{figure}

\subsection{Single Spin Initialization}

\label{section: 2.4}
ESR measurements at the single spin level require a strong initial polarization of the spin states. This section discusses two electron spin initialization methods demonstrated in ESR-STM: cryogenic cooling and spin-transfer torque from spin-polarized tunnel current. 
For the initialization of nuclear spin states in a strongly coupled electron-nuclear spin system, see section \ref{section: Cu}. Please note that while ESR-STM is used to measure a single spin, all ESR-STM measurements reported so far have been performed in a time-ensemble averaged fashion (i.e., by initializing and measuring a single spin repeatedly, see more in section \ref{section: ensemble-detection}). As a result, an ensemble description of the spin states, as used hereafter, is typically sufficient (see, however, section \ref{section: QM detection}). 
\hfill\break

Thermal initialization of electron spins relies on the cryogenic environment that hosts modern STM systems. In most cases spins on surfaces can be naturally initialized to nearly the substrate lattice temperature due to couplings to substrate phonons and conduction electrons. The thermal spin polarization (of a thermal density matrix) can be expressed using a Boltzmann ratio. Consider an effective two-level spin system in a magnetic field $\boldsymbol{B}_{\mathrm{ext}}$ at a temperature $T_{\text{spin}}$. The Boltzmann population in the lower level is
\begin{equation}
    P =  \frac{1} {\exp[-f_{\text{ESR}}/(20.8366 \text{ GHz/K} \times T_{\text{spin}})]  + 1} = \frac{1} {\exp(\Delta \boldsymbol{m} \cdot \boldsymbol{B}_{\mathrm{ext}}/k_B T_{\text{spin}})  + 1},
    \label{eq: population}
\end{equation}
where $k_B/h = 20.8366$ GHz/K is the ratio of the Boltzmann and Planck constants, and $\Delta \boldsymbol{m}$ is the magnetic moment difference between the two spin levels involved in the ESR transition of frequency $f_{\text{ESR}}$. We summarize the measured values of $\Delta \boldsymbol{m}$ for Fe and Ti spins on 2ML MgO and their conversion relations at the end of this section for the convenience of the readers. \hfill\break

Initialization beyond the thermal polarization requires additional pumping. In ESR-STM, spin-transfer torque from spin-polarized tunneling electrons offers a convenient way to achieve this goal \cite{Loth_2010,Loth:2010aa,PhysRevLett.104.026601}. As will be explained in section \ref{section: QM detection}, when spin-polarized tunneling electrons pass through a surface spin, their spins interact and may exchange angular momentum during the tunneling process, resulting in a spin flip-flop process. Consider a simple case in which the tip spin is fully polarized to the $\ket{\uparrow}$ state and a surface spin-1/2 is thermally initialized to be a mixed state with $70\%$ of the population in the $\ket{\uparrow}$ state and 30$\%$ in the $\ket{\downarrow}$ state. At positive sample bias, spin-up electrons tunnel from the tip to the sample, which can cause an inelastic tunneling process associated with a spin flip-flop. Since the tunneling electron can only be flipped from $\ket{\uparrow}$ to $\ket{\downarrow}$ due to its assumed initial polarization in $\ket{\uparrow}$, the surface spin can only go from $\ket{\downarrow}$ to $\ket{\uparrow}$ (but not the other way around) due to angular momentum conservation. As a result, the surface spin becomes more polarized than the thermal state. The ultimate achievable spin polarization can be predicted using a rate equation that depends on the spin pumping rate (as determined by the tip's spin polarization, tunnel current amplitude, the ratio of inelastic to elastic tunneling events, etc.) and the spin relaxation rate \cite{Loth:2010aa}. \hfill\break

Spin transfer torque has been used to initialize several surface spin systems. A recent experiment on spin-1/2 Ti atoms on 2ML MgO discovered that even a short current pulse (with less than one tunneling electron per pulse on average) is efficient at initializing the Ti spin under the tip (see section \ref{section:spin-spin interaction} and Ref. \cite{Veldman964}). For higher-spin systems such as spin-5/2 Mn atoms on Cu$_2$N, it was found that a low tunnel current can pump the Mn spin from $\ket{+5/2}$ to $\ket{+3/2}$, while a very high tunnel current (that pumps faster than spin relaxations) can pull the Mn spin further up in the ladder, even reaching the highest $\ket{-5/2}$ state \cite{Loth:2010aa}. For spin-2 Fe atoms on 2ML MgO, spin-transfer torque has been used in conjunction with ESR-STM to achieve stronger initialization and larger ESR signals \cite{Willkeeaaq1543}. We anticipate that future initialization schemes involving optical or RF pumping through other atomic levels will provide higher-fidelity spin initialization on surfaces. 
\hfill\break

\textit{Useful relations for calculating thermal populations of Fe and Ti on MgO}: In its ground state, Fe on 2ML MgO hosts magnetic moment $5.44 \pm 0.03 \ \mu_\mathrm{B}$ in the out-of-plane direction \cite{Choi:2017aa}. The Fe ESR frequency corresponding to the Zeeman energy is given by
\begin{equation}
  f_{\text{Fe}}= \Delta E/h = 2 \times 5.44 \  (\mu_\mathrm{B}/h) \times B_{{\mathrm{ext}}} ^\perp = 152 \text{ GHz/T} \times B_{{\mathrm{ext}}} ^\perp,
  \label{eq: f-Fe}
\end{equation}
where $\mu_\mathrm{B}/h = 13.9962$ GHz/T is used, and the resonance frequency is determined solely by the out-of-plane component of the external magnetic field, $B_{{\mathrm{ext}}} ^\perp$. The thermal population of Fe in its ground state is obtained by combining Equation \ref{eq: population} and \ref{eq: f-Fe}
\begin{equation}
    P_{\text{Fe}} = \{1 + \exp[-B_{{\mathrm{ext}}} ^\perp/(0.137 \text{ T/K} \times T_{\text{spin}})] \}^{-1} .
\end{equation}
In an out-of-plane magnetic field $B_{{\mathrm{ext}}} ^\perp$ of 0.1 T, for example, the Fe resonance frequency is $15.2$ GHz, and Fe's ground-state populations are 54.3\%, 67.5\%, 86.1\%, and 99.9\% at $T_{\text{spin}} =$ 4.2 K, 1 K, 0.4 K, and 0.1 K, respectively. This estimation demonstrates the importance of achieving a low temperature for a reasonable initial spin polarization. \hfill\break

The magnetic moment of spin-1/2 Ti on 2ML MgO depends on the magnetic field direction and deviates from 1 $\mu_\mathrm{B}$ due to orbital contributions, as shown in Table \ref{table:spins}. In an external magnetic field $B_{\mathrm{ext}}$ along the out-of-plane direction, for example, oxygen- and bridge-site Ti have ESR frequencies of
\begin{equation}
f_{\text{TiO}}= 8.54 \text{ GHz/T} \times B_{\mathrm{ext}}, \ \  f_{\text{TiO}} =  27.7 \text{ GHz/T} \times B_{\mathrm{ext}},
\end{equation}
and thermal populations of
\begin{equation}
    P_{\text{TiO}} = \{1 + \exp[-B_{\mathrm{ext}}/(2.44 \text{ T/K} \times T_{\text{spin}})] \} ^{-1}, \ \ P_{\text{TiB}} = \{ 1+ \exp[-B_{\mathrm{ext}}/(0.75 \text{ T/K} \times T_{\text{spin}})] \} ^{-1}.
\end{equation}

\subsection{Single Spin Control in ESR-STM and Its Mechanism}
\label{section:driving}

Coherent control of spins is the defining feature of ESR-STM compared to other STM-based spin measurement schemes such as IETS or magnetization curves (see section \ref{section: 5.1}). The initial idea of ESR-STM was formulated based on the well-known massive electric field (on the order of 10$^{9}$ V/m) generated by a DC bias voltage across an STM tunnel junction. It was imagined that a moderate RF voltage supplied to the tunnel junction can similarly generate a very large RF electric field, allowing single spins to be addressed locally while producing little heat (unlike when using a magnetic coil). Although an oscillating \textit{magnetic} field is required to directly induce ESR transitions with $\Delta m = \pm 1$ (see section \ref{section: 2.1}), it was previously demonstrated in other solid-state spin systems such as semiconductor quantum wells and quantum dots that an oscillating electric field can indirectly drive $\Delta m = \pm 1$ transitions by modulating spin Hamiltonian parameters such as $g$-factors \cite{Kato:2003wr} or mechanically oscillating the electrons' spin density in the presence of a magnetic field gradient \cite{PhysRevLett.96.047202,Pioro-Ladriere:2008vq}. \hfill\break 

The ESR-STM driving scheme proved to be effective, and it has now been performed on a variety of spins, including Ti, Fe, Cu, and FePc on MgO (Table \ref{table:spins}). In addition to the original scheme of applying an RF wave directly to the tip, RF driving through an antenna (that capacitively couples to the tip) has also been reported (Figure \ref{fig:ESRSTMessentials}c) \cite{Seifert2020}. Recently, a ``remote'' driving scheme was demonstrated, in which ESR driving was shown for a remote spin placed near the tip but not directly in the tunnel junction (see Figure \ref{fig:ESRSTMessentials}c, section \ref{section: ELDOR}, and Ref. \cite{2021arXiv210809880P}). It was discovered that remote driving works only when an single-atom magnet, in this case Fe, is positioned close to the remote spin (and their separation sensitively affects the driving strength) \cite{2021arXiv210809880P}. \hfill\break

These experimental observations lead us to the conclusion that a large magnetic field gradient, either from a magnetic tip or a nearby single-atom magnet, is important for ESR driving in STM. As a result, the most likely driving mechanism of ESR-STM, in our opinion, is that an RF voltage mechanically oscillates the spin-carrying electron in the presence of a static magnetic field gradient \cite{PhysRevB.96.205420}, as in the case of quantum dots \cite{PhysRevLett.96.047202,Pioro-Ladriere:2008vq}. This oscillation produces an effective RF magnetic field that can then drive the $\Delta m = \pm 1$ ESR transitions. This conclusion is consistent with a careful examination of ESR-STM spectra obtained over a wide parameter range \cite{Seifert2020}. \hfill\break


There are numerous other ESR-STM mechanisms that have been proposed (for a recent review, see Ref. \cite{DELGADO2021100625}). Direct driving from oscillating magnetic fields in the tunnel junction was estimated to be negligible compared to the observed Rabi rates \cite{Seifert2020, PhysRevLett.122.227203}, although a careful RF simulation of the tunnel junction has yet to be performed. In an early report of ESR-STM of Fe on MgO, the driving mechanism was proposed to be related to MgO's crystal field and spin-orbit coupling (combined with the mechanical oscillation of atomic spin density) \cite{Baumann417}. Further experimental and theoretical studies, however, show that this mechanism is typically weaker than the aforementioned driving through the tip's magnetic field gradient \cite{PhysRevB.96.205420} and cannot explain experimental results such as ESR driving of spin-1/2 atoms \cite{Baeeaau4159, Yang:2018aa}. In an interesting proposal of a cotunneling driving mechanism \cite{PhysRevB.100.035411}, the RF electric field periodically modifies the hopping amplitude between the surface atom and a metallic reservoir, resulting in an oscillating driving term after tracing out the reservoir's degrees of freedom. Other proposals rely on spin-transfer torque induced by spin-polarized tunnel current \cite{PhysRevB.99.054434}, RF modulation of the tunnel barrier \cite{Balatsky2012}, spin-phonon interactions \cite{PhysRevLett.99.047201, Mullegger:2015va}, or modulations of $g$-factor anisotropy \cite{PhysRevResearch.1.033185}. Nevertheless, these proposed mechanisms do not address the observed necessity of a magnetic field gradient, only work for a spin in the tunnel junction, or do not obey other experimental observations \cite{Seifert2020}, and thus are unlikely to provide the dominant driving force.

\subsection{Single Spin Detection}
\label{section: detection}

The ability to detect a single spin state is at the heart of the emerging spin-based quantum technology. Single spin detection has been achieved in optical spectroscopy, RF force microscopy, scanning magnetometry, and electrical measurements. Optically detected magnetic resonance of single spins was first demonstrated in single molecules \cite{Kohler:1993uy,Wrachtrup1993} and then extended to defects in solids such as nitrogen-vacancy centers in diamond \cite{Gruber2012}. For single spin detection, the change of an optical fluorescence signal is monitored while the frequency of an RF wave is swept across the magnetic resonance. In force microscopy, single spin detection was achieved by recording the change in the vibration frequency of an oscillating cantilever upon spin resonance, which varies due to the magnetic exchange interaction between a magnetic tip and the spin \cite{Rugar:1992wx,Rugar2004}. Scanning magnetometers with tips based on either a superconducting quantum interference device \cite{Vasyukov:2013ve} or a single nitrogen-vacancy center \cite{Grinolds:2013wq} have also reached single electron-spin sensitivity, where the tips, tens of nanometers away from the single spin, sense its stray magnetic field.   \hfill\break

Unlike the aforementioned techniques, electrical readout of single spin states hinges on the spin-to-charge conversion. In semiconductor quantum dots and donor atoms, for example, the spin-to-charge conversion typically makes use of a neighboring charge reservoir, whose chemical potential is tuned between the spin-up and spin-down states. This results in a spin-dependent tunneling event, during which the change in charge can be capacitively sensed by a second quantum dot nearby \cite{Elzerman:2004wt,Morello:2010vm}. In single-molecule magnets, electron and nuclear spins can be read out from the conductance of an effective quantum dot in a break junction configuration \cite{Moreno-Pineda:2021wt, Vincent2012,Stefan2014}. In the STM setup, early efforts to detect spin resonance used a non-magnetic tip to read an RF tunnel current at the Larmor frequency of an electron spin, where the spin sensitivity could be due to exchange coupling between the local spin and the tunneling electrons \cite{Manassen1989,Kemiktarak:2007uq}. \hfill\break

Most later studies of ESR-STM, however, rely on the use of spin-polarized STM tips. Owing to the spin sensitivity of the tip, traditional DC tunnel current detection is sufficient to directly probe the spin states of the atom in the tunnel junction, resulting in a simpler circuit design and higher data quality. Below we discuss two complementary descriptions of the spin-polarized tunneling readout. \hfill\break

\subsubsection{Ensemble Description of Readout}

\label{section: ensemble-detection}
So far, ESR-STM measurements have been carried out in a time-ensemble averaged fashion, which means that the spin initialization, control, and measurement sequences are performed repeatedly, and an averaged tunnel current readout reflects the ensemble average of the spin state being measured. As shown in Figure \ref{fig:setup}c,d, a lock-in amplifier is typically used to improve the signal-to-noise ratio, where a driven state (that one would like to measure) placed in the lock-in A cycle is contrasted with a reference state (typically the thermal state) placed in the lock-in B cycle. The lock-in output signal then yields the difference between the driven and the reference states. Similar schemes are used in pulsed measurements, but the total pulse sequence is typically much shorter than the length of each lock-in cycle and is thus repeated within each lock-in cycle. Typical measurement parameters for the readout of one spin state (i.e., one data point during a sweep) are a total data acquisition time of 1$\sim$5 seconds, a lock-in frequency of 100$\sim$500 Hz, and for pulsed measurements, a total pulse sequence length of 0.5$\sim$2 $\mu$s (which needs to be longer than the spin's $T_1$ time for re-initialization).
\hfill\break

The ensemble readout signal of a spin state by a spin-polarized tip is described by tunneling magnetoresistance (TMR), where the tunnel conductance depends on the relative alignment between the tip spin and the surface spin as \cite{Wiesendanger2009}
\begin{equation}
  G=G_{\mathrm{j}} [1+a \braket{\boldsymbol{S}_{\mathrm{tip}}} \cdot \braket{\boldsymbol{S} }].
  \label{TMRconductance}
\end{equation}
Here, $G_{\mathrm{j}}$ is the spin-averaged junction conductance, $a$ is a prefactor that describes the TMR contribution to the total conductance, and $\braket{\boldsymbol{S}_{\mathrm{tip}}}$ and $\braket{\boldsymbol{S}}$ are the expectation values of the tip and surface spins (taken in the same reference frame). The ESR-STM readout can then be understood as a modification of $\braket{\boldsymbol{S}}$ upon a near-resonant RF wave, which results in a detectable change in $G$. To detect this change in $G$ and to drive the ESR transition, a combination of a DC bias voltage ($V_{\mathrm{DC}}$) and an RF bias voltage ($V_{\mathrm{RF}}$) is applied to the STM junction, yielding the total bias voltage
\begin{equation}
  V=V_{\mathrm{DC}}+V_{\mathrm{RF}} \cos(\omega t+\phi),
\end{equation}
where $\omega$ and $\phi$ denote the angular frequency and the phase of the RF wave, respectively. The tunnel current is given by the product of the tunneling conductance and the bias voltage
\begin{equation}
  I(t) = GV = G_{\mathrm{j}} [1+a \braket{\boldsymbol{S}_{\mathrm{tip}}} \cdot \braket{\boldsymbol{S}}] [V_{\mathrm{DC}}+V_{\mathrm{RF}} \cos(\omega t+\phi)].
  \label{eq: tunnelreadout}
\end{equation}
The current signal is then read out using a preamplifier with a bandwith of around 1 kHz, resulting in a time-averaged current $\overline{I(t)}$ that contains no RF components.\hfill\break

The two contributions to the detected signal, DC and homodyne (Figure \ref{fig:ESRSTMessentials}b), are most simply illustrated in a CW measurement, which we will discuss below (in a pulsed measurement, these two contributions carry similar forms but depend on more parameters). A CW measurement uses a prolonged data acquisition time wherein the initial Rabi spin oscillations (see section \ref{section: 2.1}) have already decayed, and a steady spin state is instead reached through the balance of driving and dissipations. The steady-state solution of a spin-1/2 system is historically derived from the Bloch equation (and the same result can be obtained from modeling an open quantum spin-1/2 system \cite{Breuer:2010wp, 2021arXiv210809880P}). The Bloch equation in the rotating frame reads \cite{Weil:2007wk, Slichter:2011wo}
\begin{equation}
  \frac{d\braket{\boldsymbol{S}}}{dt} = \braket{\boldsymbol{S}} \times [-(\omega - \omega_0) \boldsymbol{e}_z  - \Omega \boldsymbol{e}_x] - \frac{\braket{S_z} - \braket{S_{z} ^ 0}}{T_1} \boldsymbol{e}_z - \frac{\braket{S_x} \boldsymbol{e}_x + \braket{S_y} \boldsymbol{e}_y}{T_2},
\end{equation}
where $T_1$ is the longitudinal relaxation time that tends to bring the system to the thermal-equilibrium spin polarization $\braket{S_{z} ^ 0}$, and $T_2$ is the transverse relaxation time that tends to diminish the transverse magnetization. The Larmor frequency is $\omega_0 = g \mu_\mathrm{B} B_{\mathrm{ext}}/\hbar$ (with the static field $B_{\mathrm{ext}}$ applied in the $-z$ direction), and the driving field strength is $\Omega = g \mu_\mathrm{B} B_1/\hbar$ (with $B_1$ applied in the $x$ direction in the rotating frame). By setting all time derivatives to zero, the steady-state solution of the surface spin in the rotating frame is obtained as \cite{Weil:2007wk, Slichter:2011wo}
\begin{gather}
  \braket{S_x} = \frac{\Omega (\omega - \omega_0) T_2^2}{1 + (\omega - \omega_0)^2 T_2^2 + \Omega^2 T_1 T_2} \braket{S_{z} ^ 0} , \ \ 
  \braket{S_y} = \frac{-\Omega T_2}{1 + (\omega - \omega_0)^2 T_2^2 + \Omega^2 T_1 T_2} \braket{S_{z} ^ 0}, \nonumber \\
  \braket{S_z} = \frac{1 + (\omega - \omega_0)^2 T_2^2}{1 + (\omega - \omega_0)^2 T_2^2 + \Omega^2 T_1 T_2} \braket{S_{z} ^ 0}  \label{Bloch-SS-z}.
\end{gather}
While the tip spin $\braket{\boldsymbol{S}_{\mathrm{tip}}}$ is stationary in the lab frame, it rotates at the angular frequency $+\omega$ around the $z$ axis in the rotating frame (of the surface spin) as
\begin{equation}
  \braket{\boldsymbol{S}_{\mathrm{tip}}}= \braket{S^{\mathrm{tip}}_{z}} \boldsymbol{e}_z + \braket{S^{\mathrm{tip}}_{xy}} [\cos{(\omega t + \psi)}\boldsymbol{e}_x +\sin{(\omega t + \psi)} \boldsymbol{e}_y],
  \label{TipSpin}
\end{equation}
where $\psi$ denotes the initial phase of the tip spin with respect to the $x$ axis. Inserting Equation \ref{Bloch-SS-z} and \ref{TipSpin} into Equation \ref{eq: tunnelreadout} and taking a time average yields the detected tunnel current
\begin{equation}
  \overline{I(t)} = G_{\mathrm{j}} V_{\mathrm{DC}} [1 + a \braket{S^{\mathrm{tip}}_{z}} \braket{S_z}] + \frac{1}{2} G_{\mathrm{j}} a V_{\mathrm{RF}} \braket{S^{\mathrm{tip}}_{xy}} [\cos{(\psi - \phi)} \braket{S_{x}} + \sin{(\psi - \phi)} \braket{S_{y}}].
  \label{TimeAveragedI}
\end{equation}
Here the low bandwidth of the current preamplifier allows us to ignore the RF components of the tunnel current. Nonetheless, a $V_{\mathrm{RF}}$-dependent term (the second term in Equation \ref{TimeAveragedI}) remains. This term, derived from the component $G_{\mathrm{j}} a (\braket{\boldsymbol{S}_{\mathrm{tip}}} \cdot \braket{\boldsymbol{S}}) V_{\mathrm{RF}} \cos(\omega t+\phi)$ in Equation \ref{eq: tunnelreadout}, arises because the relative rotation of the tip and surface spins acts as an RF rectifier, allowing the RF voltage to drive current only at specific times during an RF cycle, resulting in an average DC current. \hfill\break

As previously mentioned, in typical CW ESR-STM measurements a lock-in detection scheme is used for higher signal-to-noise ratio, which contrasts $\overline{I(t)}$ of a driven state ($V _{\mathrm{RF}}$, $ \Omega \ne 0$) with a thermal state ($V _{\mathrm{RF}} = \Omega = 0$). The final ESR signal measured by a lock-in amplifier can then be expressed as
\begin{equation}
  I_{\text{ESR}} = \Delta I_{\text{DC}} +  \Delta I_{\text{Homodyne}} = G_{\mathrm{j}} a  V_{\mathrm{DC}} \braket{S^{\mathrm{tip}}_{z}}[\braket{S_z}-\braket{S_z^0}] + \frac{1}{2} G_{\mathrm{j}} a  V _{\mathrm{RF}} \braket{S^{\mathrm{tip}}_{xy}} [\cos{(\psi - \phi)} \braket{S_x} + \sin{(\psi - \phi)} \braket{S_y}].     
  \label{ESRSignal}
\end{equation}
The measured ESR signal thus contains two contributions, labeled as DC and homodyne in the equation above. As sketched in Figure \ref{fig:ESRSTMessentials}b left, the DC ESR signal, $\Delta I_{\text{DC}}$, originates from a population change of the surface spin under RF driving (i.e., $\braket{S_z}-\braket{S_z^0}$) and is sensed by the tip magnetization along the surface spin's quantization axis $z$ ($\braket{S^{\mathrm{tip}}_{z}}$). In contrast, as shown in Figure \ref{fig:ESRSTMessentials}b right, the homodyne ESR signal, $\Delta I_{\text{Homodyne}}$, detects the transverse magnetization of the surface spin ($\braket{S_x}$ and $\braket{S_y}$). As explained above, the transverse magnetization is sensed by a combination of the rotating transverse tip magnetization ($\braket{S^{\mathrm{tip}}_{xy}}$) and an oscillating RF voltage at the driving frequency, hence the name, homodyne detection. The DC ESR signal follows $\braket{S_z}$ and has a symmetric Lorentizian-peak lineshape, while the homodyne ESR signal follows $\braket{S_x}$ and $\braket{S_y}$ and can have an asymmetric lineshape (Figure \ref{fig:ESRSTMessentials}b) \cite{Weil:2007wk, Slichter:2011wo}. \hfill\break


Finally, in pulsed ESR measurements, the homodyne ESR signal is present during the RF driving (unless for remote driving of a spin not in the tunnel junction, see section \ref{section: ELDOR}). The DC readout, if turned on after the driving, measures the spin population along the $z$ axis following spin manipulation. Some relaxation effects are included due to the finite duration of the DC readout pulse. 

\subsubsection{Single-Event Description of Readout}

\label{section: QM detection}

The future possibility of using surface spins as qubits necessitates non-averaging, single-event spin detection, which can be made possible by employing existing electrical or optical detection schemes developed in other quantum systems \cite{Devoret:2000tz, Hadfield:2009ug}. To properly describe single readout events and the post-measurement spin states, one needs to go beyond the ensemble description in the previous section and understand the microscopic interactions that occur between individual tunneling electrons and surface spins during the readout process. In addition, this description enables us to understand the initialization mechanism via spin-transfer torque (section \ref{section: 2.4}) and why a tip's spin polarization is reflected in the spin-flip IETS step heights (Figure \ref{fig:atoms}c). \hfill\break

Here we follow a microscopic model presented in Refs. \cite{Loth_2010,Loth:2010aa}, which was originally developed to describe the IETS d$I$/d$V$ spectra for Mn and Fe on Cu$_2$N. This treatment is consistent with theoretical studies in Refs. \cite{PhysRevLett.104.026601, PhysRevLett.102.256802,Fransson:2009tv,PhysRevLett.103.050801,PhysRevLett.103.176601}. A more complete model that includes higher-order tunneling events can be found in Ref. \cite{Ternes_2015}. Consider an initial state $\ket{\sigma_i m_i}$ composed of a tunneling electron in the spin state $\ket{\sigma_i}$ along the quantization axis $z$ and a local surface spin in spin state $\ket{m_i}$ along the quantization axis $z'$. During tunneling events, the tunneling electron scatters with the surface spin. The tunneling probabilities are argued to take the following form
\begin{equation}
  Y(m_f, \sigma_f, m_i, \sigma_i ) = \frac{1}{Y_0}|\braket{ \sigma_f m_f | \boldsymbol{\sigma} \cdot \boldsymbol{S} + u  | \sigma_i m_i}|^2 =  \frac{1}{Y_0}|\braket{ \sigma_f m_f | ( \sigma_z S_z + u) + \frac{1}{2} (\sigma_- S_+ + \sigma_+ S_-) | \sigma_i m_i}|^2,
  \label{Tunnel-rate}
\end{equation}
where $u$ represents spin-independent potential scattering between the tunneling electron and the surface spin, $\boldsymbol{\sigma} \cdot \boldsymbol{S}$ represents their spin-dependent Kondo-like scattering, and $Y_0$ is a normalization factor. In the following, we consider the simple situation where the tunneling electron and the surface spin are polarized in the same direction ($z \parallel z'$). In this case, elastic tunneling, mediated by $\sigma_z S_z + u$, maintains the tip and surface spin states before and after tunneling (i.e., $\ket{\sigma_f m_f} = \ket{\sigma_i m_i}$). Inelastic tunneling is created by the flip-flop terms $\sigma_- S_+ + \sigma_+ S_-$, resulting in a final state $\ket{\sigma_f m_f} = \ket{\sigma_i + 1, m_i-1}$ or $\ket{\sigma_i - 1, m_i + 1}$. \hfill\break

Using this microscopic model, post-measurement spin states can be described. Consider a surface spin originally in the state $\ket{m_i}$, a tip spin fully polarized in the $+1/2$ state, and positive sample bias. If the next tunneling event is elastic, the spin will stay in the spin-$m_i$ state. If inelastic tunneling instead happens, the surface spin will be flipped to $m_i + 1$ because the tunneling electron spin must begin in the $+1/2$ state in the tip and can only flip to $-1/2$ after flip-flop scattering. Without detailed knowledge of this tunneling event, the surface spin is described by a density matrix of a mixed state of $\ket{m_i}$ and $\ket{m_i + 1}$, whose mixing ratio is determined by the elastic vs. inelastic tunneling rates. Considered this way, the tunneling measurement appears to be not simply a projection. The post-measurement state for a general spin state measured with a partially polarized tip should carry a similar form but be dependent on more parameters. Future theoretical and experimental investigations are needed to formulate a fully fledged quantum measurement theory for spin-polarized tunneling readout in ESR-STM.  \hfill\break



There are two other interesting observations one can make from this model. First, both the elastic and the inelastic tunnel currents can be dependent on the surface spin state and hence be used as a spin readout method. Second, the difference of the IETS step heights at positive and negative bias voltages is directly related to the tip's spin polarization. The former statement can be seen by considering the elastic tunnel probability, which typically depend on the surface spin state $m_i$ because in general
\begin{equation}
 \frac{1}{Y_0}|\braket{ \sigma_i m_i | ( S_z \sigma_z + u) | \sigma_i m_i}|^2 \ne \frac{1}{Y_0}|\braket{ \sigma_i m_j | ( S_z \sigma_z + u) | \sigma_i m_j}|^2,
\end{equation}
for $m_i \ne m_j$, due to the interplay of the two interaction terms, $S_z \sigma_z$ and $u$ \cite{Loth_2010}. Similarly, the inelastic tunneling probability as governed by the spin flip-flop terms also in general depends on the surface spin state \cite{Loth_2010}. As a result, both elastic and inelastic tunnel currents can be used to read out the spin state. It is generally believed that the inelastic contribution dominates the spin readout signals of Ti on MgO, as evidenced by Ti's short $T_1$ time under a tunnel current, whereas the elastic contribution dominates for Fe on MgO because of Fe's long $T_1$ time despite tunnel current effects (partially due to its anisotropy barrier) \cite{Paul:2017aa}. A careful study of the elastic and inelastic contributions to the tunnel current, as done for Mn and Fe on Cu$_2$N \cite{Loth_2010,Loth:2010aa}, is still lacking for spins on MgO.   \hfill\break

In order to make the latter claim, we need to relate the inelastic tunnel current to the tip's spin polarization, which can be defined as 
\begin{equation}
  \eta_{\text{tip}} = \frac{\text{DOS}(+1/2) - \text{DOS}(-1/2)}{\text{DOS}(+1/2) + \text{DOS}(-1/2)},
\end{equation}
where $\text{DOS}(+1/2)$ and $\text{DOS}(-1/2)$ are the tip’s spin +1/2 and $-$1/2 density of states near the Fermi level, respectively. We can also define an effective ``sample'' spin polarization \cite{Loth_2010}
\begin{equation}
  \eta_{\text{s}} = \frac{\sum_{\sigma_f} [Y(m_f, \sigma_f, m_i, +1/2 ) - Y(m_f, \sigma_f, m_i, -1/2 )]}{\sum_{\sigma_f} [Y(m_f, \sigma_f, m_i, +1/2 ) + Y(m_f, \sigma_f, m_i, -1/2 )]},
\end{equation}
which ranges between $-$1 and +1 and describes the ``spin-filtering'' effect during tunneling. When $\eta_{\text{s}} = 1$, for example, $Y(m_f, \sigma_f, m_i, +1/2 )$ is much greater than $Y(m_f, \sigma_f, m_i, -1/2 )$, and so only spin-$+1/2$ tunneling electrons can tunnel through the surface spin. Using these definitions of spin polarizations, the inelastic tunnel conductance from the tip to the sample and from the sample to the tip can be loosely written as \cite{Loth_2010}
\begin{gather}
  G_{\text{inelastic, t} \to \text{s}} =  G/4 \times [ (1  + \eta_{\text{tip}})  (1 + \eta_{\text{s}} ) + (1  - \eta_{\text{tip}}) (1 - \eta_{\text{s}} )],  \label{eq: Isp-inelastic1} \\
  G_{\text{inelastic, s} \to \text{t}} =  G/4 \times [ (1  + \eta_{\text{tip}})  (1 - \eta_{\text{s}}) + (1  - \eta_{\text{tip}}) (1 + \eta_{\text{s}} )], \label{eq: Isp-inelastic2}
\end{gather}
The difference between the IETS steps at positive and negative sample bias voltages then yields the tip's spin polarization
\begin{equation}
  P = \frac{G_{\text{Inelastic, t} \to \text{s}} - G_{\text{Inelastic, s} \to \text{t}}}{G_{\text{Inelastic, t} \to \text{s}} + G_{\text{Inelastic, s} \to \text{t}}} = \eta_{\text{tip}} \cdot \eta_{\text{s}}.
  \label{eq: SP}
\end{equation}
This result is analogous to Equation \ref{TMRconductance} in conventional TMR formulation and indicates that the IETS step asymmetry generally characterizes the tip's spin polarization $\eta_{\text{tip}}$ (as stated in section \ref{section: 2.2} and Figure \ref{fig:atoms}). When the quantization axes of the tunneling electron and the surface spin are not aligned, the interpretation becomes less transparent and requires a careful simulation \cite{Loth:2010aa}. \hfill\break

\subsection{Spin-Spin Interactions and Coupled Spin Systems}

\label{section:spin-spin interaction}

To go beyond single spin quantum control, it is necessary to magnetically couple multiple spins in a controllable way. Using vertical and lateral atom manipulation techniques of STM, artificial spin structures with well-defined interactions can be built on a surface \cite{Khajetoorians:2019wq, Spinelli_2015}. These multi-spin systems constructed atom-by-atom can then be employed for quantum control (section \ref{section: 3}), sensing (section \ref{section: 4}), and simulation (section \ref{section: 5}).\hfill\break

\begin{figure}[ht]
  \centering
  \includegraphics[page=1,width=1\linewidth]{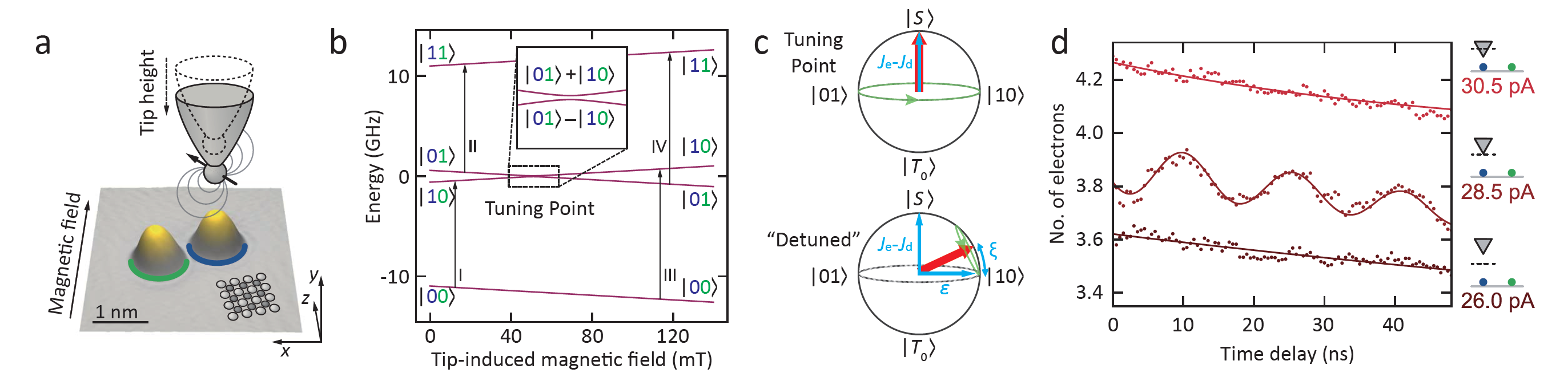}\hspace*{0\textwidth}%
  \caption{Tuning eigenstates of two coupled spins using a magnetic tip. a) STM image of two Ti atoms on MgO. b) Eigenstates and eigenenergies as a function of the tip's magnetic field. An avoided level crossing appears at the tuning point, where the tip field compensates for the $g$-factor difference. c) Eigenstates of two coupled spins represented by Bloch spheres in the singlet-triplet subspace. When $\varepsilon$ $\sim$ 0, $\ket{S}$ and $\ket{T_{0}}$ states are the eigenstates. With increasing $\abs{\varepsilon}$, the spin eigenstates become closer to the Zeeman product states, as characterized by the mixing parameter $\xi$. d) Pump-probe measurements of free coherent evolution observed at a ``tuning point'' when singlet-triplet states are formed (28.5 pA). The tuning point corresponds to $\xi$ $\sim$ $\pi/2$ in (c), and the free coherent evolution is a Larmor precession in the singlet-triplet subspace with frequency $(J_{\mathrm{e}} - J_{\mathrm{d}})/h$. (a,b,d) Reproduced with permission.\textsuperscript{\cite{Veldman964}} 2021, AAAS.}
  \label{fig:Detuning}
\end{figure}

To realize these functionalities, a good understanding of spin-spin interactions and their influences on spin states is required. In this section, we summarize experimental findings by focusing on two coupled spin-1/2 atoms, such as Ti on 2ML MgO (Figure \ref{fig:Detuning}a) \cite{Veldman964, Baeeaau4159, PhysRevLett.119.227206}. Spins on a thin insulating layer are mostly coupled via dipolar and exchange interactions. The spin Hamiltonian for two coupled spins, $\boldsymbol{S}_{1}$ and $\boldsymbol{S}_{2}$, is given by

\begin{equation}
  H = \mu_{\mathrm{B}} (\boldsymbol{B}_{\mathrm{ext}}+\boldsymbol{B}_{\mathrm{tip}}) \cdot \mathbf{g}_{1} \cdot \boldsymbol{S}_{1} + \mu_{\mathrm{B}} \boldsymbol{B}_{\mathrm{ext}} \cdot \mathbf{g}_{2} \cdot \boldsymbol{S}_{2} + \boldsymbol{S}_{1} \cdot  \mathbf{J}_{\text{e}} \cdot  \boldsymbol{S}_{2} + \boldsymbol{S}_{1} \cdot  \mathbf{J}_{\text{d}} \cdot  \boldsymbol{S}_{2}.
  \label{eq: CoupledSpin-Hamiltonian}
\end{equation}

Here, we assume that the tip acts on the first spin ($\boldsymbol{S}_{1}$) with an effective magnetic field ($\boldsymbol{B}_{\mathrm{tip}}$) (see section \ref{section: tipfield}), and $\mathbf{J}_{\text{e}}$ and $\mathbf{J}_{\text{d}}$ represent the exchange and dipolar couplings, respectively. In typical ESR-STM measurement conditions, $\boldsymbol{B}_{\mathrm{ext}}$ is greater than $\boldsymbol{B}_{\mathrm{tip}}$ and determines the quantization axis (taken as the $z$ axis below). The exchange coupling term is typically well approximated by $J_{\mathrm{e}} \boldsymbol{S}_{1} \cdot \boldsymbol{S}_{2}$, where the exchange coupling tensor is taken to be symmetric and described by a coupling strength $J_e$. In most cases, the dipolar coupling is small compared to the Zeeman energy (determined by $\boldsymbol{B}_{\mathrm{ext}}$), so we can further adopt the secular approximation of the dipolar coupling \cite{Slichter:2011wo}. Under these approximations the Hamiltonian becomes

\begin{gather}
  H = \mu_{\mathrm{B}} g_{1} (-{B}_{\mathrm{ext}}+{B}_{\mathrm{tip}})S_{1z} - \mu_{\mathrm{B}} g_{2} {B}_{\mathrm{ext}} S_{2z} + J_{\mathrm{e}} \boldsymbol{S}_{1} \cdot \boldsymbol{S}_{2} + J_{\mathrm{d}}(3S_{1z}S_{2z}-\boldsymbol{S}_{1} \cdot \boldsymbol{S}_{2}) \nonumber \\
 = \mu_{\mathrm{B}} g_{1} (-{B}_{\mathrm{ext}}+{B}_{\mathrm{tip}}) S_{1z} - \mu_{\mathrm{B}} g_{2} {B}_{\mathrm{ext}} S_{2z} + (J_{\mathrm{e}}+2J_{\mathrm{d}})S_{1z}S_{2z} + (J_{\mathrm{e}}-J_{\mathrm{d}}) (S_{1x}S_{2x}+S_{1y}S_{2y}), 
  \label{eq: CoupledSpin-Hamiltonian2}
\end{gather}
where $g_{1}$ and $g_{2}$ are the g-values projected along the external magnetic field direction, and 

$J_{\mathrm{d}} = \mu_{0} \mu_{\mathrm{B}}^{2} g_{1} g_{2} (1-3\cos^{2}\theta)/2 \pi r^{3}$ is the dipolar coupling constant with $\theta$ being the angle between the quantization axis ($z$) and the connecting vector of two spins ($\boldsymbol{r}$). The last two terms show the effects of exchange and dipolar couplings on the spin states, i.e., an energy splitting by $J_{\mathrm{e}}+2J_{\mathrm{d}}$ and a mixing of the spin states by $J_{\mathrm{e}}-J_{\mathrm{d}}$ because the last term can be written as spin flip-flop terms $(J_{\mathrm{e}}-J_{\mathrm{d}})/2 \times (S_{1}^{+}S_{2}^{-}+S_{1}^{-}S_{2}^{+})$. \hfill\break

It is also known that the exchange and dipolar couplings of spins have distinct dependence on distance and direction. The exchange interaction is typically spatially isotropic (i.e., direction independent) but it decays exponentially over distance according to $J_{\mathrm{e}}=J_{\mathrm{0}} \exp{[-(r-r_{0})/d_{\mathrm{e}}]}$, where $d_{\mathrm{e}}$ is the decay constant and $J_{\mathrm{0}}$ is the exchange coupling strength at $r=r_{0}$. The dipolar interaction, on the other hand, decays more slowly and is direction-dependent, as evidenced by the $\theta$-dependence in its coupling constant $J_{\mathrm{d}}$. These two magnetic interactions can be disentangled by measuring the ESR spectra for a series of spin pairs at different separations and orientations. Roughly speaking, for pairs at very close distance, the isotropic exchange interaction dominates and can be determined at any orientation, while farther-spaced pairs need to be constructed with different orientations in order to determine the anisotropic dipolar interactions. Using this procedure, the interactions between bridge-bridge Ti pairs \cite{Baeeaau4159}, bridge-oxygen Ti pairs \cite{Baeeaau4159}, oxygen-oxygen Ti pairs \cite{PhysRevLett.119.227206}, and the dipolar coupling of Fe-Fe pairs \cite{Choi:2017aa} have been obtained.\hfill\break

Spin-spin interactions in a spin pair can change the spin eigenstates and result in the creation of singlet-triplet states. In a basis composed of Zeeman product states ($\ket{00},\ket{01},\ket{10},\ket{11}$), the eigenstates of the coupled-spin Hamiltonian (Equation \ref{eq: CoupledSpin-Hamiltonian2}) are
\begin{equation}
\begin{gathered}
  \ket{T_{-1}} = \ket{00}, \ket{S(\xi)} = \cos\frac{\xi}{2} \ket{01} - \sin\frac{\xi}{2} \ket{10} 
  \\ \ket{T_{0}(\xi)} = \sin\frac{\xi}{2} \ket{01} + \cos\frac{\xi}{2} \ket{10}, \ket{T_{+1}} = \ket{11}.
  \label{eq: CoupledSpin-eigenstates}
\end{gathered}
\end{equation}
Here, $\xi$ is a mixing parameter (Figure \ref{fig:Detuning}c) determined by the ratio of the spin flip-flop coupling energy ($J_{\mathrm{e}}-J_{\mathrm{d}}$) to the Zeeman energy difference of the two spins ($\varepsilon$) as
\begin{equation}
  \tan\xi = \frac{J_{\mathrm{e}}-J_{\mathrm{d}}}{\varepsilon} = \frac{J_{\mathrm{e}}-J_{\mathrm{d}}}{\mu_\mathrm{B} [(g_{2}-g_{1}) B_{\mathrm{ext}} + g_{1} B_{\mathrm{tip}}]}.
  \label{eq: DetuningEnergy}
\end{equation}

When the Zeeman energy difference, $\varepsilon$, is small compared to the coupling energy ($\varepsilon \ll J_{\mathrm{e}}-J_{\mathrm{d}}$), the mixing parameter $\xi \sim \pi/2$, and the eigenstates correspond to the singlet and triplet states, $\ket{S} = (\ket{01}-\ket{10})/\sqrt{2}$ and $\ket{T_{0}} = (\ket{01}+\ket{10})/\sqrt{2}$. In contrast, when the Zeeman energy difference is large ($\varepsilon \gg J_{\mathrm{e}}-J_{\mathrm{d}}$), the mixing paramter $\xi \sim 0$, and the eigenstates are Zeeman product states $\ket{01}$ and $\ket{10}$ (Figure \ref{fig:Detuning}b,c). Note that the tip's magnetic field enters the Zeeman energy difference of the spins (Equation \ref{eq: DetuningEnergy}) and can thus affect the spin eigenstates. Two strongly coupled spins ($J_{\mathrm{e}}-J_{\mathrm{d}} \gg \varepsilon$), for example, are robust against the local magnetic field fluctuations and can be used to enhance spin coherence (see section \ref{section: coherence} and Ref. \cite{Baeeaau4159}). \hfill\break

On the other hand, if two spins are weakly coupled, the tip's magnetic field can sensitively affect the formation of the singlet-triplet states (Figure \ref{fig:Detuning}b). The authors of Ref. \cite{Veldman964} recently demonstrated this by first creating the singlet-triplet states using a carefully tuned tip's magnetic field and then observing a ``Larmor precession'' in the singlet-triplet subspace. As shown in Figure \ref{fig:Detuning}a, two Ti atoms with slightly different Zeeman energies ($\sim$1 GHz) were used, with a magnetic tip located over one of the atoms to provide a local, tunable magnetic field. ESR-STM was then used to identify a specific tip-atom separation, dubbed the ``tuning point'', where the tip's magnetic field is such that the two spins have the same Zeeman energies ($\varepsilon = 0$). This results in a perfect mixing of $\xi = \pi/2$ and the creation of the singlet-triplet states.  \hfill\break

To show the existence of the singlet-triplet states at the tuning point, electrical pump-probe experiments were performed, where a DC pump pulse is used to initialize the spin state followed by a DC probe pulse that measures the evolved spin state after some time delay. 
In the weakly-coupled Ti spin pair, the Zeeman energy is the dominating energy scale, and so the ground state is $\ket{00}$. A DC pump pulse from a spin polarized tip is utilized to locally excite the spin underneath the tip apex (say, the first spin), which, to some fidelity, flips it and creates the spin state $\ket{10}$ (this spin flipping is due to spin-transfer torque, see section \ref{section: 2.4}). At the tuning point, the state $\ket{10}$ can be regarded as a superposition of the singlet state $\ket{S}$ and the triplet state $\ket{T_{0}}$, and a free evolution then occurs in the $xy$ plane of the singlet-triplet subspace (green trajectory in upper panel of Figure \ref{fig:Detuning}c and the 28.5 pA curve of Figure \ref{fig:Detuning}d). This precession can be detected using a DC probe pulse that measures along the $\ket{10}$-$\ket{01}$ axis in the singlet-triplet subspace because the first spin in the $\ket{10}$ and $\ket{01}$ states has a different spin orientation during the Lamor precession. In contrast, away from the tuning point (lower panel of Figure \ref{fig:Detuning}c), the excited spin state $\ket{10}$ is essentially an excited eigenstate, with only an exponential decay towards the ground state $\ket{00}$ but no coherent evolution (26 pA/30.5 pA of Figure \ref{fig:Detuning}d), in some sense similar to the relaxation of a single-spin excited state. This work showcases the power of combined high energy and temporal resolutions in STM-based ESR and electrical pump-probe measurements. In principle, pulsed DC and RF measurements can be extended to larger spin structures or magnetic materials where local spin flips may induce interesting dynamics. \hfill\break


\subsection{Creating a Tunable Local Magnetic Field Using the STM Tip}
\label{section: tipfield}

\begin{figure}[ht]
  \centering
  \includegraphics[page=1,width=0.6\linewidth]{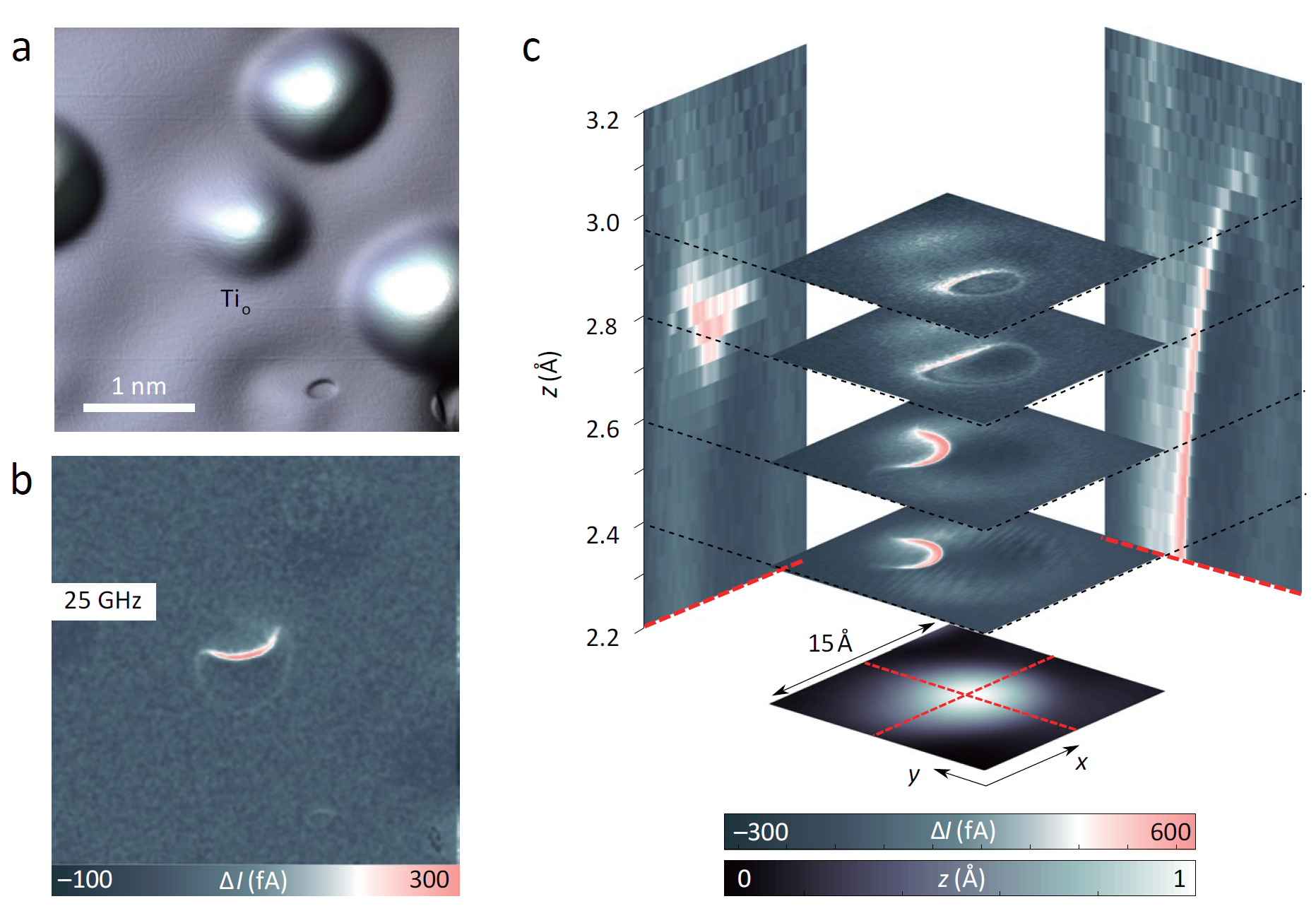}\hspace*{0\textwidth}%
  \caption{Magnetic resonance imaging (MRI) in STM. a) Constant-current STM image of an oxygen-site Ti on MgO. b) An MRI scan in the same area with a spin-polarized tip showing a spatial pattern of resonance signals. The RF frequency is fixed at $f$= 25 GHz, and the spatial variations originate from magnetic interactions between the tip spin and the surface spin. c) More MRI scans at different tip-atom distances.  
  Reproduced with permission.\textsuperscript{ \cite{Willke:2019aa}} 2019, Springer Nature.}
  \label{fig:PhilipNatPhys2019}
\end{figure}

A unique feature of ESR-STM is the presence of a nearby magnetic tip. The tip's magnetic moment is largely classical because of its proximity to the metallic tip body and hence fast decoherence. While magnetic tips with only one Fe atom on the apex were found to be paramagnetic \cite{Loth_2010}, ESR tips commonly used (with three to ten Fe atoms) often have large enough anisotropy to keep the tip spin fixed along a certain direction (typically at a 15--60$^\circ$ angle to the external field applied during the preparation of the spin-polarized tip \cite{PhysRevB.104.174408}). Similar to two coupled surface spins as discussed in section \ref{section:spin-spin interaction}, the tip spin $\braket{\boldsymbol{S}_{\mathrm{tip}}}$ and the surface spin $\boldsymbol{S}$ are also coupled via dipolar and exchange interactions as
\begin{equation}
  H = g \mu_\mathrm{B} \boldsymbol{B}_{\mathrm{tip}} \cdot \boldsymbol{S} = g \mu_\mathrm{B} (\boldsymbol{B}_{\mathrm{tip-dip}} + \boldsymbol{B}_{\mathrm{tip-ex}}) \cdot \boldsymbol{S} =\frac{g_{\mathrm{tip}}g\mu_{0}\mu_{\mathrm{B}}^{2}}{4\pi r^{3}} [\braket{\boldsymbol{S}_{\mathrm{tip}}} - 3(\braket{\boldsymbol{S}_{\mathrm{tip}}} \cdot \hat{\boldsymbol{r}})\hat{\boldsymbol{r}}] \cdot \boldsymbol{S} + J_0 e^{-(r - r_0)/d_\mathrm{e}} \braket{\boldsymbol{S}_{\mathrm{tip}}} \cdot \boldsymbol{S},
  \label{TipField}
\end{equation}
where $g_{\mathrm{tip}}$ and $g$ are the $g$-factors of the tip spin and the surface spin, respectively, $\boldsymbol{r}$ and $\hat{\boldsymbol{r}}$ are the connecting vector from the tip spin to the surface spin and the corresponding unit vertor, and $J_0 e^{-(r - r_0)/d_e}$ characterizes the exponential decay of the exchange coupling strength. The tip's magnetic field (from the magnetic interactions between the tip and the surface spins) was found to be continuously tunable from 1 mT to 10 T, covering 4 orders of magnitude \cite{PhysRevLett.122.227203}. \hfill\break 

The dipolar and exchange contributions in Equation \ref{TipField} can be disentangled by performing ESR-STM measurements over a wide range of tip-atom separations \cite{Seiferteabc5511}. Another interesting way to visualize the tip's magnetic interactions with a surface magnetic atom is by obtaining ESR-STM signals while scanning the tip laterally over the atom. An example of this so-called nano-MRI (magnetic resonance imaging) method is shown in Figure \ref{fig:PhilipNatPhys2019}. MRI scans contain unique signatures of the magnetic tips used for scanning, which, upon a careful analysis, can reveal the dipolar and exchange couplings between the tip and the surface spins \cite{Willke:2019aa}.



\begin{table}
 \caption{Summary of ESR-STM experiments performed on MgO/Ag substrates}
  \begin{tabular}[bthp]{@{}lllll@{}}
    \hline
    \textbf{System studied} & \textbf{Short description}  & \textbf{Section} & \textbf{Year}& \textbf{Ref.}  \\
    \hline
    Isolated Fe  &  First demonstration of ESR-STM & 2  & 2015 & \cite{Baumann417}\\
    Isolated Fe  &  Measurement of $T_1$ time at high $B$-field (pump-probe, no ESR) & 2 & 2017 & \cite{Paul:2017aa}  \\
    Isolated Fe  &  Measurement of $T_2$ time; spin-transfer torque initialization & 4 & 2018 & \cite{Willkeeaaq1543} \\
    Isolated Fe  &  Optimization of Fe ESR in a vector $B$-field; tip $B$-field sweep & 2 & 2019 & \cite{Willke:2019ab} \\
    Isolated Ti, Ti-Ti  & Demonstration of ESR-STM on spin-1/2; Ti-Ti interaction & 2,3  & 2017 & \cite{PhysRevLett.119.227206} \\
    Isolated Ti & Measuring the exchange interaction between tip and surface spins   & 2  & 2019 & \cite{PhysRevLett.122.227203} \\
    Isolated Ti  & $g$-factor anisotropy in a vector magnetic field; external $B$-field sweep & 2 & 2021 & \cite{PhysRevB.103.155405}  \\
    Isolated Ti  & Tip spin direction; $g$-factor anisotropy in a vector magnetic field & 2,3 & 2021 & \cite{PhysRevB.104.174408}  \\
    Isolated Ti, Fe  & Magnetic resonance imaging of tip spin-surface spin magnetic interactions & 2 & 2019 & \cite{Willke:2019aa} \\
    Isolated Ti, Fe  & On ESR-STM driving mechanism; external $B$-field sweep & 2 & 2020 & \cite{Seiferteabc5511}  \\
    Isolated Ti, Fe & Measurement and binding-site control of hyperfine couplings & 3 & 2018 & \cite{Willke336} \\
    Isolated Cu  & Manipulation of nuclear spin states through hyperfine coupling & 3,4 & 2018 & \cite{Yang:2018aa} \\
    Fe-Fe pair   &  Sensing of Fe's effective magnetic field & 2,3 & 2017 & \cite{Choi:2017aa} \\   
    Fe-Ho pair   &  Sensing of Ho's effective magnetic field; Ho initialization & 2,3 & 2017 & \cite{Natterer:2017aa}  \\  
    Fe-Dy pair  &  Sensing of Dy's effective magnetic field; tip $B$-field sweep & 2,3 & 2021 & \cite{Singha:2021tj} \\ 
    Ti-Ti pair & Enhanced coherence in singlet-triplet states & 2,4 & 2018 & \cite{Baeeaau4159} \\
    Ti-Ti pair & Spin-flip initialization; coherent evolution in the singlet-triplet basis & 2 & 2021 & \cite{Veldman964}  \\
    Ti-Ti-Fe & Simultaneous and individual driving of two Ti spins using one tip &  2,4  & 2021 & \cite{2021arXiv210809880P} \\
    Ti-Ti-Ti-Ti & Quantum simulation of a resonating valence-bond state & 5 & 2021 & \cite{Yang:2021aa}   \\
    FePc, FePc-FePc  & ESR-STM study of molecular spins and their interactions  & 2,3 &  2021 & \cite{Zhang:2021te} \\
    \hline
  \end{tabular}
  \label{table:experiment}
\end{table}

\section{Quantum Sensing Using Individual Atomic Spins on Surfaces}
\label{section: 3}

Quantum sensing, in broad terms, encompasses all measurement approaches using quantum systems as sensors and harnessing the high sensitivity of quantum states to external perturbations \cite{RevModPhys.89.035002}. Ideal quantum sensors respond exclusively to the desired external signals, where their sensitivity is determined by the coupling strength to external signals as well as the quantum sensors' coherence qualities. Different quantum systems excel at sensing different physical quantities. Trapped ions \cite{RevModPhys.87.1419} and Rydberg atoms \cite{Herrmann_1986}, for example, are good electric-field sensors, while some superconducting circuits \cite{PhysRev.140.A1628, Fagaly2006} and spin-based systems such as semiconductor quantum dots \cite{Elzerman:2004wt} and color centers in diamond \cite{Balasubramanian2008, Degen2008, Maze2008, Taylor2008} respond sensitively to magnetic fields. \hfill\break

The detection of small signals with high spatial resolution is a primary goal of quantum sensing. In magnetic field sensing, a challenging objective is to achieve the sensing of single spins and their interactions at the atomic scale \cite{Fuechsle2012, MaminH.2013, Toyli2010}. Traditional STM-based spectroscopy, such as IETS, attains atomic resolution but has an energy resolution limited by the thermal Fermi-Dirac broadening of tunneling electrons \cite{PhysRev.165.821, Song2010, Ast2016, PhysRevLett.107.076804}. The energy resolution of ESR-STM, on the other hand, is not limited by electronic thermal broadening because the energy of the tunneling electrons is not the measured quantity. Instead, the energy in ESR-STM is measured against the frequency of a supplied RF wave, which resonantly drives a surface spin state and regulates the tunnel current flow through the TMR effect (see section \ref{section: ensemble-detection}). The energy resolution of ESR-STM is thus limited only by the performance of this TMR-based ``spin regulator'', given by a combination of $T_{1}$ and $T_{2}$ of the surface spin depending on the specific detection mechanism (for CW ESR measurement, the energy resolution is related to the linewidth specified in Equation \ref{ESR_Decoherence}). Due to their distinct energy resolutions, traditional IETS spectra can be used to quantify strong exchange coupling or Ruderman-Kittel-Kasuya-Yosida (RKKY) interaction (see section \ref{section: 5.1}), while ESR-STM can detect considerably weaker electronic dipole-dipole interactions and hyperfine couplings. \hfill\break

In this section, we mainly describe the use of single magnetic atoms on surfaces as the smallest quantum sensors. Spins on surfaces satisfy the basic criteria for quantum sensing, such as the existence of discrete energy levels, the ability to initialize, manipulate, and read out these levels, and the interaction with external signals \cite{RevModPhys.89.035002}. Surface spins are mostly used to sense atomic-scale local magnetic fields, either from a nearby magnetic atom (section \ref{section: 3.1}) or a nuclear spin (section \ref{section: nuclear-sensing}). The outcome is a shift in ESR frequencies with an energy resolution of about 40 neV under current CW ESR-STM measurement conditions. In section \ref{section: 3.3}, we briefly discuss recent developments of tip-based sensing that might soon enter the quantum regime. \hfill\break

\subsection{Single Spin Sensing of Dipolar Fields at the Atomic Scale}

\label{section: 3.1}

\begin{figure}[ht]
  \centering
  \includegraphics[page=1,width=0.9\linewidth]{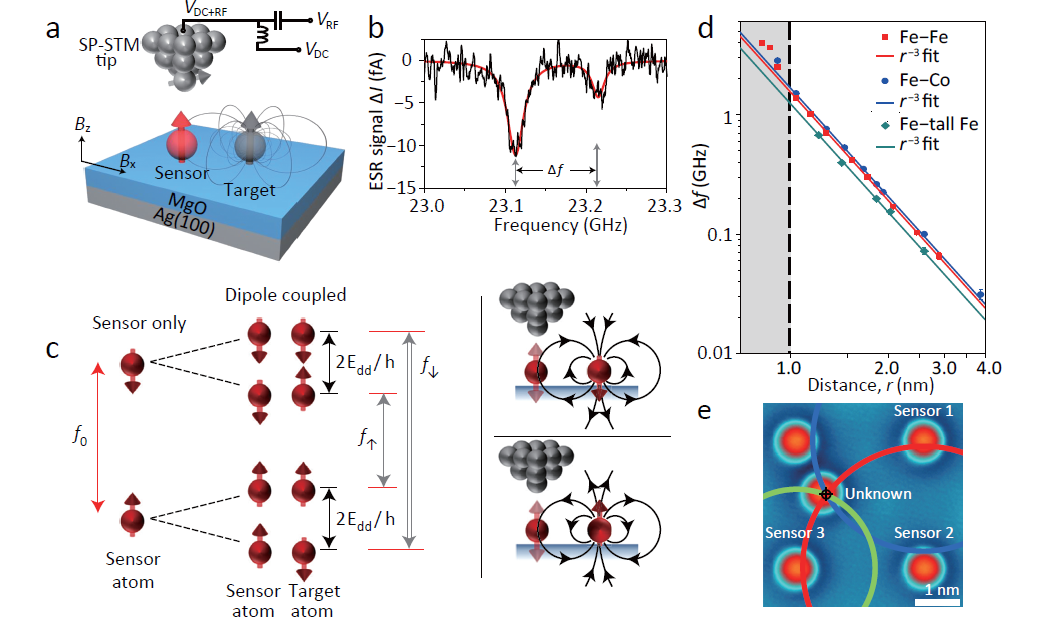}\hspace*{0\textwidth}%
  \caption{Atomic-scale sensing of dipolar fields from single-atom magnets. a) Schematic showing the sensing of dipole fields from a target atom using a sensor spin. b) ESR spectrum obtained from an Fe atom (sensor) separated by 2.46 nm from another Fe atom (target). c) Schematic energy diagram of the sensor atom under the influence of the dipole fields of the target atom. Two ESR peaks in (b) correspond to the two transitions marked by grey arrows in (c). The Zeeman energy of the target spin is not included in the energy level diagram. d) ESR splitting as a function of atomic separations for pairs of Fe-Fe, Fe-Co, and Fe-`tall' Fe. The observed $r^{-3}$-distance dependence indicates that dipole-dipole interactions dominate at these separations, while deviations in the grey area (with $r$ less than 1 nm) result from additional exchange interactions. The magnetic moments extracted from the fits are $5.44 \pm 0.03 \ \mu_{\mathrm{B}}$ for Fe, $5.88 \pm 0.06 \ \mu_{\mathrm{B}}$ for Co, and $4.35 \pm 0.08 \ \mu_{\mathrm{B}}$ for `tall' Fe. e) An STM image of five Fe atoms used in ``nano-GPS''. Through trilateration using 3 sensor atoms, the location and the magnetic moment of a target, unknown atom can be determined. 
  Reproduced with permission.\textsuperscript{\cite{Choi:2017aa}} 2017, Springer Nature.}
  \label{fig:DipoleSensing}
\end{figure}

Fe atoms on MgO serve as a good sensor for local out-of-plane magnetic fields, as their magnetic moments are fixed perpendicular to the surface plane by a large magnetic anisotropy (Table \ref{table:spins}) \cite{Baumann417, PhysRevLett.115.237202}. As shown in Figure \ref{fig:DipoleSensing}a, using vertical atom manipulation, an Fe atom (sensor) was positioned close to another magnetic atom (target, which is also Fe in this case) to probe the magnetic field emanating from it. During the lock-in measurement time (about 1--10 ms, see section \ref{section: ensemble-detection}), due to perturbations, the target Fe atom occasionally flips its spin and hence the magnetic field emanating from it, resulting in two ESR peaks observed on the sensor atom (Figure \ref{fig:DipoleSensing}b,c). The ESR peak height ratio thus corresponds to the Boltzmann occupations of the two lowest spin states of the target Fe atom at this temperature, while the ESR peak splitting directly yields the sensor-target magnetic interaction \cite{Choi:2017aa}. The ESR peak splitting was found to increase at reduced separations ($r$) between the two Fe atoms (see Figure \ref{fig:DipoleSensing}d) with a trend that follows an $r^{-3}$-power law, which is characteristic of the magnetic dipole-dipole interaction (section \ref{section:spin-spin interaction}). Fitting to the measured splitting (Figure \ref{fig:DipoleSensing}d) reveals the magnetic moment of Fe atom as $5.44 \pm 0.03 \ \mu_\mathrm{B}$ \cite{Choi:2017aa}, a remarkable precision for an atomic-scale measurement. \hfill\break

The accurate determination of the Fe's magnetic moment on MgO allows its use as a sensor of magnetic fields from other single-atom magnets. This technique has been used to determine the magnetic moments of Co \cite{Choi:2017aa}, Ho \cite{Natterer:2017aa}, Dy \cite{Singha:2021tj}, and some unknown magnetic species (Figure \ref{fig:DipoleSensing}d,e), all with 1\% or better accuracy. For lanthanide atoms (Ho and Dy), Fe sensing was proven to be non-invasive and does not perturb the magnetic stability \cite{Natterer:2017aa, Singha:2021tj}. \hfill\break

An interesting use of Fe sensors is the trilateration measurement of the position of a target spin, dubbed nano-GPS (Global Positioning System). Figure \ref{fig:DipoleSensing}e illustrates the concept. Using three Fe sensors, a unique location (better than 0.1 nm) and magnetic moment (better than 0.1 $\mu_{\mathrm{B}}$) of an unknown spin can be determined.\hfill\break

Other sensors such as spin-1/2 Ti and FePc have also been utilized. Ti sensors have been used to determine the Ti-Ti interactions at different binding sites \cite{PhysRevLett.119.227206,Baeeaau4159} and the magnetization direction of a magnetic tip \cite{PhysRevB.104.174408}. FePc molecules have been used to determine FePc-FePc and FePc-Ti couplings \cite{Zhang:2021te}. The scheme described in this section is in principle applicable to the sensing of any magnetic objects that can stay on or near a surface. We expect that with its high spatial and energy resolutions, ESR-STM will enable the sensing of complex magnetic molecules, magnetic nanostructures, and various magnetic materials. \hfill\break

\subsection{Single Spin Sensing of Hyperfine Interactions on a Surface}
\label{section: nuclear-sensing}

Nuclear spins have been widely exploited as probes to the chemical environment and electronic characteristics of atoms, molecules, and crystals due to their well-characterized interactions with electron spins. While single nuclear spins have been detected and controlled using other approaches \cite{Koehler1995, Neumann542, Vincent2012, Pla2013}, directly correlating hyperfine spectra with atomic-scale surroundings of the nucleus had not been demonstrated. Ref. \cite{Willke336} achieved this goal by using ESR-STM to resolve hyperfine interactions and modify them by changing the adatom's binding site. \hfill\break

Before discussing the experimental results, we first introduce some theoretical concepts related to hyperfine interactions. In the presence of a nuclear spin $\boldsymbol{I}$, the spin Hamiltonian becomes

\begin{equation}
  H = H_{\mathrm{EZ}} + H_{\mathrm{NZ}} + H_{\mathrm{HF}} + H_{\mathrm{NQ}} 
    = \mu_{\mathrm{B}} \boldsymbol{B}_{\mathrm{ext}} \cdot \mathbf{g} \cdot \boldsymbol{S} + \mu_{\mathrm{n}} \boldsymbol{B}_{\mathrm{ext}} \cdot \mathbf{g}_{n} \cdot \boldsymbol{I} + \boldsymbol{S} \cdot \mathbf{A} \cdot \boldsymbol{I} + \boldsymbol{I} \cdot \mathbf{P} \cdot \boldsymbol{I},
  \label{HyperfineHamiltonian}
\end{equation}

where $\boldsymbol{S}$ and $\boldsymbol{I}$ represent the electron and nuclear spin operators, respectively, and $\mu_{\mathrm{B}}$ and $\mu_{\mathrm{n}}$ are the Bohr magneton and nuclear magneton, respectively. The $g$-factors of electron and nuclear spin ($\mathbf{g}$ and $\mathbf{g}_{n}$), the hyperfine constant ($\mathbf{A}$), and the electric quadrupole constant ($\mathbf{P}$) are written in the general tensor forms. The first two terms in the Hamiltonian are the Zeeman energies of electron and nuclear spins, the third term represents the hyperfine interaction, and the last term corresponds to the nuclear electric quadrupole interaction. Note that the magnetocrystalline anisotropy is not included in Equation \ref{HyperfineHamiltonian} for simplicity (which is valid in the cases discussed below). In a typical ESR-STM environment, the electronic Zeeman interaction is on the order of $\sim$0.1 meV, the magnetic hyperfine interaction is $\sim$1 $\mu$eV, and the nuclear Zeeman energy is $\sim$10 neV. As a result, the nuclear Zeeman energy can often be ignored, as we shall do later, which means that the quantization axis of the nuclear spin is predominantly determined by the electron spin through the hyperfine coupling rather than the external magnetic field.  \hfill\break

Since the nuclear and electron spins share similar chemical environment, the $\mathbf{g}$, $\mathbf{A}$, and $\mathbf{P}$ tensors often have collinear principal axes and can be diagonalized simultaneously. The diagonal form of the hyperfine interaction tensor reads \cite{abragam2012electron}:

\begin{equation}
  H_{\mathrm{HF}} = A_{x}S_{x}I_{x} + A_{y}S_{y}I_{y} + A_{z}S_{z}I_{z} =  \sum_{i=x,y,z} A_{i}S_{i}I_{i}.
  \label{HFterm}
\end{equation}

The hyperfine interaction, $\mathbf{A}$, is well-known to have two contributions: the nucleus can couple to the electron spin cloud outside the nucleus via the dipolar interaction (the dipolar hyperfine term) or it can directly interact with the finite electron spin density at the nuclear site (the contact hyperfine term). The nuclear electric quadrupole ($\mathbf{P}$) interaction, on the other hand, originates from the electric field gradient present at the site of the nucleus, and its diagonal form is given by: 

\begin{equation}
  H_{\mathrm{NQ}} = P_{x}I_{x}^{2} + P_{y}I_{y}^{2} + P_{z}I_{z}^{2} = \sum_{i=x,y,z} P_{i}I_{i}^{2}.
  \label{NQterm}
\end{equation}

\begin{figure}[ht]
  \centering
  \includegraphics[page=1,width=0.95\linewidth]{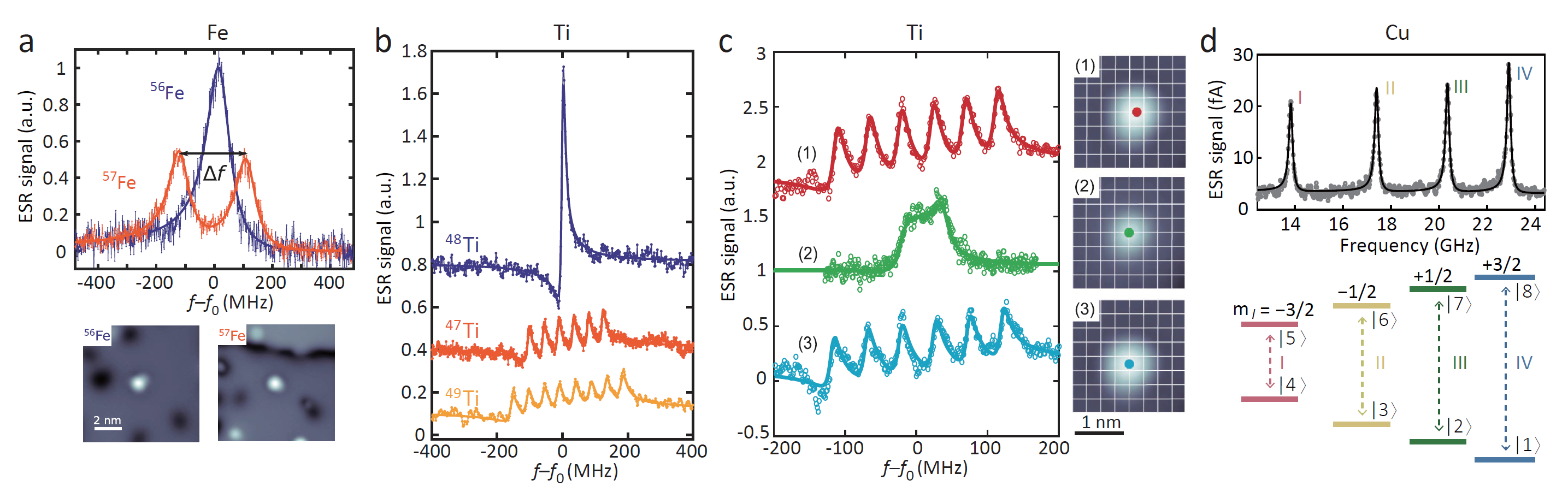}\hspace*{0\textwidth}%
  \caption{Hyperfine interactions of individual atoms on a surface. a) ESR spectra measured on \textsuperscript{56}Fe with zero nuclear spin (blue) and \textsuperscript{57}Fe with $I$=1/2 (orange). The bottom panels show STM images of \textsuperscript{56}Fe and \textsuperscript{57}Fe. b) ESR spectra measured on bridge-site Ti atoms with different nuclear spins. A single ESR peak was observed on \textsuperscript{48}Ti that carries no nuclear spin, while 6 and 8 ESR peaks were seen for \textsuperscript{47}Ti ($I$=5/2) and \textsuperscript{49}Ti ($I$=7/2), respectively. c) Binding-site-dependent ESR spectra of the same \textsuperscript{47}Ti atom moved by lateral atom manipulation. Right: STM images showing the Ti binding sites where the spectra on the left are taken. The Ti atom was changed from (1) a bridge site to (2) an oxygen site then back to (3) a bridge site. The interception points of the grid correspond to oxygen sites of MgO. d) ESR spectrum of \textsuperscript{65}Cu ($I$=3/2) on MgO. Lower panel: Schematic energy diagram of Cu with $S$=1/2 and $I$=3/2. (a--c) Reproduced with permission.\textsuperscript{\cite{Willke336}} 2018, AAAS. (d) Reproduced with permission.\textsuperscript{\cite{Yang:2018aa}} 2018, Springer Nature.}
  \label{fig:Hyperfine}
\end{figure}

We are now ready to discuss the experimental ESR-STM studies of the hyperfine and nuclear electric quadrupole interactions in Fe, Ti, and Cu atoms on MgO, as summarized in Figure \ref{fig:Hyperfine}. Fe atoms on MgO are typically seen to host only one ESR peak, which corresponds to \textsuperscript{56}Fe and \textsuperscript{54}Fe isotopes (both with $I = 0$) that have a combined abundance of 98\%. In a small fraction of Fe atoms, however, two split ESR peaks were observed with a splitting $\Delta f = 231 \pm 5$ MHz (Figure \ref{fig:Hyperfine}a). This splitting originates from the nuclear spin of \textsuperscript{57}Fe ($I=1/2$ with $\sim$2\% natural abundance). The two observed ESR peaks correspond to two nuclear spin states, and their almost equal ESR peak heights indicate that the thermal populations of the two nuclear spins are nearly identical. The simultaneous appearance of two ESR peaks in one spectrum implies that the nuclear spin relaxes faster than the measurement duration of $\sim$1 ms within a lock-in cycle (see section \ref{section: ensemble-detection}). To relate the ESR splitting to the hyperfine coupling strength, first note that the large electron spin anisotropy of Fe on MgO sets its electron spin perpendicular to the surface. Through the hyperfine interaction, the nuclear spin is also aligned to this direction. The Hamiltonian of \textsuperscript{57}Fe on MgO can then be written as
\begin{equation}
  H = g_{z} \mu_{\mathrm{B}} B_{z} S_{z} + A_{z} I_{z} S_{z},
\end{equation}
where the $z$-axis is the out-of-plane direction, and the nuclear electric quadrupole interaction is disregarded. Since we are only concerned with the two lowest electron spin states of Fe (which are approximately the $m_{S} = \pm 2$ states) \cite{Baumann417}, here we treat Fe as an effective two-level system by ignoring the other three spin states (which justifies the absence of the magnetic anisotropy term in the Hamiltonian in Equation \ref{HyperfineHamiltonian}). Because the Fe hyperfine spectrum involves $m_{S} = \pm 2$ and $m_{I} = \pm 1/2$ states, the ESR splitting in Figure \ref{fig:Hyperfine}a corresponds to $\Delta f = \Delta m_{S} \Delta m_{I} A_{z} = 4A_{z}$, which allows the determination of the $z$-component of the hyperfine coupling constant as $A_{z} = 58$ MHz. \hfill\break

The hyperfine spectra of Ti on MgO have more diversity than Fe in that Ti's hyperfine interactions are nuclei- and site-dependent. Figure \ref{fig:Hyperfine}b shows three different ESR spectra obtained for bridge-site Ti. Because one ESR peak splits into $2I+1$ sub-peaks in the presence of a nuclear spin $I$, the observed one-, six-, and eight-peak ESR spectra correspond to \textsuperscript{48}Ti ($I=0$), \textsuperscript{47}Ti ($I=5/2$) and \textsuperscript{49}Ti ($I=7/2$), respectively. The ESR peak splittings were found to be equal ($\sim$47 MHz) for \textsuperscript{47}Ti and \textsuperscript{49}Ti, indicating their indistinguishable nuclear gyromagnetic ratios. \hfill\break

Interestingly, STM atom manipulation can be used to alter Ti's hyperfine interaction by modifying its binding-site-dependent local environment. Figure \ref{fig:Hyperfine}c shows that the ESR peak splitting was significantly reduced when \textsuperscript{47}Ti was moved from a bridge site to an oxygen site. The splitting was restored when the atom was relocated to another bridge site. In ref. \cite{Willke336}, a higher-resolution ESR spectrum shows that the broad ESR peak for oxygen-site Ti is composed of several not-fully-resolved peaks with irregular intervals and nonuniform intensities. To elucidate why this is the case, consider the full Hamiltonian in Equation \ref{HyperfineHamiltonian}

\begin{equation}
  H = H_{\mathrm{EZ}} + H_{\mathrm{HF}} + H_{\mathrm{NQ}} 
    = \mu_{\mathrm{B}} \sum_{i=x,y,z} g_{i} B_{i} S_{i} + \sum_{i=x,y,z} A_{i}S_{i}I_{i} + \sum_{i=x,y,z} P_{i}I_{i}^{2}.
\end{equation}

For oxygen-site Ti, the smaller ESR peak splittings reflect a reduction of the hyperfine constant, $A_i$, making it comparable to the nuclear quadrupole interaction strength, $P_i$. Therefore, ill-defined eigenstates in the electron and nuclear spin space result in irregularly spaced ESR peaks seen on oxygen-site Ti. \hfill\break

A more careful, orbital-based analysis reveals the origin of the very different hyperfine coupling strengths between bridge- and oxygen-site Ti. Recall that the hyperfine interaction contains two contributions: an isotropic contact term and an anisotropic dipolar term. The contact contribution for bridge-site Ti ($+50$ MHz) was found to be already larger than oxygen-site Ti ($+19$ MHz). The dipolar contribution additionally enhances the hyperfine coupling of bridge-site Ti but cancels it for oxygen-site Ti due to their different orbital occupations \cite{Willke336}. \hfill\break

Individual Cu atoms on MgO \cite{Yang:2018aa} exhibit hyperfine interactions that are two orders of magnitude greater than those of Ti. Cu atoms have two stable isotopes, \textsuperscript{63}Cu and \textsuperscript{65}Cu, with natural abundances of 69\% and 31\%, respectively. On MgO, both isotopes have electron spins of 1/2 and nuclear spins of 3/2, and the eight spin eigenstates are labeled as $\ket{i}$ ($i=$1 to 8) in Figure \ref{fig:Hyperfine}d. The ESR spectra of \textsuperscript{65}Cu show four hyperfine-split ESR peaks. The hyperfine constant, $A$, extracted from the four ESR frequencies \cite{Yang:2018aa} is about 2.86 GHz for \textsuperscript{63}Cu and about 3.05 GHz for \textsuperscript{65}Cu, which highlights how ESR-STM can be used to distinguish different isotopes atom-by-atom. The large hyperfine constants of Cu on MgO, compared to Cu in other environments or other atoms on MgO, stem from the significant $s$-electron contribution ($\sim$60\% occupation of the 4$s$ orbital) to the electron spin, which produces a large contact hyperfine coupling. Due to the strong hyperfine interaction of Cu on MgO, considerable mixing of electron and nuclear spin states occurs and creates uneven spacing of the four ESR peaks. The different ESR frequencies of the four peaks also cause a thermally-induced nuclear polarization along its quantization axis $z$ ($\braket{I_{z}}/I \sim$ 1.7\% at 1.2 K and 0.65 T), which is reflected by the different ESR amplitudes of the four peaks in Figure \ref{fig:Hyperfine}d. \hfill\break

As shown in this section, ESR-detected hyperfine interactions are sensitive to the atomic surroundings. This type of hyperfine spectroscopy, combined with STM's imaging and atom manipulation capabilities, can be utilized to extract information from the local chemical and electronic environments of a single spin. 

\subsection{Functionalized-Tip-Based Magnetic-Field Sensing}

\label{section: 3.3}

Aside from using spin sensors on a surface as described in the previous sections, attaching a quantum spin to the STM tip enables scanning magnetometry at the atomic scale. The difficulty here is to identify a suitable spin carrier that can preserve its magnetic properties at the tip and allow a way to detect the magnetic signal. Spin-1 nickelocene molecules were found to be one such spin carrier, of which the spin states and magnetic anisotropy are robust in a variety of metallic environments including at a metallic tip apex, likely related to the protection by the cyclic $\pi^*$ orbital of the Cp rings that sandwiches the Ni atom \cite{doi:10.1021/acs.nanolett.6b05204}. Given the resolvable spin states of nickelocene at the tip apex, a local magnetic field can alter the spin state energies and be detected through IETS or ESR spectroscopy. This is a major advantage compared to conventional spin-polarized STM, where the magnetic structure and excitations of the magnetic tip apex are mostly unknown. IETS-based magnetic-field imaging with nickelocene on an STM tip was recently accomplished in pioneering works in Refs. \cite{Czap:2019vi,Verlhac:2019tw}, where the magnetic exchange interactions from individual magnetic atoms or molecules on surfaces were spatially mapped by tracking the variations in IETS spectra of the nickelocene molecule on the tip \cite{Czap:2019vi,Verlhac:2019tw}. In a similar measurement scheme, a spin-1/2 cobaltocene molecule attached to an STM tip was found to exhibit a Kondo resonance, and the Kondo resonance splitting was used to extract the exchange field of an Fe atom on the surface \cite{Garnier:2020vt}.
\hfill\break

Albeit impressive realizations of atomic-scale magnetic-field imaging, mappings performed using conventional STM spectroscopy are limited to the sub-meV energy resolution due to thermal broadening effects (see the introduction in section \ref{section: 3}). We expect that future investigations using an ESR-active molecule attached to an STM tip can allow for the combination of high energy and spatial resolutions, thus enabling sensitive scanning quantum sensing in three spatial dimensions at the atomic scale.

\section{Quantum Control}
\label{section: 4}

Over the last two decades, global efforts have been made to promote the coherence of solid-state qubits to a level suitable for quantum information processing. Compared with ions or atoms trapped in a vacuum, solid-state spins are easier to scale up thanks to modern fabrication techniques, but they suffer from additional decoherence sources in the solid-state environment. One way to reduce decoherence is by using nuclear spins which can possess coherence times as long as hours \cite{Zhong2015} due to their insensitivity to magnetic noise, but this also means that nuclear spins are more difficult to manipulate via magnetic fields. \hfill\break

Quantum control of well-characterized single spins and coupled spin systems on surfaces has recently been demonstrated using ESR-STM \cite{Yang509}. Spins on surfaces have the potential to form bottom-up quantum nanodevices and shed light on decoherence mechanisms in solid-state environments. In this section, we first discuss the sources of decoherence of surface spins. We then discuss recent developments on the coherent control of a single spin and the extensions to multi-spin quantum control. \hfill\break
  
\subsection{Quantum Coherence of Spins on Surfaces}
\label{section: coherence}

Decoherence, the loss of a quantum system's phase information, occurs from its interactions with the surrounding environment. Ensuring a long coherence time is of practical importance to achieve high-fidelity quantum-information processing, high-sensitivity quantum sensing, and high-efficiency quantum communications. Improving coherence relies on a good understanding of the decoherence sources. \hfill \break


While the energy relaxation time ($T_{1}$) of surface spins has been intensively studied using STM \cite{Natterer:2017aa, Loth1628, Yan:2015tp, Paul:2017aa} and was found to exceed 10 ms for Fe on MgO in a high magnetic field \cite{Paul:2017aa}, the coherence time ($T_{2}$) has barely been explored until recently. The development of ESR-STM has prompted interest in identifying the decoherence sources of surface spins both experimentally \cite{Willkeeaaq1543, Baeeaau4159} and theoretically \cite{Delgado2017, Delgado2012, Gauyacq2015, Shakirov2016, Shakirov2017, IbanezAzpiroz2017, Vorndamme2020}. Section \ref{Decoherence_Single} discusses this experimental endeavor. Section \ref{Decoherence_Coupled} discusses the use of singlet-triplet states for the enhancement of quantum coherence. \hfill\break

\subsubsection{Decoherence of Single Spins on Surfaces}
\label{Decoherence_Single}

\begin{figure}[ht]
  \centering
  \includegraphics[width=1\linewidth]{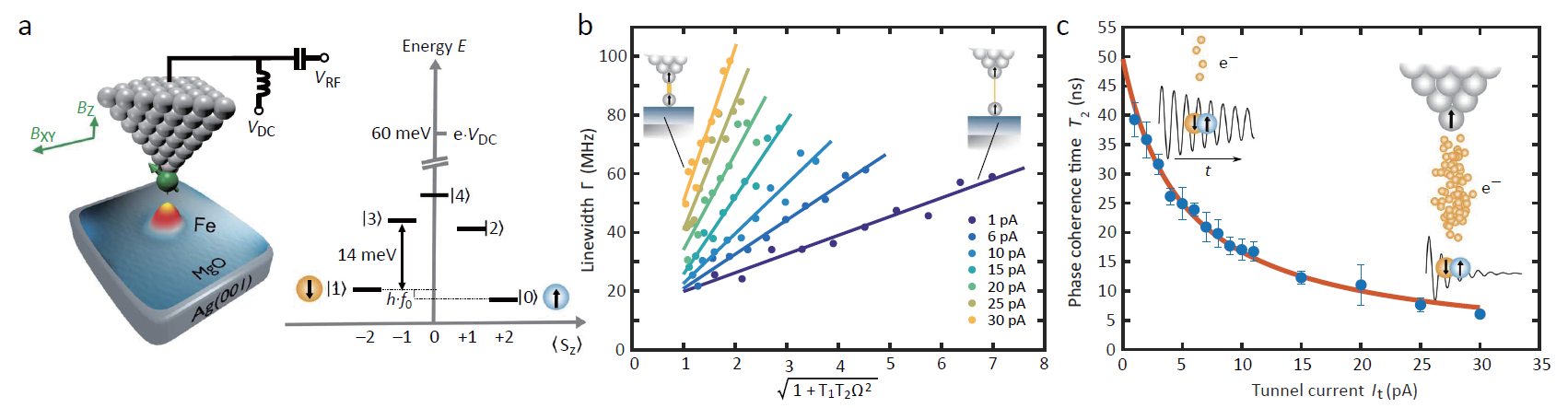}\hspace*{0\textwidth}%
  \caption{Spin coherence and decoherence in ESR-STM. a) Experimental setup for ESR measurements and the energy-level diagram of the five lowest energy levels of an Fe atom on MgO/Ag(100). The external magnetic field was applied mostly along the in-plane direction. The two lowest levels, $\ket{0}$ and $\ket{1}$, were split by $f_{0}$ $\approx$ 21 GHz (87 $\mu$eV) due to the small out-of-plane component of the external magnetic field ($B_{z}$). b) ESR linewidth as a function of the renormalized drive amplitude (see the $x$ axis label). Each curve corresponds to a different tunnel current (ranging from 1 to 30 pA). c) Extracted spin coherence time from (b) plotted against the tunnel current. 
  Reproduced with permission.\textsuperscript{\cite{Willkeeaaq1543}} 2018, AAAS.}
  \label{fig:PhilipSciAdv2017}
\end{figure}

The first systematic experimental study of the coherence time ($T_{2}$) of a surface spin was performed on isolated Fe atoms on 2ML MgO (Figure \ref{fig:PhilipSciAdv2017}a) \cite{Willkeeaaq1543}. Under typical ESR-STM measurement conditions, two dominant decoherence sources were identified: the tunneling electrons and the local magnetic field fluctuations from the magnetic tip. 

The influence of tunneling electrons on the relaxation and decoherence of an Fe spin can be shown by measuring the ESR peak intensity and broadening as a function of RF power ($V_{\mathrm{RF}}$) at different current setpoints ($I_{\mathrm{t}}$) (Figure \ref{fig:PhilipSciAdv2017}b). 
The ESR peak intensity ($I _{\mathrm{peak}}$) and the linewidth ($\Gamma$) due to DC readout can be described by the steady-state solution to the Bloch equations (Equation \ref{Bloch-SS-z}) as

\begin{equation}
\Delta I_{\mathrm{DC}} = \frac{I _{\mathrm{peak}}}{1+(f-f_0)^2/(\Gamma/2)^2}  \propto (\braket{S_z} - \braket{S_z^0}), \ \ 
  I _{\mathrm{peak}} = I _{\mathrm{sat}} \cdot \frac{\Omega ^{2} T_{1} T_{2}}{\Omega ^{2} T _{1} T _{2} + 1}, \ \ \Gamma = \frac{1}{\pi T _{2}} \sqrt{1 + \Omega ^{2} T_{1} T_{2}},
  \label{ESR_Decoherence}
\end{equation}
where $f-f_0$ is the frequency detuning, and $\Omega$ is the on-resonance Rabi rate, which is typically proportional to $V_{\mathrm{RF}}$. When the two spin states are equally populated due to strong RF driving, the ESR peak intensity detected by $V_{\mathrm{DC}}$ is saturated to $I _{\mathrm{sat}}$ \cite{Willkeeaaq1543}. By fitting $I _{\mathrm{peak}}$ and $\Gamma$ to Equation \ref{ESR_Decoherence}, we can extract $T_{2}$ under different tunnel conditions from the slope of the curves shown in Figure \ref{fig:PhilipSciAdv2017}b. As plotted in Figure \ref{fig:PhilipSciAdv2017}c, $T_{2}$ rapidly increases with decreasing tunnel current $I_{\mathrm{t}}$ in a roughly inversely proportional fashion, showing that tunnel current has a considerable influence on $T_{2}$. A careful analysis reveals that the decoherence rate has both tunnel-current-dependent and tunnel-current-independent terms:

\begin{equation}
  T_{2}^{-1} = \frac{P_{T_{2}}}{e} \cdot I_{\mathrm{t}} + T_{2}^{-1}(0).
\end{equation}
The average probability of a tunnel electron that causes decoherence, quantified by $P_{T_{2}}$, was found to be as high as 64\% \cite{Willkeeaaq1543}. In contrast, only a very small fraction of tunneling electrons (0.5\%) causes the energy relaxation ($T_1$ process) of the Fe spin due to its large anisotropy barrier \cite{Paul:2017aa}. 
By extrapolating $T_{2}$ to the zero-current limit, one can obtain the current-independent decoherence rate, $T_{2}^{-1}(0)$, that corresponds $T_{2}(0)=50 \pm 6 \, \mathrm{ns}$ as shown in Figure \ref{fig:PhilipSciAdv2017}c. The value of $T_{2}^{-1}(0)$ and its temperature dependence were found to vary widely from one magnetic tip to the other, indicating that the current-independent decoherence is caused by thermally induced changes of the tip's magnetic moment. The tunnel current and tip's magnetic fluctuations are thus detrimental to spin coherence, which can be partially mitigated by using a remote driving and sensing scheme as discussed in section \ref{section: ELDOR}. \hfill\break

\subsubsection{Enhanced Quantum Coherence in Singlet-Triplet States}
\label{Decoherence_Coupled}

The decoherence in ESR-STM can be significantly suppressed by using specific two-level transitions that are designed to be insensitive to magnetic field fluctuations \cite{PhysRevLett.54.1000, PhysRevA.72.062316, Wolfowicz2013, PettaJ.2005}. These so-called ``atomic clock transitions'' can be constructed with spins on surfaces by using singlet and triplet states that carry zero magnetic quantum numbers, as we shall discuss in this section. \hfill\break

\begin{figure}[ht]
  \centering
  \includegraphics[width=1\linewidth]{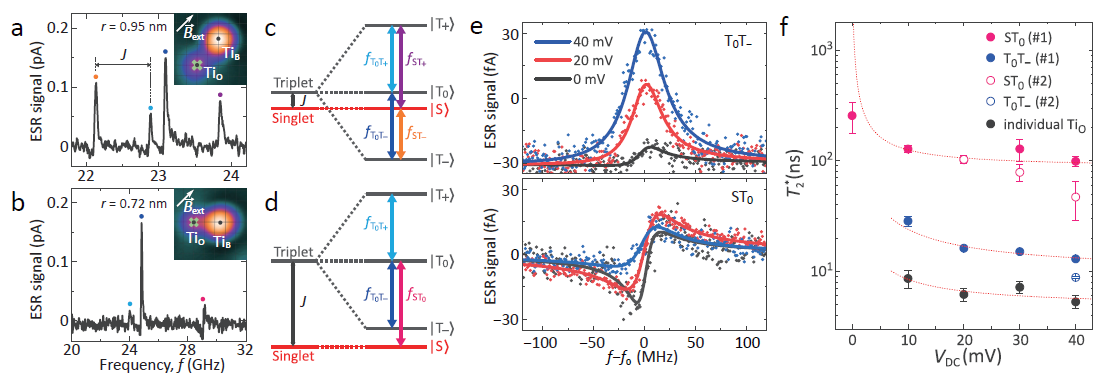}\hspace*{0\textwidth}%
  \caption{Enhanced spin coherence using singlet-triplet transitions. a,b) ESR spectrum for Ti-Ti pairs with atomic separations of (a) $r$ = 0.92 nm and (b) 0.72 nm in a nearly in-plane magnetic field of 0.9 T. Each pair composes of an oxygen-site Ti (Ti\textsubscript{O}) and a bridge-site Ti (Ti\textsubscript{B}). For the pair in (a) at $r$ = 0.92 nm, four ESR peaks were observed (the splitting corresponding to the coupling energy $J$ is labelled). For the closer pair in (b) at $r$ = 0.72 nm, three ESR peaks were observed where the magenta one arises from the ST\textsubscript{0} transition. Insets: STM images of Ti\textsubscript{O}-Ti\textsubscript{B} pairs. Grid line intersections represent the positions of oxygen atoms in the MgO lattice. The ``x'' marks represent the tip positions while obtaining the ESR spectra. c,d) Schematic energy level diagrams for Ti pairs in (a,b). In (c), the Zeeman energy is the dominant energy scale, and so the triplet state T\textsubscript{--} is the ground state. In (d), the coupling energy $J$ exceeds the Zeeman energy, and so the singlet state S becomes the ground state. The corresponding resonance peaks in (a,b) and ESR transitions in (c,d) are marked by the same colors. e) DC bias voltage dependence of ESR signals of the T\textsubscript{0}T\textsubscript{--} (top) and the ST\textsubscript{0} (bottom) transitions measured for the Ti pair in (b). The tip-atom separation was maintained at a constant value during this measurement. The ST\textsubscript{0} transition is measurable at zero $V_{\mathrm{DC}}$ due to the homodyne detection. f) The spin coherence time $T_{2}^{*}$ as a function of the DC bias voltage $V_{\mathrm{DC}}$ at a constant tip-atom separation. Red curves are fits to $1/V_{\mathrm{DC}}$. $\sharp 1$ and $\sharp 2$ label the results from two different Ti\textsubscript{O}-Ti\textsubscript{B} pairs (having the same separations but measured with two different tips). Reproduced with permission.\textsuperscript{\cite{Baeeaau4159}} 2018, AAAS.}
  \label{fig:YujeongSciAdv2018}
\end{figure}

In Ref. \cite{Baeeaau4159}, two strongly coupled spin-1/2 Ti atoms on MgO were used to create singlet and triplet states with zero magnetic quantum numbers ($m=0$) (Figure \ref{fig:YujeongSciAdv2018}). In general, two coupled spin-1/2 atoms yield four eigenstates denoted as $\ket{\text{S}(\xi)}$, $\ket{\text{T}_{0}(\xi)}$, $\ket{\text{T}_{-}}$, and $\ket{\text{T}_{+}}$ (see section \ref{section:spin-spin interaction}). $\ket{\text{S}(\xi)}$ and $\ket{\text{T}_{0}(\xi)}$ become the singlet and triplet states (S and T\textsubscript{0}) at the perfect mixing angle $\xi = \pi/2$. This occurs when the spin-spin interaction substantially exceeds the Zeeman energy difference between them (see section \ref{section:spin-spin interaction}). Using atom manipulation, the spacing between two Ti atoms can be precisely controlled. At atomic separation below 1.3 nm, the magnetic interaction between Ti atoms in the pair exponentially increases due to their exchange coupling, and the resulting spin states become closer to the S and T\textsubscript{0} states. \hfill\break

The ESR spectra for two different Ti-pair configurations are shown in Figure \ref{fig:YujeongSciAdv2018}a,b. When the Ti pair is separated relatively far by 0.92 nm, four ESR transitions of $\Delta m = \pm 1$ appear in the measurable frequency range (Figure \ref{fig:YujeongSciAdv2018}a,c). The four peaks can be divided into two groups, each split by the coupling energy $J \sim$ 0.8 GHz. Because in this pair the Ti-Ti coupling energy does not exceed their Zeeman energy difference at 0.9 T, the singlet (S) and triplet (T\textsubscript{0}) states are not well developed. When two Ti atoms are placed closer (0.72 nm, Figure \ref{fig:YujeongSciAdv2018}b,d), the Ti-Ti magnetic exchange energy increases to $\sim$30 GHz and dominates over their Zeeman energy difference at 0.9 T. As a result, nearly perfect S and T\textsubscript{0} states appear, with the singlet state S being the ground state. In the frequency range of 5--32 GHz, three ESR transitions were observed for the 0.72 nm pair, two of which are the triplet-triplet transitions of $\Delta m = \pm 1$ and the other is from the desired singlet-triplet transition of $\Delta m = 0$ (magenta curve in Figure \ref{fig:YujeongSciAdv2018}b). \hfill\break

Interestingly, the ESR driving of the singlet-triplet ST\textsubscript{0} transition differs from that of other transitions. First, we should note that unlike traditional ensemble ESR where the RF magnetic fields are designed to be maximized along a transverse axis $x$ (perpendicular to the quantization axis, $z$), the RF magnetic field in ESR-STM reflects the direction of the magnetic field gradient generated by the STM tip (see section \ref{section:driving}) and typically contains components both perpendicular and parallel to the quantization axis. We can include the RF driving magnetic fields into Equation \ref{eq: CoupledSpin-Hamiltonian2}. Assuming that the tip's magnetic field acts only on the first spin, the Hamiltonian becomes
\begin{gather}
  H = -\hbar \omega_{\mathrm{0,1}} S_{1z} \otimes \mathcal{I} - \hbar \omega_{\mathrm{0,2}} \mathcal{I} \otimes S_{2z} + (J_{\mathrm{e}} + 2J_{\mathrm{d}})  S_{1z} \otimes S_{2z} + (J_{\mathrm{e}} - J_{\mathrm{d}}) \times (S_{1x} \otimes S_{2x} + S_{1y} \otimes S_{2y})
  \nonumber \\
  +  2 \hbar \Omega_{1x} \cos (\omega t + \phi) S_{1x} \otimes \mathcal{I}  +  2 \hbar \Omega_{1z} \cos (\omega t + \phi)  S_{1z} \otimes \mathcal{I} \nonumber \\
  = \frac{1}{2} \Bigg [ \begin{smallmatrix}
    -\hbar \omega_{\mathrm{0,1}} - \hbar \omega_{\mathrm{0,2}} + \frac{J_{\mathrm{e}}+2J_{\mathrm{d}}}{2} + \hbar \tilde{\Omega}_{1z} & 0 & \hbar \tilde{\Omega}_{1x} & 0 \\
    0  & -\hbar \omega_{\mathrm{0,1}} + \hbar \omega_{\mathrm{0,2}} - \frac{J_{\mathrm{e}}+2J_{\mathrm{d}}}{2} + \hbar \tilde{\Omega}_{1z} & J_{\mathrm{e}}-J_{\mathrm{d}} & \hbar \tilde{\Omega}_{1x} \\
    \hbar \tilde{\Omega}_{1x}  & J_{\mathrm{e}}-J_{\mathrm{d}} & \hbar \omega_{\mathrm{0,1}} - \hbar \omega_{\mathrm{0,2}} - \frac{J_{\mathrm{e}}+2J_{\mathrm{d}}}{2} - \hbar \tilde{\Omega}_{1z} & 0    \\
    0 & \hbar \tilde{\Omega}_{1x}  &  0   &  \hbar \omega_{\mathrm{0,1}} +  \hbar \omega_{\mathrm{0,2}} + \frac{J_{\mathrm{e}}+2J_{\mathrm{d}}}{2} - \hbar \tilde{\Omega}_{1z} 
  \end{smallmatrix}\Bigg ].
\end{gather}
Here $\omega_{\mathrm{0,1}}$ and $\omega_{\mathrm{0,2}}$ are the Larmor frequencies of spins 1 and 2, respectively, $\omega$ is the applied RF frequency, and $\tilde{\Omega}_{1x,z}=2 \Omega_{1x,z} \cos (\omega t + \phi)$ are the shorthand expressions for the driving terms, where $\Omega_{1x}$ and $\Omega_{1z}$ are the Rabi rates that characterize the RF field strengths for spin 1 along the $x$ and $z$ directions in the lab frame. \hfill\break

When the two spins have the same Zeeman splitting, $\omega_{\mathrm{0,1}} = \omega_{\mathrm{0,2}}$, the eigenstates $\ket{\text{S}(\xi=\pi/2)}$, $\ket{\text{T}_{0}(\xi=\pi/2)}$, $\ket{\text{T}_{-}}$, and $\ket{\text{T}_{+}}$ can be connected to the original Zeeman state basis by a unitary transformation $U$ 
\begin{equation}
  \begin{bmatrix}
    \ket{00} \\ \frac{1}{\sqrt{2}}(\ket{01} - \ket{10}) \\ \frac{1}{\sqrt{2}}(\ket{01} + \ket{10}) \\  \ket{11}
  \end{bmatrix}
  = \begin{bmatrix}
    1 & 0 & 0 & 0\\
    0 & \frac{1}{\sqrt{2}} & -\frac{1}{\sqrt{2}} & 0\\
    0 & \frac{1}{\sqrt{2}} & \frac{1}{\sqrt{2}} & 0\\
    0 & 0 & 0 & 1\\
  \end{bmatrix}
  \begin{bmatrix}
    \ket{00} \\ \ket{01}  \\ \ket{10} \\  \ket{11}
  \end{bmatrix}
  = U \begin{bmatrix}
    \ket{00} \\ \ket{01}  \\ \ket{10} \\  \ket{11}
  \end{bmatrix}.
\end{equation}

In the new singlet-triplet basis, the Hamiltonian becomes
\begin{equation}
  U H U^\dagger = 
  \frac{1}{2} \Bigg [ \begin{smallmatrix}
    -\hbar \omega_{\mathrm{0,1}} - \hbar \omega_{\mathrm{0,2}} + \frac{J_{\mathrm{e}}+2J_{\mathrm{d}}}{2} + \hbar \tilde{\Omega}_{1z} & \frac{-\hbar \tilde{\Omega}_{1x}}{\sqrt{2}} & \frac{\hbar \tilde{\Omega}_{1x}}{\sqrt{2}}  & 0 \\
    \frac{-\hbar \tilde{\Omega}_{1x}}{\sqrt{2}}  & -\frac{3J_{\mathrm{e}}}{2} & -\hbar \omega_{\mathrm{0,1}} + \hbar \omega_{\mathrm{0,2}} + \hbar \tilde{\Omega}_{1z} & \frac{\hbar \tilde{\Omega}_{1x}}{\sqrt{2}} \\
    \frac{\hbar \tilde{\Omega}_{1x}}{\sqrt{2}}  & -\hbar \omega_{\mathrm{0,1}} + \hbar \omega_{\mathrm{0,2}} + \hbar \tilde{\Omega}_{1z} & \frac{J_{\mathrm{e}}}{2}-2J_{\mathrm{d}}  &  \frac{\hbar \tilde{\Omega}_{1x}}{\sqrt{2}} \\
    0 & \frac{\hbar \tilde{\Omega}_{1x}}{\sqrt{2}}  &  \frac{\hbar \tilde{\Omega}_{1x}}{\sqrt{2}}  &  \hbar \omega_{\mathrm{0,1}} +  \hbar \omega_{\mathrm{0,2}} + \frac{J_{\mathrm{e}}+2J_{\mathrm{d}}}{2} - \hbar \tilde{\Omega}_{1z} 
  \end{smallmatrix}\Bigg ].
  \label{eq: STHamiltonian}
\end{equation}
Focusing on the central $2 \times 2$ block of Equation \ref{eq: STHamiltonian} that corresponds to the S and T\textsubscript{0} states, we can see that the energy levels for the ST\textsubscript{0} transition (diagonal terms) are determined only by the coupling energies ($J_{\mathrm{e}}$ and $J_{\mathrm{d}}$), and so the S and T\textsubscript{0} states are insensitive to magnetic field fluctuations. The off-diagonal terms of the central $2 \times 2$ ST\textsubscript{0} matrix represents the Zeeman energy difference between the two spins. As a result, in order to resonantly drive the ST\textsubscript{0} transition, a local RF field ($\tilde{\Omega}_{1z}$) needs to be supplied to only one of the two spins along its quantization axis, $z$ (to supply a Zeeman energy difference). 
Other transitions with $\Delta m = \pm 1$, in contrast, have energy splittings that are sensitive to magnetic field fluctuations and can be driven by both global and local transverse RF magnetic fields (as represented by the $\tilde{\Omega}_{1x}$ terms in Equation \ref{eq: STHamiltonian}).  \hfill\break

The noise immunity of the singlet-triplet transition was found to greatly enhance its coherence time. Under typical ESR conditions of $V_{\mathrm{DC}}=40 \, \mathrm{mV}$ and $I_{\mathrm{t}}=10 \, \mathrm{pA}$, the spin coherence time for the singlet-triplet (ST\textsubscript{0}) transition was found to be $T_{2}^{*} \sim 100 \, \mathrm{ns}$. This value is $\sim$8 times longer than the $T_{2}^{*}$ for the T\textsubscript{0}T\textsubscript{--} transition (where $\Delta m \ne 0$) in the same atomic structure and $\sim$20 longer than that of a single Ti atom (Figure \ref{fig:YujeongSciAdv2018}f). Note that the term $T_{2}^{*}$ is used here because the measured coherence times include the effect of inhomogeneous line broadening caused by temporal fluctuations of time-varying magnetic fields during the time-ensemble averaged ESR measurement (see section \ref{section: ensemble-detection}). \hfill\break

Another interesting observation is the very different ESR lineshapes between the singlet-triplet (ST\textsubscript{0}) transition and other transitions with $\Delta m = \pm 1$ (Figure \ref{fig:YujeongSciAdv2018}e), which implies their different detection mechanisms. In Figure \ref{fig:YujeongSciAdv2018}e, the ESR lineshapes of the ST\textsubscript{0} transition are highly asymmetric (lower panel), while they are much more symmetric for the T\textsubscript{0}T\textsubscript{--} transition (upper panel). The symmetric ESR lineshape of the T\textsubscript{0}T\textsubscript{--} transition originates from the DC detection of the spin populations by the DC bias voltage $V_{\mathrm{DC}}$ (see Figure \ref{fig:ESRSTMessentials}b and section \ref{section: ensemble-detection}), and hence the peak amplitudes decrease with lowering $V_{\mathrm{DC}}$ (upper panel of Figure \ref{fig:YujeongSciAdv2018}e). In contrast, the CW measurement of the singlet-triplet transition is not from DC detection \cite{Baeeaau4159} because the time-averaged populations of the spin states do not vary and thus no time-averaged DC signal is present (see, however, the pulsed DC readout in Figure \ref{fig:Detuning}d). The ST\textsubscript{0} detection instead relies on the CW homodyne detection of the precessing transverse magnetization along the $\ket{10}$-$\ket{01}$ axis in the singlet-triplet subspace. During the precession, $\ket{10}$ and $\ket{01}$ states have different spin orientations for the spin under the tip, yielding an RF TMR that is detectable through $V_{\mathrm{RF}}$ (similar to the homodyne detection scheme of a single spin in section \ref{section: ensemble-detection}).  \hfill\break

As a result of the homodyne detection of the singlet-triplet (ST\textsubscript{0}) transition, the spin coherence time can be further enhanced by decreasing the DC bias voltage and the DC current. During this measurement, the STM feedback loop was kept open to maintain the same tip-atom separation and consequently the same driving strength. The ESR lineshapes of the ST\textsubscript{0} transition was found to be sharper with reduced $V_{\mathrm{DC}}$ and became best resolved at $V_{\mathrm{DC}}=0$ (lower panel of Figure \ref{fig:YujeongSciAdv2018}e). This is because the coherence time $T_{2}^{*}$ was much increased due to the absence of the tunnel current, while the driving amplitude remains the same. The spin coherence time of the ST\textsubscript{0} transition was found to be increased to $T_{2}^{*} \sim$ 260 ns at $V_{\mathrm{DC}}=0$, an order-of-magnitude longer than single Ti atoms. The remaining deviation of $T_{2}^{*}$ from the intercept in Figure \ref{fig:YujeongSciAdv2018}f can be attributed to the RF tunnel current generated by $V_{\mathrm{RF}}$, the finite sample temperature, and the relatively short spin relaxation time $T_{1}$. The spin coherence times $T_{2}^{*}$ of other transitions also benefit from lower $V_{\mathrm{DC}}$ as shown in Figure \ref{fig:YujeongSciAdv2018}f (see section \ref{Decoherence_Single}), but their signal strengths are also greatly reduced because their detection in this case requires a DC bias voltage.  \hfill\break

Further improvements to spin coherence can be potentially obtained by remotely sensing a surface spin far from the STM tip, using new atomic species with longer spin relaxation times, lowering the sample temperature, or increasing the thickness of MgO layer for better isolation from Ag's conduction electrons. Artificial spin structures with higher-order insensitivity to field fluctuations can also be designed and constructed. \hfill\break

\subsection{Coherent Manipulation of Individual Spins on a Surface}
\label{PulsedESR}


Coherent manipulation of spin states can be achieved via pulsed spin resonance, in which a sequence of control pulses is designed to generate the desired spin state and provide filtering of the environmental noise (through Hahn echo and other dynamical decoupling sequences) \cite{RevModPhys.89.035002, PhysRevA.58.2733, Bluhm2011}. The limitations of CW-ESR can be largely overcome by pulsed schemes, and the results of pulsed ESR measurements are often easier to interpret. \hfill\break

\begin{figure}[ht]
  \centering
  \includegraphics[page=1,width=1\linewidth]{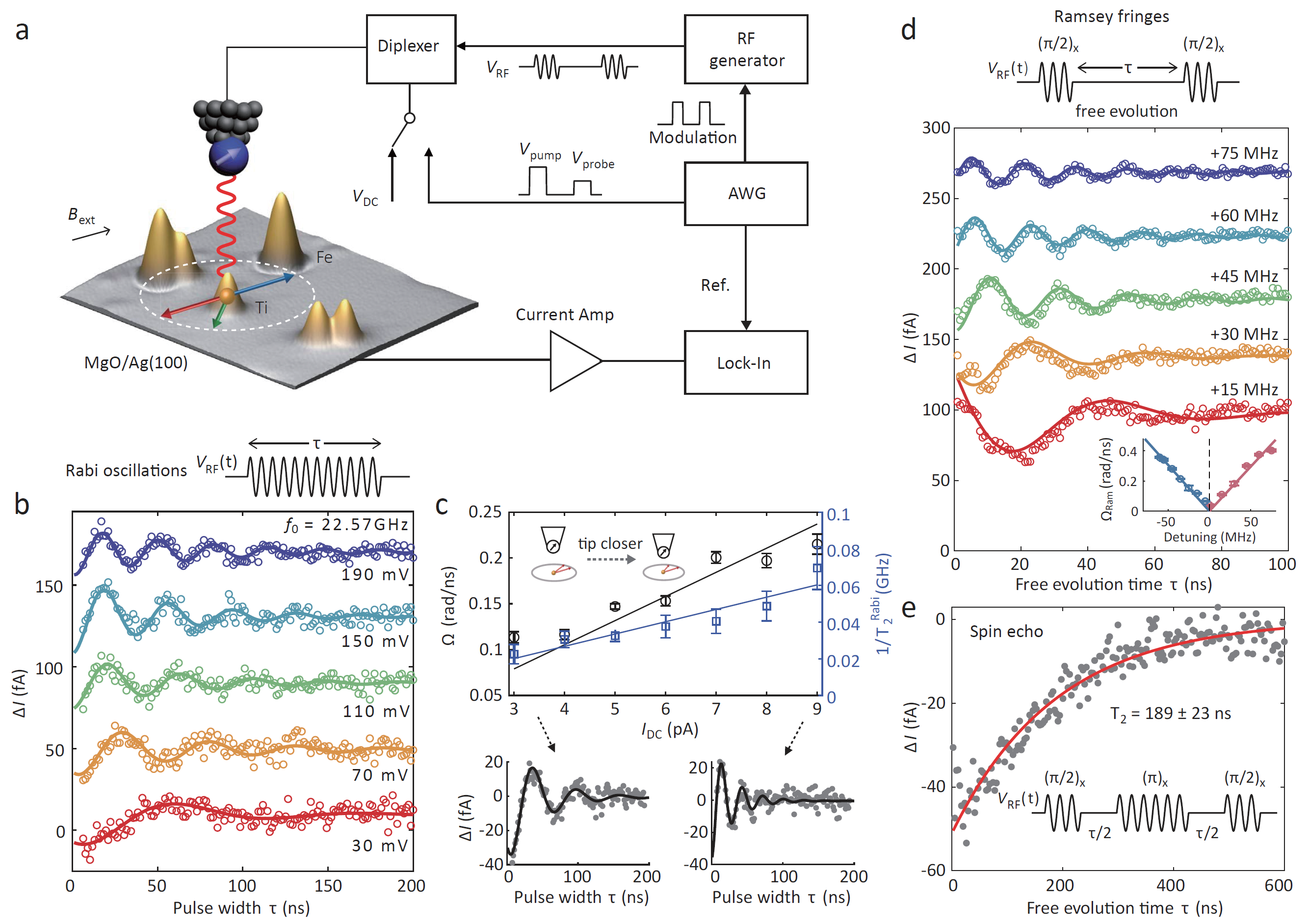}\hspace*{0\textwidth}%
  \caption{Coherent spin manipulation of a single Ti atom. a) Experimental setup for pulsed ESR-STM. The STM is equipped with an RF generator and an arbitrary waveform generator (AWG). A sequence of RF and DC pulses are delivered to an oxygen-site Ti atom on MgO/Ag(100) via the STM tip. b) Rabi oscillations of the Ti spin at different RF powers ($V_{\mathrm{RF}}$). c) Rabi rates and coherence times measured at different tip-atom separations. d) Ramsey fringe measurement of the Ti spin. e) Hahn echo measurements of the Ti spin, yielding $T_2$ = 189 $\pm$ 23 ns. Reproduced with permission.\textsuperscript{\cite{Yang509}} 2019, AAAS.}
  \label{fig:KaiScience2019}
\end{figure}

Recently, the coherent control of individual atomic spins on a surface has been achieved using ESR-STM \cite{Yang509}. The experimental setup for pulsed ESR-STM is shown in Figure \ref{fig:KaiScience2019}a. An arbitrary waveform generator (AWG) gates an RF signal generator to create short RF pulses with a constant amplitude and frequency. The pulse sequences are then applied to the surface spin through the tip, and the resultant modification in the surface spin state is measured using the tunnel current in a time-ensemble averaged fashion (see section \ref{section: ensemble-detection}). \hfill\break

The pulsed ESR-STM experiments were performed on spin-1/2 oxygen-site Ti atoms on MgO (Figure \ref{fig:KaiScience2019}a) (it is also possible to perform pulsed ESR-STM measurements on bridge-site Ti atoms and Fe atoms on 2ML MgO, despite somewhat lower Rabi rates) \cite{Yang509}. Figure \ref{fig:KaiScience2019}b shows the Rabi oscillations of an oxygen-site Ti atom at different RF driving powers. When an RF pulse is applied at the resonance frequency, the Ti spin coherently rotates between its two spin states, $\ket{0}$ and $\ket{1}$, at the Rabi rate $\Omega$ (see section \ref{section: 2.1}). This coherent spin rotation can be detected by incrementally increasing the RF pulse width and measuring the tunnel current at each width (upper panel of Figure \ref{fig:KaiScience2019}b). The resultant Rabi signals show oscillatory behavior, as expected, and the oscillation amplitude decays exponentially over time due to decoherence with a rate of $T_{2}^{\mathrm{Rabi}}=40$ ns (Figure \ref{fig:KaiScience2019}b). Note that in these measurements the DC readout voltage has been applied either in a pulsed fashion after the RF pulses or continuously during the entire sequence. The former method's signal is more straightforward to interpret, while the latter can be seen as its time integral, yielding stronger signals and allowing an easier operation of the STM's feedback loop \cite{Yang509}.  \hfill\break

As shown in Figure \ref{fig:KaiScience2019}b and c, increasing the tunnel current at constant DC and RF voltages induces a linear increase of the Rabi rate $\Omega$, as a result of a larger RF electric field and an increased tip's magnetic field gradient at the Ti atom (see section \ref{section:driving}). Increasing the tunnel current, on the other hand, decreases the spin coherence time (see section \ref{Decoherence_Single}). It is found that the spin decoherence rate and the Rabi rate scale similarly with the tip-atom separation for a given tip (Figure \ref{fig:KaiScience2019}c), and so the number of Rabi cycles within the coherence time is almost independent of the tip-atom separation (at a fixed RF power). \hfill\break

Ramsey fringe and spin-echo measurements have also been carried out on single oxygen-site Ti spins \cite{Yang509}. The Ramsey fringe measurement is composed of two phase-coherent RF pulses, each corresponding to a $\pi$/2-rotation, with a waiting time $\tau$ in between (upper panel of Figure \ref{fig:KaiScience2019}d). When the spin is initially in its ground state, $\ket{\psi_{\mathrm{i}}} = \ket{0}$, applying the first $\pi$/2-pulse to the spin causes it to enter a superposition state of $\ket{0}$ and $\ket{1}$: $\ket{\psi(t=0)}=(\ket{0}+\ket{1})/\sqrt{2}$. During the waiting time $\tau$, the spin undergoes free evolution, where the superposition state accumulates a relative phase $\ket{\psi(t=\tau)}=(\ket{0}+e^{-i  2 \pi \Delta f \tau} \ket{1}$)/$\sqrt{2}$ with $\Delta f$ being the frequency detuning. With the second $\pi$/2-pulse applied, the spin state becomes $\ket{\psi_{\mathrm{f}}}=\frac{1}{2}(1+e^{-i  2 \pi \Delta f \tau})\ket{0}+\frac{1}{2}(1-e^{-i  2 \pi \Delta f \tau})\ket{1}$ and is subsequently measured. Here we assumed that the first $\pi$/2 rotation pulses is about the $+y$ axis in the rotating frame and the second rotation is about the $-y$ axis. When the two $\pi$/2-pulses are exactly on resonance, there is no accumulation of phase, and the spin state stays in the ground state $\ket{0}$ after the sequence. If we intentionally provide a frequency detuning ($\Delta f$) to the RF pulses, depending on the accumulated relative phase, the observed spin state can be in either the ground or excited state. The probability that the spin ends up in the excited state is given by

\begin{equation}
  P_{\ket{1}}(\tau) = \abs{\braket{1|\psi_{\mathrm{f}}}}^{2} = \sin^{2}(\pi \Delta f \tau) = \frac{1}{2}[1-\cos(2 \pi \Delta f \tau)].
\end{equation}

A population readout, such as in the DC readout of ESR-STM, should then show an oscillatory signal as a function of waiting time $\tau$, with the oscillation frequency determined by the frequency detuning $\Delta f$. This behavior is known as the Ramsey fringe and has been measured on a single Ti spin as shown in Figure \ref{fig:KaiScience2019}d. In this phase-sensitive measurement, the oscillations decay due to decoherence. In Figure \ref{fig:KaiScience2019}d, the envelope of the decaying oscillations yields a coherence time $T_{2}^{*} \approx 40 \, \mathrm{ns}$, which is comparable to the one obtained from the Rabi oscillation measurement despite the shorter $\pi/2$-pulses used in Ramsey measurements.\hfill\break

In a time-ensemble averaged measurement of an individual spin, slow temporal magnetic field variations within the second-long data acquisition time can cause inhomogeneous broadening and an effective dephasing (see section \ref{section: ensemble-detection}). Spin echo experiments can be performed to reduce this dephasing effect. Here, an intermediate $\pi$-pulse is inserted at the middle of the two $\pi$/2-pulses in the Ramsey sequence (inset of Figure \ref{fig:KaiScience2019}e). The inhomogeneous dephasing induced by a slow magnetic field variation has the opposite dephasing contributions in the first and second periods of free evolution, thus canceling its effect. By removing these decoherence sources, the spin coherence time extracted from the spin echo measurements (Figure \ref{fig:KaiScience2019}e) is about 190 ns, which is several times longer than the spin coherence time obtained from Rabi and Ramsey fringe measurements. \hfill\break

This section discusses the coherent spin manipulation of a surface spin. A relatively large Rabi rate (one cycle per 20 ns), enabled by the strong RF electric field in the tunnel junction, has been measured. Upon future improvements on spin coherence, higher spin controllability might be achieved.

\subsection{Multi-Spin Control}
\subsubsection{Electron-Nuclear Spin Control}
\label{section: Cu}


\begin{figure}[ht]
  \centering
  \includegraphics[page=1,width=1 \linewidth]{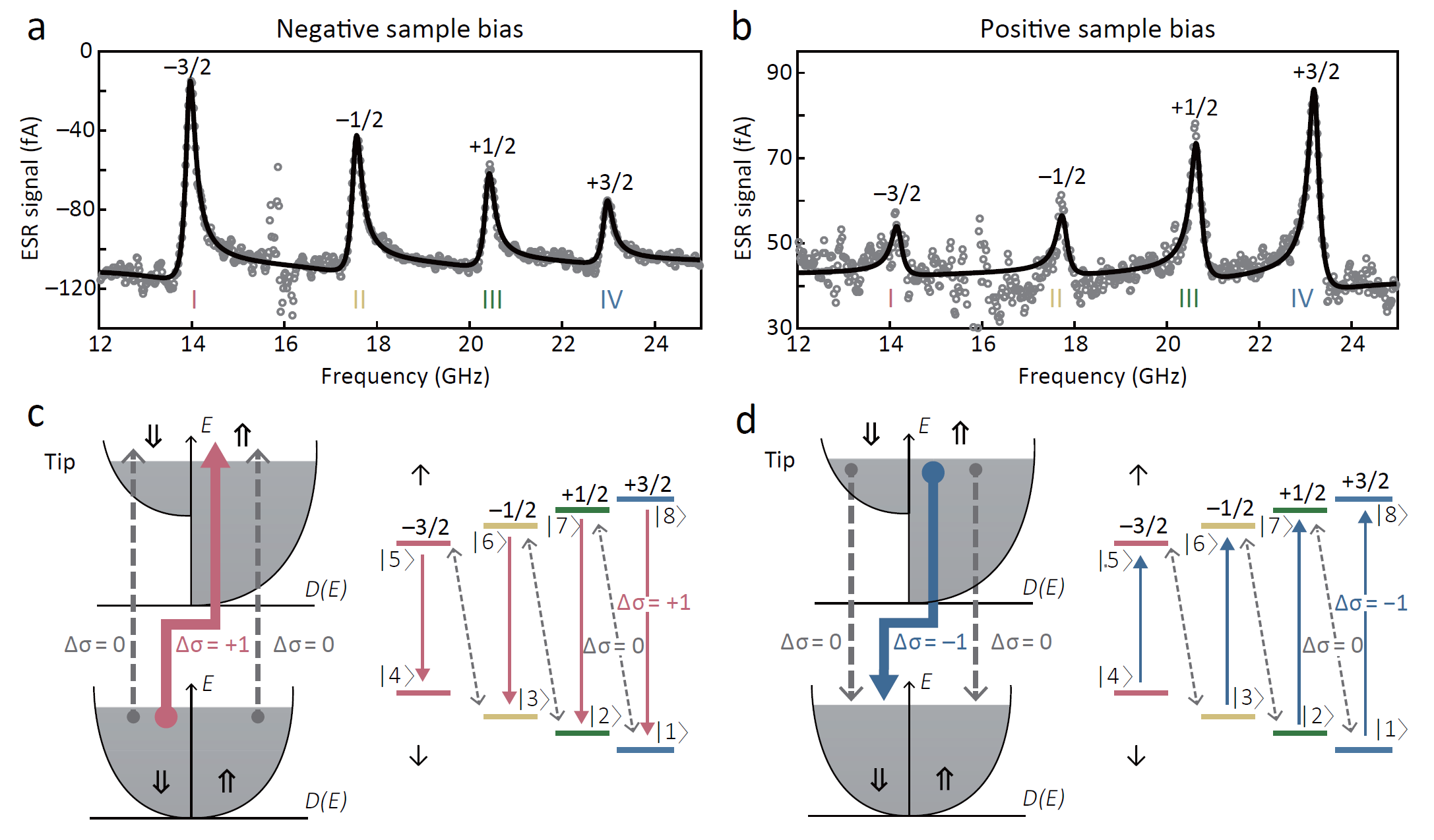}\hspace*{0\textwidth}%
  \caption{Electrically controlled nuclear spin polarization of a single Cu atom on MgO/Ag(100). a,b) ESR spectra of a \textsuperscript{65}Cu atom measured at (a) $V_{\mathrm{DC}}$$=$$-22$  mV and (b) +22 mV using the same tip. c,d) Nuclear spin polarization by tunneling electrons at (c) negative and (d) positive bias voltages. $\Delta \sigma$ is the change of the tunneling electron spin after the scattering with Cu's electron or nuclear spin. The left panels in (c) and (d) show how spin-dependent tunneling electron can be used to pump the Cu electron spin (see section \ref{section: 2.4}). The right panels in (c) and (d) show how the pumping of Cu electron and nuclear spins leads to spin state accumulation at states $\ket{4}$ (at negative bias) and $\ket{8}$ (at positive bias). Eigenstates $\ket{1}$--$\ket{8}$ are not purely Zeeman product states, but for convenience their dominant nuclear and electron spin states are marked by $\pm$3/2, $\pm$1/2 and $\uparrow$, $\downarrow$, respectively. Reproduced with permission.\textsuperscript{\cite{Yang:2018aa}} 2018, Springer Nature.}
  \label{fig:KaiNatNano2018}
\end{figure}

In ESR-STM, the control of a nuclear spin has been achieved in a strongly coupled electron-nuclear spin system, single Cu atoms on 2ML MgO/Ag(100). Both \textsuperscript{63}Cu and \textsuperscript{65}Cu show large hyperfine interactions of about 2.86 GHz and 3.05 GHz, respectively, as seen from ESR-STM measurements shown in section \ref{section: nuclear-sensing} and Ref. \cite{Yang:2018aa}. Despite the large hyperfine interactions, the nuclear spin polarization of Cu is still only about 1.7\% at thermal equilibrium at 1.2 K and 0.65 T. \hfill\break

The nuclear spin polarization of Cu can be driven beyond this level by first polarizing Cu's electron spin using spin-polarized current (i.e., the spin-transfer torque) then transferring this polarization to Cu's nuclear spin (with the help of tunneling electron scatterings, see below). The idea here, known as hyperpolarization in general \cite{https://doi.org/10.1002/chem.201405253, Albert1994}, is based on an effective angular momentum transfer from a more spin-polarized source (Cu's electron spins in this case) to a less spin-polarized target (Cu's nuclear spins). To understand this process, first consider Cu's spins states. Since a Cu atom has $S=1/2$ and $I=3/2$, it hosts eight eigenstates, denoted by $\ket{i}$ ($i=$1 to 8) in Figure \ref{fig:KaiNatNano2018}. Under a strong magnetic field where the electron Zeeman energy is much larger than the hyperfine interaction energy, the eigenstates are close to the Zeeman product states $\ket{m_{\mathrm{S}},m_{\mathrm{I}}}$ (with $m_{\mathrm{S}}=\downarrow,\uparrow$ and $m_{\mathrm{I}}=\pm\frac{1}{2},\pm\frac{3}{2}$). The flip-flop terms of the hyperfine coupling ($S^{+}I^{-}$ and $S^{-}I^{+}$) induce further hybridization between $\ket{\downarrow,m_{\mathrm{I}}}$ and $\ket{\uparrow,m_{\mathrm{I}}-1}$, where $m_{\mathrm{I}}$$=$$+\frac{3}{2},\pm\frac{1}{2}$, resulting in three pairs of hybridized states ($\ket{i}$ and $\ket{8-i}$ with $i=$1, 2, and 3, shown in the right panels of Figure \ref{fig:KaiNatNano2018}c,d). \hfill\break

Populations in the eight spin eigenstates of Cu can be manipulated by scatterings with spin-polarized tunneling electrons, which are essential for Cu's nuclear spin hyperpolarization. Two types of Cu spin transitions can be induced by tunneling electron scatterings, marked by dashed and solid arrows in the right panels of Figure \ref{fig:KaiNatNano2018}c,d. The first type of transitions happens between eigenstates $\ket{i}$ and $\ket{8-i}$ (dashed arrows in the right panels of Figure \ref{fig:KaiNatNano2018}c,d). Due to the nonzero $S_z$ components between $\ket{i}$ and $\ket{8-i}$ (i.e., with $\abs{\braket{i|S_{z}|8-i}}^{2} \ne 0$, $i = 1$--7), their transitions can be triggered by the potential or $S_z$ scattering term in Equation \ref{Tunnel-rate} that does not flip the tunneling electron spin. The second type of scattering happens between the eigenstates $\ket{i}$ and $\ket{9-i}$ ($i = 1$--8) due to spin flip-flops between the tunneling electron spin and Cu's electron spin (solid arrows in the right panels of Figure \ref{fig:KaiNatNano2018}c,d). This process can occur because the eigenstates $\ket{i}$ and $\ket{9-i}$ have nonzero $S_x$ matrix elements (i.e., $\abs{\braket{i|S_{x}|9-i}}^{2} \ne 0$), where $x$ represents a direction perpendicular to the quantization axis $z$ (cf. Equation \ref{Tunnel-rate}). \hfill\break

The combination of the two types of scatterings allows nuclear hyperpolarization via spin-polarized tunnel electrons. When electrons tunnel from the sample to the tip at a negative sample bias voltage, nuclear spin populations are pumped to the $m_{I}=-\frac{3}{2}$ state by spin-polarized tunneling electrons (Figure \ref{fig:KaiNatNano2018}a, c). This is due to the fact that the tip spin here is predominantly polarized towards the spin-up state with $\sigma = +1/2$, thus at a negative sample bias voltage (where a tunnel electron ends up in the tip), an angular momentum transfer of $\Delta \sigma = +1$ occurs (Figure \ref{fig:KaiNatNano2018}c, left panel). As a result, the solid arrows in the right panel of Figure \ref{fig:KaiNatNano2018}c go downwards, which causes population accumulation in the state $\ket{4}$ with $m_{I}=-\frac{3}{2}$, and hence its stronger ESR peak in Figure \ref{fig:KaiNatNano2018}a. The situation is the opposite at a positive sample bias voltage because here the tunnel electron predominantly starts from the $\sigma = +1/2$ spin state in the tip and so $\Delta \sigma = -1$ transitions occur, causing population accumulation in the state $\ket{8}$ with $m_{I}=+\frac{3}{2}$ (Figure \ref{fig:KaiNatNano2018}b,d). Using high tunnel current, the maximal nuclear polarization achieved in Ref. \cite{Yang:2018aa} was about 17 times higher than the Boltzmann distribution at 1 K, which corresponds to an effective temperature of 200 mK. \hfill\break

This section discusses how the direction and magnitude of the nuclear spin polarization can be controlled by the polarity and amplitude of the spin-polarized tunnel current, respectively. Recent integration of ESR-STM with dilution refrigerators \cite{PhysRevB.103.155405} allows for higher thermal nuclear spin polarization, which can be further enhanced by hyperpolarization. The use of ESR-STM for coherent control and readout of nuclear spins is an interesting future direction.  \hfill\break

\subsubsection{Electron-Electron Spin Control}
\label{section: ELDOR}

The performance of useful quantum protocols requires the independent and simultaneous driving of multiple spins. STM-based spin driving relies on the strong electric field under the tip, and so it has been unclear whether multi-spin driving could ever be possible. A recent work tackles this issue \cite{2021arXiv210809880P} and presents exciting results demonstrating the driving and detection of two atomic spins. \hfill\break

This work uses a weakly-coupled pair of Ti with one ``local'' Ti placed in the tunnel junction and the other, ``remote'' Ti, located not directly under the tip apex. Next to the remote Ti is an single-atom magnet, Fe, which supplies a large magnetic field gradient (Fe can be treated as a single-atom magnet because the supplied RF frequencies are far from Fe's ESR frequencies). The idea here is that the ESR driving requires an oscillating magnetic field, which for a conventional ``local'' spin in the tunnel junction, necessitates (1) a strong oscillating electric field under the tip and (2) the inhomogeneous tip's magnetic field (see section \ref{section:driving}). Out of these two conditions, the oscillating electric field can at least propagate across microscopic distances, while the tip's magnetic field is highly confined to only the spin in the tunnel junction \cite{Yan:2015tp}. Therefore, the key requirement for driving a remote spin is to supply it with a sufficiently large magnetic field gradient, which in this case is provided by the nearby Fe atom. Indeed, through two types of double resonance spectroscopy, a clear change in the ESR signal of the local spin sensor was observed when the remote spin was driven at its resonance frequencies, demonstrating the realization of remote driving \cite{2021arXiv210809880P}. This remote driving scheme can be easily generalized to multiple coupled spins on a surface, paving the way for future quantum control and simulation protocols based on global unitary transformations of spins on a surface.

\section{Quantum Simulation}
\label{section: 5}
A bottom-up approach to understanding quantum materials can be obtained by constructing and measuring artificial atomic structures. Tailored spin nanostructures assembled atom-by-atom provide an attractive platform for the simulation of many-body quantum and classical spin Hamiltonians \cite{Khajetoorians:2019wq, RevModPhys.91.041001, Spinelli_2015}. A general interacting spin Hamiltonian can be written as
\begin{equation}
  H = \sum_{\braket{ij}} \boldsymbol{S}_i \cdot \mathbf{J}_{ij} \cdot  \boldsymbol{S}_j + \sum_{i = 1}^N \boldsymbol{S}_i \cdot \mathbf{D}_i \cdot  \boldsymbol{S}_i   + \sum_{i = 1}^N \mu_\mathrm{B} \boldsymbol{B}_i \cdot \mathbf{g}_i \cdot \boldsymbol{S}_i,
  \label{eq: NSpinH}
\end{equation}
where the definitions of the quantities follow those in Equation \ref{eq: spin-Hamiltonian}.
Here $\boldsymbol{B}_i$ represents the total magnetic field applied to spin $i$, which can include the external magnetic field $\boldsymbol{B}_{\mathrm{ext}}$, the tip's static magnetic field (see section \ref{section: tipfield}), and the RF driving fields. The atomically-precise fabrication capability of STM enables individual control of spin parameters $\boldsymbol{S}_i$ and $\mathbf{J}_{ij}$. Substrate engineering allows for a global control of $\mathbf{D}_i$ and, in the case of substrate-mediated interactions, $\mathbf{J}_{ij}$. Special substrates such as superconductors, heavy-element systems, and magnetic materials can induce electron pairing interactions, spin-orbit coupling, and magnetic interactions, respectively, which can be non-trivial to achieve in other quantum simulators \cite{Bloch:2012ty, Schafer:2020vg, RevModPhys.93.025001}. Another advantage of STM-based simulation is the flexibility of both the atomic species and the constructed patterns, which allows for the building of regular or irregular multi-species structures by design \cite{Khajetoorians:2019wq}. Spins with large $S$ or fast decoherence behave like classical magnetic moments and can be positioned on bipartite lattices to simulate a ferromagnet or an antiferromagnet, on frustrated lattices to simulate a spin ice \cite{Bramwell:2001tp}, or in randomized configurations to simulate a spin glass \cite{RevModPhys.58.801}. Spins with small $S$ and long coherence times, on the other hand, are quantum mechanical in nature and can be exploited to construct exotic quantum phases such as quantum spin liquids \cite{RevModPhys.89.025003} where intricate many-body entanglement exists. A hybrid use of classical and quantum spins adds to the variety of the simulated Hamiltonians. \hfill\break

In the following, we discuss some highlights of quantum simulations using spins on surfaces. Section 5.1 shows simulations using 1D spin chains detected by IETS spectroscopy, and section 5.2 presents some latest simulation results in 2D spin arrays probed by ESR-STM. 




\subsection{1D Spin Chain}
\label{section: 5.1}

Theoretical studies of 1D spin chains date back to 1925, when Ising solved a 1D spin chain with a preferred direction, notwithstanding his failure to detect any spontaneous magnetic order at any nonzero temperature \cite{Ising:1925tw}. In 1928, Heisenberg proposed the spin model that now bears his name \cite{1928ZPhy...49..619H}. Following an initial trial by Bloch \cite{1930ZPhy...61..206B}, in 1931, Bethe devised a famous ansatz for solving the quantum spin-1/2 antiferromagnetic Heisenberg chain, and he identified a disordered ground state at zero temperature and gapless excitations therein \cite{Bethe:1931ux}. 
In the early eighties, Faddeev and Takhtajan revealed that the gapless excitations of a spin-1/2 antiferromagnetic Heisenberg chain are exotic spin-1/2 particles termed spinons \cite{FADDEEV1981375}.
For integer-spin chains, Haldane made a surprising discovery that they act substantially differently than half-integer spins and host gapped rather than gapless excitations \cite{HALDANE1983464,PhysRevLett.50.1153}. 
1D spin chains have piqued the theoretical interest of many researchers until today and are the subject of various analytical and numerical investigations \cite{Schollwock:2008wr}. \hfill\break

Experimentally, research of 1D spin chains in real materials did not take off until the 1970s, when it was discovered that the coupling between magnetic moments in particular crystals can be highly anisotropic. The magnetic properties of these materials can thus be understood as quasi-1D chains or 2D sheets in a 3D crystal, allowing measurements to be made using bulk probes such as thermal transport, neutron scattering, and nuclear magnetic resonance. In many of these compounds, the spin-carrying particles are spin-1/2 Cu$^{2+}$ ions or spin-1 Ni$^{2+}$ ions. Among Cu-based compounds, KCuF$_3$ and Sr$_2$CuO$_3$ realize spin-1/2 antiferromagnetic Heisenberg chains, whereas SrCu$_2$O$_3$ and similar compounds realize spin-1/2 ladders with two or more legs \cite{Schollwock:2008wr}.
Among Ni-based coupounds, Ni(C$_2$H$_8$N$_2$)$_2$NO$_2$(ClO$_4$) and

Ni(C$_5$H$_{14}$N$_2$)$_2$N$_3$(PF$_6$) realize spin-1 Haldane chains but have some anisotropy \cite{Schollwock:2008wr}.  An interesting recent discovery shows quantum entanglement between two spins tens of nanometers apart in a SrCuO-family spin chain material \cite{Sahling:2015uw}, which could be potentially exploited as channels for short-range quantum communications \cite{doi:10.1080/00107510701342313}. A more detailed introduction to quasi-1D magnetic materials can be found in Refs. \cite{Schollwock:2008wr, Vasiliev:2018ue,  doi:10.1080/00018739100101492}. \hfill\break

The advent of local probes such as STM opens up a new avenue for studying 1D spin chains. After early pioneering works on self-assembled spin chains \cite{Gambardella:2002wv}, STM-based atom manipulation enables the construction of highly tunable artificial spin chains that are not limited by thermodynamics. STM-created spins chains on metals or thin insulators have significantly different characteristics because of their distinct coupling to the holding substrates, as we shall discuss next. \hfill\break

\subsubsection{Spin Chains on Metals}

Spins directly sitting on metals, if not fully Kondo screened \cite{Madhavan567, PhysRevB.61.9990}, tend to decohere fast. They can thus be largely regarded as classical magnetic moments. In addition to the relatively weak direct exchange, spins on metals experience indirect RKKY (Ruderman–Kittel–Kasuya–Yosida) coupling due to the mediation by conduction electrons \cite{PhysRev.96.99,10.1143/PTP.16.45,PhysRev.106.893}. The RKKY interaction is often manifested as an isotropic spin interaction that tends to align or anti-align spins depending on their distances (for the anisotropic component, see the next paragraph) \cite{PhysRev.96.99,10.1143/PTP.16.45,PhysRev.106.893}. The RKKY interaction for spins on metals has been determined by building spin pairs of different distances and measuring the interactions using various methods. These methods include quantifying the Kondo resonance lineshapes in Co on Cu(100) \cite{PhysRevLett.98.056601}, measuring the Kondo resonance splitting in Fe on Pt(111) \cite{Khajetoorians:2016ud, Steinbrecher:2018ux}, and acquiring single-atom magnetization curves in Co on Pt(111) \cite{Zhou:2010uy} and Fe on Cu(111) \cite{Khajetoorians:2012us}. The RKKY interaction was used to construct antiferromagnetically coupled spin chains on metals, where an even-odd effect was observed \cite{Khajetoorians:2012us}. Here an odd-length chain with a net magnetic moment can be stabilized by a uniform external field, while an even-length chain flips between two degenerate configurations and requires stabilization by a local magnetic field gradient (typically from a magnetic island) \cite{Meier:2008uz, PhysRevLett.108.197204, Khajetoorians1062}. A neat application along these lines is a spin-based logic unit, where two magnetic islands as inputs are connected to an output spin using antiferromagnetic spin chain leads \cite{Khajetoorians1062}.\hfill\break

In addition to the aforementioned isotropic RKKY interaction, in heavy substrates with significant spin-orbit coupling such as Pt, the metal-mediated spin coupling can also contain a Dzyaloshinskii-Moriya term that tends to create a spin spiral \cite{PhysRevLett.44.1538, Smith1976NewMF}. The isotropic and Dzyaloshinskii-Moriya parts of the RKKY interaction were disentangled in Fe on Pt(111) by comparing the field-dependent Kondo and IETS spectra with theoretical simulations \cite{Khajetoorians:2016ud}. Using the measured interactions, artificial spin chains with spin chirality have been constructed \cite{Steinbrecher:2018ux, PhysRevLett.108.197204,PhysRevLett.112.047204}. This shows the utility of substrates in producing controlled interactions between spins on a surface.
\hfill\break

Besides noble metal substrates, spin chains on superconductors are being heavily investigated for potential usage in topological quantum computing. Readers are referred to dedicated reviews on this topic \cite{Jack:2021vo, PAWLAK20191, RevModPhys.91.041001}. 

\subsubsection{Spin Chains on thin insulators}

\begin{figure}[p]
  \centering
  \includegraphics[width=0.78 \linewidth]{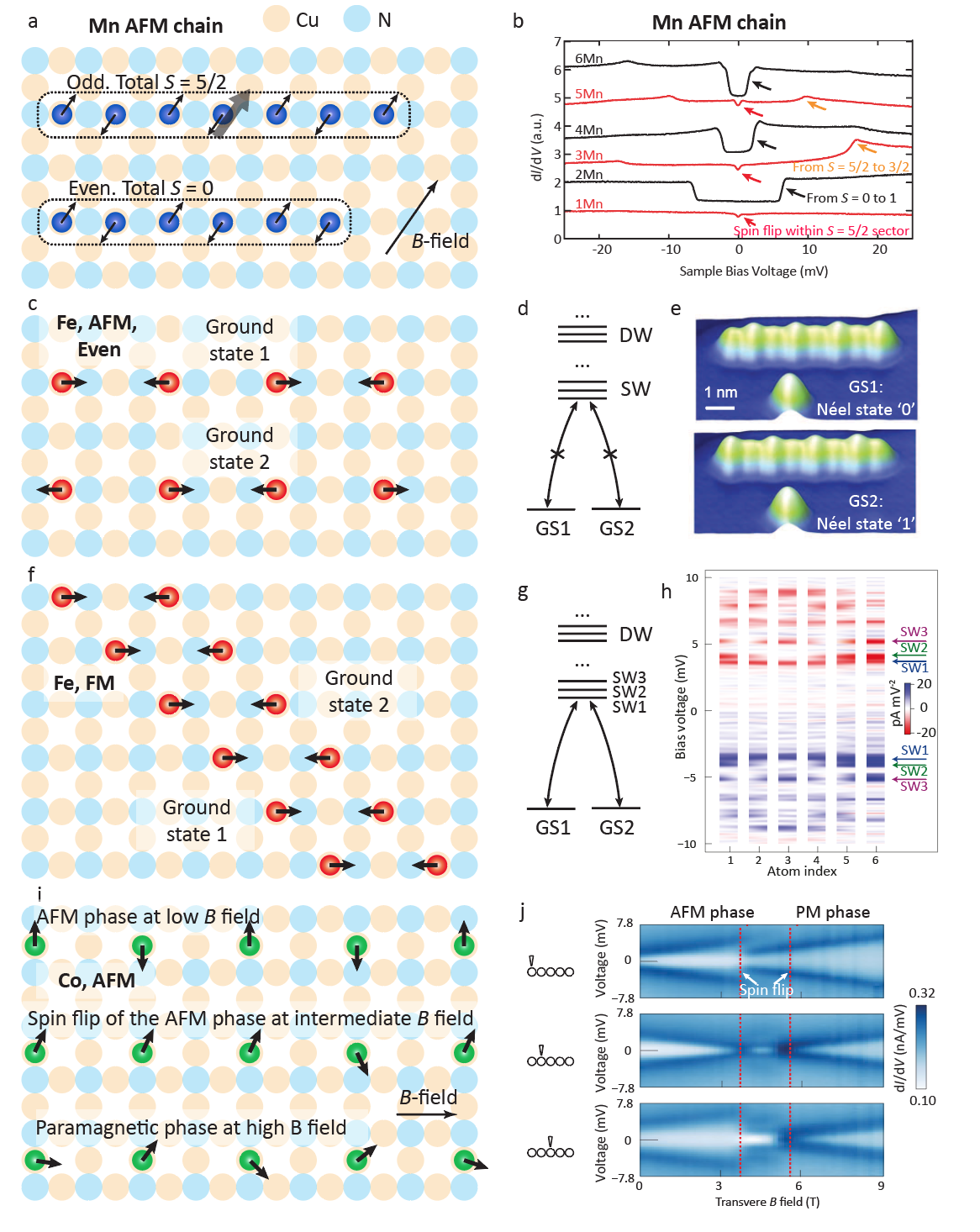}\hspace*{0\textwidth}%
  \caption{Quantum simulation using artificial spin chains on Cu$_2$N. a,b) Simulation of composite quantum magnets using strongly antiferromagnetically (AFM) coupled Mn spins. An odd (even) number of Mn spins form a composite magnet of spin-5/2 (spin-0). In (b), IETS spectra of odd-length (red) and even-length (black) Mn chains show distinct spin excitations, whose nature is determined by their magnetic-field dependence. c--e) Simulation of bistable N\'eel antiferromagnets using AFM-coupled even-length Fe chains. The ground states are bistable because low temperature and low bias voltage quench spin wave (SW) excitations and domain wall (DW) formations as sketched in (d). Panel (e) shows the spin-resolved STM topograph of the bistable N\'eel states. f--h) Simulation of a ferromagnet and visualization of SW excitations using ferromagnetically (FM) coupled Fe chains. The spatially-dependent IETS spectra in (h) can be seen to host zero, one, or two spatial nodes at increasing energies, which correspond to the SW1, SW2, and SW3 excited states, respectively. i,j) Simulation of a transverse-field Ising model using a weakly-AFM-coupled Co chain. The IETS spectra in (j) show a discontinuity with each spin flip, including the final flip that results in a phase transition to the paramagnetic (PM) state. In (i), only one spin configuration out of a superposition state is illustrated, and the transverse $B$-field has been changed to an in-plane direction for illustration purposes. (b) Reproduced with permission.\textsuperscript{\cite{Hirjibehedin1021}} 2006, AAAS. (e) Reproduced with permission.\textsuperscript{\cite{Loth196}} 2012, AAAS. (h) Reproduced with permission.\textsuperscript{\cite{Spinelli:2014vm}} 2014, Springer Nature. (i) Reproduced with permission.\textsuperscript{\cite{Toskovic:2016wg}} 2016, Springer Nature.}
  \label{fig:Cu2N}
\end{figure}

The quantum behavior of artificial spin chains can be harnessed with spins on thin insulators. Except for a recent study that used Ti spins on 2ML MgO on Ag (see the next section and Ref. \cite{Yang:2021aa}), most spin structures on passivated substrates were built with Mn, Co, or Fe atoms on 1ML Cu$_2$N grown on Cu(100). When these magnetic impurities are deposited on Cu$_2$N, they are incorporated into the surface by pushing down the Cu atom beneath and pulling side nitrogen atoms towards the impurities \cite{Hirjibehedin1199}, typically resulting in some in-plane anisotropy (see Table \ref{table:spins}). Mn on Cu$_2$N, however, experiences very small anisotropy owing to the near cancellation of orbital angular momenta in Mn's 3$d^5$ configuration \cite{abragam2012electron, PhysRevB.83.014413}. Figure \ref{fig:Cu2N} summarizes different magnetic phases simulated by spin chains on Cu$_2$N, as we shall discuss next. \hfill\break 

The simulation of composite quantum magnets was carried out using strongly antiferromagnetically coupled spin-5/2 Mn spins 
(Figure \ref{fig:Cu2N}a,b) \cite{Hirjibehedin1021}. Due to the small anisotropy of Mn on Cu$_2$N, the Mn spin chain simulates a nearly isotropic antiferromagnetic Heisenberg Hamiltonian
\begin{equation}
    H = \sum_{i = 1}^{N-1} J \boldsymbol{S}_i \cdot \boldsymbol{S}_{i+1},
    \label{eq: Mnchain}
\end{equation}
where $J =$ 6.2 meV is the dominant energy scale due to the close spacing between Mn in this chain (0.36 nm, sitting on the nearest-neighbor Cu atoms along the N-Mn-N row as shown in Figure \ref{fig:Cu2N}a). Due to the strong antiferromagnetic coupling, the entire chain can be viewed as a composite quantum magnet, as evidenced by the uniformity of IETS spectra along the chain \cite{Hirjibehedin1199}. Odd-length chains thus have the same spin-5/2 as a single Mn atom and share similar IETS steps, where spin flips within the $S = 5/2$ sector can be seen at about zero bias (red arrows in Figure \ref{fig:Cu2N}b) and spin flips from the $S = 5/2$ sector to the $S = 3/2$ sector appear at higher energies (orange arrows in Figure \ref{fig:Cu2N}b). The quantum nature of Mn spin chains is best reflected in even-length chains, where the ground state has no net spin (i.e., $\ket{S = 0, S_z = 0}$) and the lowest excitations are transitions to the triplet states (i.e., $\ket{S = 1, S_z = -1, 0, 1}$) (black arrows in Figure \ref{fig:Cu2N}b). The singlet-triplet transitions of even-length chains can be clearly identified as their IETS steps split into three mini-steps under an external magnetic field (not shown, see Ref. \cite{Hirjibehedin1199}). \hfill\break

Spin-2 Fe atoms on Cu$_2$N can simulate spin Hamiltonians with a preferred spin direction \cite{Hirjibehedin1199}
\begin{equation}
    H = \sum_{i = 1}^{N-1} J \boldsymbol{S}_i \cdot \boldsymbol{S}_{i+1} + \sum_{i = 1}^N [D S_{i, z}^2 + E (S_{i, x}^2 - S_{i, y}^2)  + \mu_\mathrm{B} g \boldsymbol{B}_i \cdot \boldsymbol{S}_i],
    \label{eq: Fechain}
\end{equation}
where $z$ is along the easy-axis anisotropy direction of Fe on Cu$_2$N (the in-plane N-Fe-N bond direction, see Table \ref{table:spins}), and the inter-spin coupling $J$ can be tuned to be either antiferromagnetic or ferromagnetic depending on the chain configuration. Ref. \cite{Loth196} reported the first Fe chains constructed on Cu$_2$N, which simulates a bistable antiferromagnet. As shown in Figure \ref{fig:Cu2N}c, an even number of Fe atoms were placed at 0.72 nm intervals along the N-Fe-N row. This Fe geometry produces an antiferromagnetic coupling of $J =$ 1.2 meV, which is small enough that different Fe spins can be seen as individuals (rather than a combined magnet as in the case of Mn), as evidenced by the position-dependent IETS spectra along the Fe chain \cite{Loth196}. Quite interestingly, as shown in the spin-polarized STM images in Figure \ref{fig:Cu2N}e, two N\'eel-like states (i.e., states that can be represented as $\ket{\uparrow \downarrow \uparrow \downarrow}$ and $\ket{\downarrow \uparrow \downarrow \uparrow}$) were found to be stable in these Fe chains at a low temperature (0.5 K) and a low bias voltage (below 7 mV). Increasing either the temperature or the bias voltage induces switching between these bistable ground states, and the switching rate is also sensitively dependent on the length of the Fe chains \cite{Loth196}. As illustrated in Figure \ref{fig:Cu2N}d, calculations \cite{PhysRevLett.110.087201} suggest three switching regimes that are dependent on the bias voltage: (1) at low bias (below 6 mV), the switching is limited to direct resonant flipping between the two N\'eel states, but at essentially negligible rates (due to their tiny overlap), (2) at intermediate bias (between 6 and 12 mV), inelastic spin-wave-like excitations occur by flipping the spin under the tip, and further relaxations can bring the spin chain to the other N\'eel state, and (3) at high bias (above 12 mV), another kind of inelastic excitation-relaxation process dominates, where domain walls are excited in one N\'eel state, followed by relaxation into the other N\'eel state. \hfill\break

In contrast, \textit{odd-length} Fe chains of a similar geometry exhibit a single stable N\'eel ground state ($\ket{\uparrow \downarrow \uparrow}$) due to the finite Zeeman energy of odd-length spin chains. 
Electrical pump-probe measurements were used to directly probe the $T_1$ relaxation rate from the excited state $\ket{\downarrow \uparrow \downarrow}$ to the ground state $\ket{\uparrow \downarrow \uparrow}$ \cite{Yan:2015tp}. The $T_1$ time was found to have an interesting, nonlinear dependence on the tip-atom separation. At a critical tip-atom separation, the tip's magnetic field counteracts the external field and brings the $\ket{\uparrow \downarrow \uparrow}$ and $\ket{\downarrow \uparrow \downarrow}$ states into degeneracy, which dramatically reduces the spin state lifetime \cite{Yan:2015tp}.  \hfill\break


Different from the previously mentioned antiferromagnetic Fe chains, Fe atoms placed along the [110] direction of Cu$_2$N simulate a \textit{ferromagnet} with $J$ = $-$0.7 meV as illustrated in Figure \ref{fig:Cu2N}f \cite{Spinelli:2014vm}. The IETS spectra of these ferromagnetic Fe chains are highly location dependent, allowing direct real-space identification of the low-lying spin-wave excitations. As shown in Figure \ref{fig:Cu2N}h, the first, second, and third spin-wave excitations carry zero, one, and two spatial nodes along the chain, respectively. Additional measurements and calculations of the switching rates between the two ground states confirmed this interpretation \cite{Spinelli:2014vm}.
\hfill\break

A particularly interesting simulation of a quantum phase transition was performed using spin-3/2 Co atoms on Cu$_2$N. The antiferromagnetic interaction between Co was engineered to be small (0.24 meV, Figure \ref{fig:Cu2N}i) to make the chain susceptible to an external transverse magnetic field \cite{Toskovic:2016wg}. Due to the large easy-plane anisotropy and half-integer spins of Co on Cu$_2$N (Table \ref{table:spins}), the low-lying spin doublets of Co are well separated from higher spin states and can thus be considered as pseudospin-1/2 (similar to Co$^{2+}$-based quasi-1D spin compounds such as CsCoCl$_3$ \cite{Tellenbach_1977}). The low-energy Hamiltonian of a Co chain on Cu$_2$N thus resembles a spin-1/2 transverse-field XXZ Heisenberg model
\begin{equation}
    H = \sum_{i = 1}^{N-1} J_{x} (S_i^{x} S_{i+1}^x + S_i^{y} S_{i+1}^y) + J_z S_i^{z} S_{i+1}^z - \mu_\mathrm{B} B_x g \sum_{i = 1}^{N} S_i^x.
    \label{eq: XXZ}
\end{equation}
Here the $z$ axis is chosen to be along the uniaxial anisotropy direction of Co spin-3/2 (i.e., the hollow-Co-hollow direction of Cu$_2$N, see Table \ref{table:spins}), and this anisotropy term is responsible for the difference between $J_z$ and $J_x$ when spin-3/2 is projected to pseudospin-1/2 \cite{Toskovic:2016wg}. This projection also generates some next-nearest-neighbor coupling terms which we ignored in Equation \ref{eq: XXZ} because they have no effect on the qualitative results \cite{Toskovic:2016wg}. At $B_x = 0$, under the experimental condition $J_z/J_{x} \approx 1/8 < 1$ (which is determined by the chain properties), the XXZ Hamiltonian in Equation \ref{eq: XXZ} exhibits a N\'eel ground state with spins polarized along the $\pm$y directions \cite{PhysRevB.65.172409} (the N\'eel state does not prefer the $x$ direction due to the applied $B_x$ field, nor the $z$ direction due to the weak coupling along it). As the transverse field $B_x$ is increased, as shown in Figure \ref{fig:Cu2N}i,j, spins in the antiferromagnetic state flip one by one, yielding new excitation modes and thus abrupt jumps in IETS spectra. The final spin flip marks the quantum phase transition from the antiferromagnetic phase to a paramagnetic phase, which has a large net magnetization along $x$ but no magnetic order.\hfill\break

Hybrid spin chains using different atomic species have also been constructed. Antiferromagnetic Fe$_n$Mn and Mn$_n$Fe chains terminated by one Mn and one Fe atom, respectively, have been built and measured \cite{PhysRevB.94.085406}. Given the Mn spin of $S = 5/2$ and the Fe spin of $S = 2$, these hybrid chains should host a total spin of $S = 1/2$ for odd $n$'s and hence exhibit a Kondo resonance. On the other hand, IETS measurements show that Mn$_n$Fe, but not Fe$_n$Mn, exhibits the expected Kondo resonance \cite{PhysRevB.94.085406}. This behavior is explained by the much larger anisotropy barrier of Fe on Cu$_2$N (Table \ref{table:spins}), which limits the rate of the chain's spin flip events (that are essential for Kondo screening to occur) \cite{PhysRevB.94.085406}. \hfill\break

Although spin chains on surfaces have primarily been studied using IETS spectroscopy, we anticipate that future experiments using ESR-STM will provide significant advantages in terms of higher energy resolution and coherent controllability. Another intriguing future direction is to build and measure topological integer-spin chains. For a more detailed introduction to STM-based spin chain works, readers are referred to a dedicated review \cite{RevModPhys.91.041001}. 



\subsection{2D Spin Array}
\label{section: 5.2}

2D spin systems with antiferromagnetic coupling are ideal testbeds for quantum magnetism. Unlike in 1D where fractionalization has been rigorously demonstrated, the exploration of exotic ground states and excitations in 2D spin models is still under active development. Certain classes of 2D quantum spin liquids (QSLs) have been demonstrated or gained sufficient credibility \cite{RevModPhys.89.025003, Savary_2016}, including (1) gapped Z$_2$ spin liquids as shown by exactly solving the toric code model \cite{KITAEV20032}, mapping to the known Ising lattice gauge theory \cite{PhysRevB.62.7850,PhysRevLett.86.292}, or numerical simulations of certain resonance valence bond (RVB) states \cite{BASKARAN1987973, PhysRevLett.86.1881}, (2) gapless Z$_2$ spin liquids as shown by exactly solving the Kitaev honeycomb model \cite{KITAEV20062}, and (3) chiral spin liquids that are spin counterparts of the Laughlin fractional quantum Hall states \cite{PhysRevLett.59.2095}. Other classes of 2D QSLs, especially U(1) QSLs predicted with gapless spin-1/2 excitations (spinons) on triangular or Kagome lattices, are difficult to demonstrate either analytically (through field theoretical methods by expressing spins with slave particles) or numerically (through density-matrix renormalization group, quantum Monte Carlo, etc.) \cite{RevModPhys.89.025003}. These challenges highlight the value of quantum simulators in revealing exotic QSL states, particularly U(1) QSLs \cite{2021arXiv210401180S,2021arXiv210404119S}. Here Z$_2$ or U(1) refers to a redundant gauge symmetry of the QSL (known as the invariant gauge group) that results from a slave-particle mean-field treatment and is a conventional way to classify QSLs \cite{Wen:2004wz}.  \hfill\break

\begin{figure}[thb]
  \centering
  \includegraphics[width=1 \linewidth]{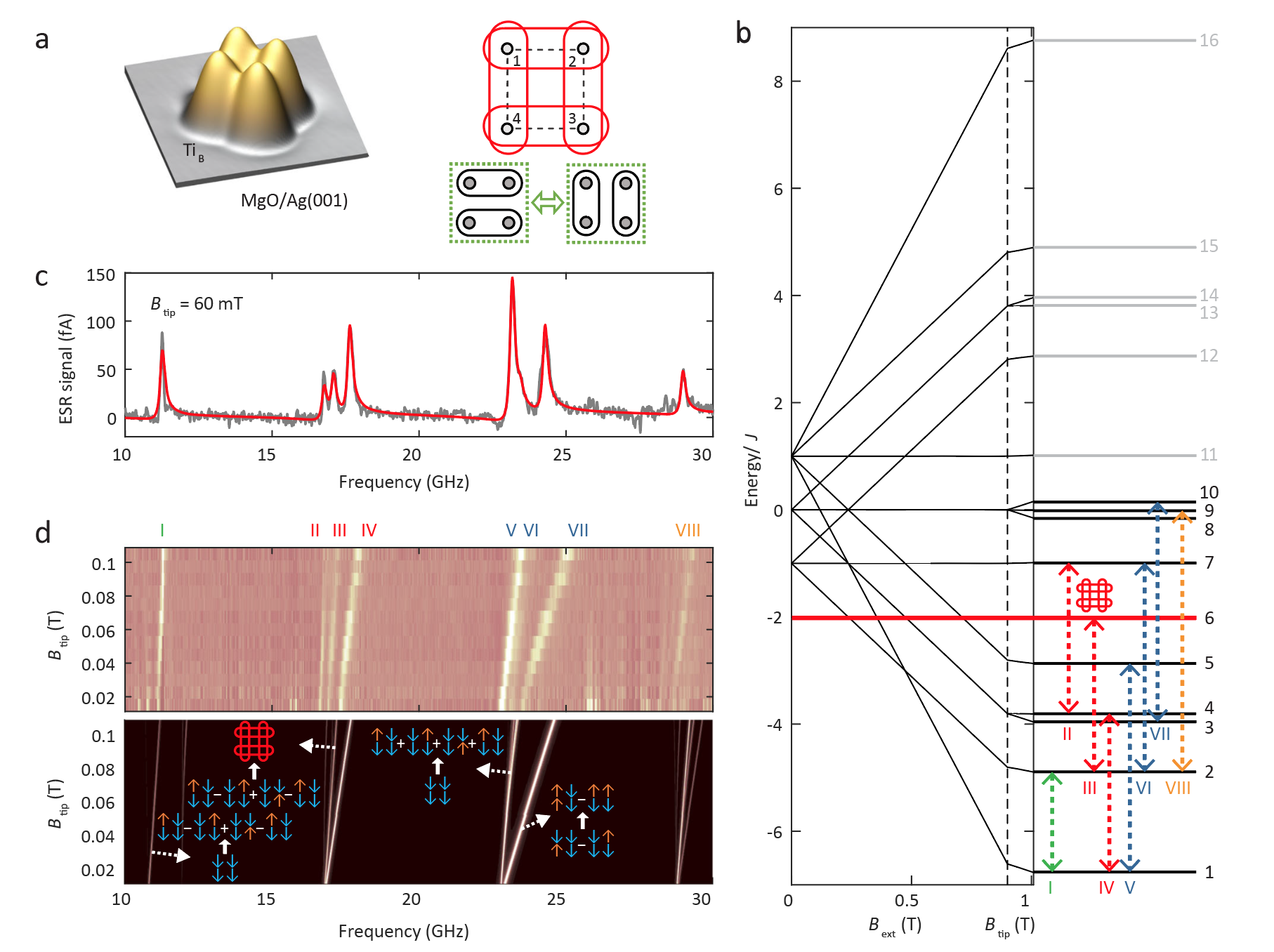}\hspace*{0\textwidth}%
  \caption{Quantum simulation of a resonating valence bond (RVB) state in a 2-by-2 spin plaquette. a) Left: STM image of a spin plaquette made of four bridge-site Ti atoms. Right: An RVB state (red knot) is formed by the superposition of two spin-singlet configurations (green boxes). b) 16 energy levels of the spin plaquette as a function of the external and tip's magnetic fields. The RVB excited state is indicated by the thick solid red line. Dashed arrows represent the measured ESR transitions in (c) and (d). c) An example of measured ESR spectra (on spin $\sharp 1$ with the labelled tip's magnetic field). d) Upper: Evolution of measured ESR intensity as a function of the tip's magnetic field. Lower: calculated ESR intensity with the corresponding transitions marked. Reproduced with permission.\textsuperscript{\cite{Yang:2021aa}} 2021, Springer Nature.}
  \label{fig:RVB}
\end{figure}

Artificial 2D spin structures assembled by STM and detected by ESR provide a new platform for probing elusive many-body magnetic ground states. A recent work \cite{Yang:2021aa} shows a preliminary step towards this goal by constructing a 2-by-2 spin-1/2 array using bridge-site Ti atoms on MgO (with a coupling $J \approx 6$ GHz, Figure \ref{fig:RVB}a). By measuring ESR spectra as a function of the tip's magnetic field and using model calculations, it is possible to distinguish the nature of all excitations as depicted in Figure \ref{fig:RVB}b--d. Importantly, the presence of an RVB excited state (red knot in Figure \ref{fig:RVB}b and d) was shown, and the RVB state becomes the ground state in another 2-by-2 spin-1/2 array with a different spacing \cite{Yang:2021aa}. We anticipate that future quantum simulations using geometrically-frustrated lattices \cite{Khajetoorians:2012us} or innovative measurement schemes will allow for more direct and in-depth investigations of many-body entanglement in correlated 2D spin systems.


\section{Conclusion and Outlook}

By incorporating electron spin resonance capability in spin-polarized scanning tunneling microscopy, quantum states of individual spins on surfaces can be initialized, controlled, and read out. Individual spins and artificial spin structures built atom-by-atom provide new platforms for sensing magnetic interactions, performing quantum operations, and simulating spin Hamiltonians at the atomic scale.
\hfill\break

The subjects presented in this review are nurtured by recent exciting developments in the areas of quantum science, condensed matter physics, and nanoscience. We believe that the quantum behavior of spins on surfaces, in return, can be harnessed to provide new perspectives to these research fields. Here we provide an outlook on some of these possibilities. \hfill\break

\textit{Quantum sensing at the atomic scale} is a promising direction that can be employed in either a sensor-on-a-surface or a sensor-on-a-tip configuration. In the former case, ESR-STM spectroscopy of a local magnetic moment on a surface (from an adatom, a defect, or a dopant) can be used to investigate local magnetic interactions in a variety of material systems. Some particularly interesting systems include novel spin centers in bulk semiconductors \cite{Wolfowicz:2021uc} and 2D materials \cite{Gottscholl:2020vz}, magnetic impurities in strongly correlated materials \cite{RevModPhys.81.45}, and quantum magnets \cite{Diep:2004wv, Vasiliev:2018ue}. A perhaps more ambitious direction is to lift an ESR-active spin to an STM tip, i.e., to form a scanned quantum sensor. The great benefit of this approach would be the combination of the high energy sensitivity of ESR spectroscopy (as demonstrated in scanning nitrogen-vacancy magnetometry \cite{Grinolds:2013wq, RevModPhys.89.035002}) with the Angstrom-scale spatial resolution of STM-based magnetic field imaging \cite{Czap:2019vi,Verlhac:2019tw}. Two challenges in this direction are the identification of appropriate spin carriers capable of maintaining their spin properties on a metallic tip, as well as the development of an appropriate detection scheme. \hfill\break

\textit{A high-spatial-resolution characterization tool for other quantum platforms} is another possible application of ESR-STM by using either spins on surfaces or on a tip. Understanding the microscopic origins of magnetic and electrical noises, for example, is critical for improving the coherence properties of solid-state spin qubits \cite{Wolfowicz:2021uc, RevModPhys.85.961}, superconducting qubits \cite{Singha:2021tj}, and trapped-ion qubits \cite{Siddiqi:2021uh}.
\hfill\break

\textit{Quantum computation based on bottom-up spin structures} created on surfaces is under active development. The advantage of using spins on surfaces is the atomically precise construction of the spin structures with well-controlled, easily-characterizable spin-spin interactions. The challenges, on the other hand, are multi-fold. First and foremost, it has been unclear how multiple spins can be controlled and detected in a tip-based setup. A possible solution to this challenge was presented in section \ref{section: ELDOR}, where a remote spin not in the tunnel junction can still be driven by an single-atom magnet and read out using double resonance spectroscopy through a sensor spin under the tip. We expect that future spin nanodevices will be similarly composed of sensor and qubit spins, where additional detection and driving might be provided by auxiliary contacts in addition to the STM tip, for example, through bottom-up atomic wires \cite{Khajetoorians1062, Weber:2012vg}. The second challenge is the relatively low quantum coherence of spins on surfaces. Aside from the approaches discussed in section \ref{section: 4} such as using the singlet-triplet clock transition, it is critical to investigate different spin systems on different substrates. We anticipate that substrates with a reduced number of low-energy excitations, such as bulk semiconductors, will be beneficial. \hfill\break

\textit{Quantum simulations with artificial spin structures} are another venue for future exploration. Bottom-up construction of 1D chains and 2D frustrated lattices, together with sensitive ESR-STM measurements, can potentially shed light on the nature of exotic quantum ground states and excitations \cite{Khajetoorians:2019wq, Spinelli_2015, Khajetoorians:2012us}. Although so far the use of surface spins for quantum simulation has been limited to studies of stationary eigenstates and eigenenergies, the recently discovered possibility to control spins not in the tunnel junction \cite{2021arXiv210809880P} can in principle allow the use of global unitary transformations of multiple spins and thus simulations of time-evolving many-body quantum states \cite{RevModPhys.86.153}. This type of simulation has proven fruitful in many quantum simulators, including those employing Rydberg atoms \cite{Endres:2016tn,Bernien:2017to,Keesling:2019tp,Bluvstein:2021ta}. Furthermore, mappings from spin to fermionic or bosonic Hamiltonians such as the Jordan–Wigner transformation can allow simulations of a broader class of Hamiltonians in condensed matter physics. \hfill\break


Overall, we believe that the field of surface-based quantum nanoscience is still in its infancy, with many exciting developments on the horizon.

\medskip

\medskip
\textbf{Acknowledgements} \par 
All authors acknowledge support from the Institute for Basic Science (IBS-R027-D1). We thank Robbie J. G. Elbertse for the careful proofreading of the manuscript.

\medskip

%

\bibliographystyle{MSP}
\bibliography{ESR-STM-20211119}





\newpage

\begin{figure}
\centering
  \includegraphics[width=0.2 \linewidth]{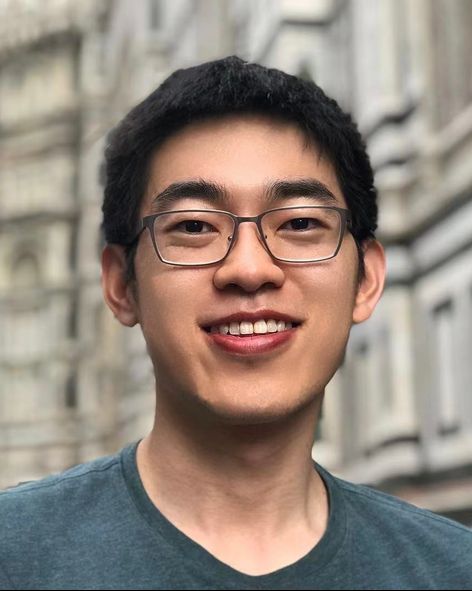}
  \caption*{Yi Chen received his B.S. from Peking University in 2014 and Ph.D. in physics from University of California at Berkeley in 2020 under the supervision of Michael Crommie. He is currently a postdoctoral fellow at Center of Quantum Nanoscience. Yi’s research interests lie in the quantum world, ranging from many-body-entangled quantum materials to few-body-entangled individual quantum systems, as well as their intersections.}
\end{figure}

\begin{figure}
\centering
  \includegraphics[width=0.2 \linewidth]{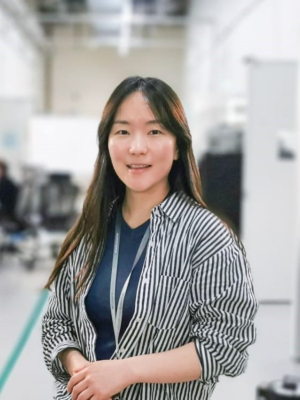}
  \caption*{Yujeong Bae received her Ph. D from Ewha Womans University in 2016. After that, she worked at the Institute of Basic Science Center for Quantum Nanoscience and IBM Almaden Reserach Center as a postdoc for 3 years. She is currently a research professor at Institute of Basic Science and Ewha Womans University. Her research interests include coherent control of spins on surfaces and spin-dependent transport in spintronic devices.}
\end{figure}

\begin{figure}
\centering
  \includegraphics[width=0.2 \linewidth]{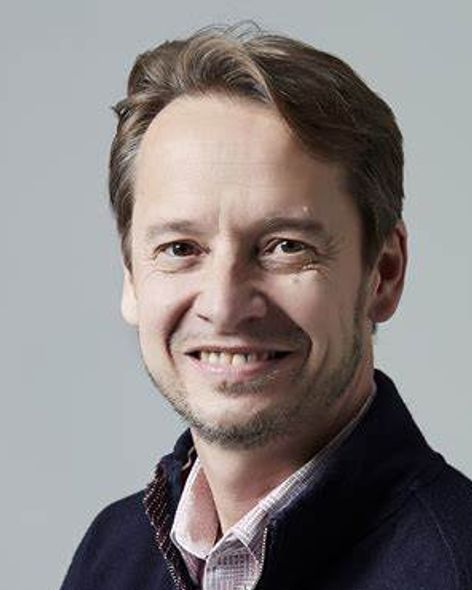}
  \caption*{Andreas Heinrich received his Ph.D. from University	of G\"{o}ttingen in 1998. After that he spent 18 years at IBM Almaden first as a researcher then a group leader. Since 2016 he became the director	of the Institute of Basic Science Center for Quantum Nanoscience and a Distinguished Professor at Ewha Womans University. His present research interests focus on the exploration of quantum functionalities of individual spins on surfaces.}
\end{figure}




\end{document}